\documentclass[twocolumn,aps,prd,amsmath,amssymb,floatfix,showkeys,nofootinbib,superscriptaddress,showpacs]{revtex4}

\usepackage{graphicx}
\graphicspath{{Plots/}}

\def\units#1{\hbox{$\,{\rm #1}$}}
\def\degrees{\hbox{$^\circ$}}

\begin{document}


\title{A model-independent analysis of the Fermi Large Area Telescope gamma-ray data from the Milky Way dwarf galaxies and halo to constrain dark matter scenarios}

\author{M.~N.~Mazziotta}
\email{mazziotta@ba.infn.it}
\affiliation{Istituto Nazionale di Fisica Nucleare, Sezione di Bari, 70126 Bari, Italy}
\author{F.~Loparco}
\email{loparco@ba.infn.it}
\affiliation{Istituto Nazionale di Fisica Nucleare, Sezione di Bari, 70126 Bari, Italy}
\affiliation{Dipartimento di Fisica ``M. Merlin" dell'Universit\`a e del Politecnico di Bari, I-70126 Bari, Italy}
\author{F.~de~Palma}
\email{Francesco.depalma@ba.infn.it}
\affiliation{Istituto Nazionale di Fisica Nucleare, Sezione di Bari, 70126 Bari, Italy}
\author{N.~Giglietto}
\affiliation{Istituto Nazionale di Fisica Nucleare, Sezione di Bari, 70126 Bari, Italy}
\affiliation{Dipartimento di Fisica ``M. Merlin" dell'Universit\`a e del Politecnico di Bari, I-70126 Bari, Italy}
  
\date{\today}

\begin{abstract}

We implemented a novel technique to perform the collective spectral analysis of sets of multiple gamma-ray point sources using the data collected by the Large Area Telescope onboard the Fermi satellite. The energy spectra of the sources are reconstructed starting from the photon counts and without assuming any spectral model for both the sources and the background. In case of faint sources, upper limits on their fluxes are evaluated with a Bayesian approach. This analysis technique is very useful when several sources with similar spectral features are studied, such as sources of gamma rays from annihilation of dark matter particles. We present the results obtained by applying this analysis to a sample of dwarf spheroidal galaxies and to the Milky Way dark matter halo.
The analysis of dwarf spheroidal galaxies yields upper limits on the product of the 
dark matter pair annihilation cross section and the relative velocity of annihilating particles that 
are well below those predicted by the canonical thermal relic scenario in a mass range 
from a few $\units{GeV}$ to a few tens of $\units{GeV}$ for some annihilation channels.

\end{abstract}



\maketitle

\section{Introduction}

Milky Way dwarf spheroidal (dSph) galaxies are candidate targets for dark matter (DM) studies through annihilation signatures. This is because their mass-to-light ratio is predicted to be of the order of $10-10^{3}$~\cite{wolf, simon}, implying that they could be largely DM dominated. Moreover, since no significant gamma-ray emission of astrophysical origin is expected (these systems host few stars and no hot gas), the detection of a gamma-ray signal could provide a clean DM signature. 

The Milky Way halo is another promising candidate for DM searches. An approach to search for DM emission from annihilation in the Galactic halo is to study the gamma-ray flux from sky positions distant from known astrophysical gamma-ray sources. The diffuse emission from unresolved sources and from the interaction of charged particles with the interstellar medium constitutes a background for this approach.

Weakly Interacting Massive Particles (WIMPs) have long been considered as well-motivated 
candidates for DM that could contribute to the $80\%$ of the non-baryonic mass density in the universe~\cite{wmap}. 

At a given energy $E$, the differential gamma-ray flux $\Phi_{\gamma}(E, \Delta \Omega)$ (in units of $\units{photons~cm^{-2}~s^{-1}~GeV^{-1}}$) from WIMP annihilation in a region covering a solid angle $\Delta \Omega$ and centered on a DM source, can be factorized as~\cite{charb}:

\begin{equation}
\Phi_{\gamma}(E, \Delta \Omega) = J(\Delta \Omega) \times \Phi^{PP}(E)
\label{eq:DMflux}
\end{equation}
where $J(\Delta \Omega)$ (in units of $\units{GeV^{2}~cm^{-5}~sr}$) is the {\em ``astrophysical factor''} (hereafter, $J$-factor), i.e., the line of sight (l.o.s.) integral of the  DM density squared in the direction of observation over the solid angle $\Delta \Omega$:

\begin{equation}
J(\Delta \Omega) = \int_{\Delta \Omega} d\Omega \int_{l.o.s.} dl \rho^{2}(l, \Omega)
\label{eq:jfactor}
\end{equation}

The term $\Phi^{PP}(E)$ (in units of $\units{GeV^{-3}~cm^{3}~s^{-1}~sr^{-1}}$) is the {\em ``particle physics factor''},
that encodes the particle physics properties of the DM, and for a given WIMP mass $m_{\chi}$ is given by: 

\begin{equation}
\Phi^{PP}(E) =  
\frac{1}{2} \frac{\langle \sigma v \rangle}{4 \pi m^{2}_{\chi}} \sum_{f} N_{f}(E, m_{\chi}) B_{f}
\label{eq:PPfactor}
\end{equation}
where $\langle \sigma v \rangle$ is the WIMP pair annihilation cross section times the relative velocity of the two annihilating particles, while $B_f$ and $N_{f}(E, m_{\chi})$ are respectively the branching ratio and the differential photon spectrum of each pair annihilation final state $f$. 

We note that the particle physics factor in Eq.~\ref{eq:DMflux} is independent of the spatial distribution of the DM, and hence independent of the particular DM source under investigation. Eq.~\ref{eq:DMflux} can be rewritten as:

\begin{equation}
\Phi^{PP}(E) = \frac{1}{J(\Delta \Omega)} \Phi_{\gamma}(E, \Delta \Omega) 
\label{eq:DMflux2}
\end{equation}
showing that the ratio between the photon flux and the $J$-factor is expected to be independent on the source if the DM annihilation mechanism is the same for all the sources. Starting from a measurement of the gamma-ray flux from a candidate DM source, if the $J$-factor is known, Eq.~\ref{eq:DMflux2} allows us to obtain a measurement of the particle physics factor. If the kinematic terms of the summation in Eq.~\ref{eq:PPfactor} are known, this measurement will yield an estimate of $\langle \sigma v \rangle$ as a function of $m_{\chi}$. Moreover, since $\Phi^{PP}(E)$ is independent of the source, the results from individual sources can be combined, thus improving the measurement.

Recently two analysis approaches were developed to constrain DM models using the Fermi LAT data~\cite{lat_DMdwarf_paper, geringer_DMdwarf}. In Ref.~\cite{lat_DMdwarf_paper}, a binned Poisson likelihood fit was used to fit both the spatial and the spectral information for the reconstructed photon events collected in a sky region around the target source. The data from 10 dSphs were also combined using a joint likelihood analysis that takes into account the uncertainties on the $J$-factors. The upper limits on 
$\langle \sigma v \rangle$ were evaluated implementing an approach based on a profile likelihood function, that incorporates 
the nuisance parameters. In Ref.~\cite{lat_DMdwarf_paper} a two-year photon data sample was analyzed with the P6\_V3\_DIFFUSE Instrument Response Functions (IRFs) in the energy range from $200 \units{MeV}$ to $100 \units{GeV}$. In Ref.~\cite{geringer_DMdwarf} a three-year photon data sample was analyzed with the P7SOURCE\_V6 IRFs in the energy range from $1$ to $100 \units{GeV}$. Photons from a sky region with an angular radius of $0.5\degrees$ from each dSph were selected and the background was evaluated by sampling positions within an angular distance of $10\degrees$ from each dSph and counting the number of events in a cone of $0.5\degrees$ angular radius. The upper limits were evaluated using a fully frequentist approach that takes in account the different $J$-factors of each source. The authors also took the uncertainties on the $J$-factor into account with a semi-Bayesian approach as well. The results of these two analyses were used to set upper limits on the annihilation cross section below the canonical value for the thermal relic WIMP scenario of $3 \times 10^{-26} \units{cm^{3}~s^{-1}}$~\cite{jungman,wmap} up to masses of about $30\units{GeV}$ for the $ b \bar{b} $ and $\tau^+ \tau^-$ channels. This limit could represent a serious challenge to the conventional WIMP dark matter hypothesis.


In this work we present the results obtained with a model-independent data analysis method~\cite{poster} applied to DM searches. This
method can be applied to the analysis of individual sources (Sect.~\ref{sec:individualanalysis}) as well as to the combined analysis of multiple sources (Sect.s \ref{sec:stacking} and \ref{sec:composite}), and does not introduce degrees of freedom in the calculation of confidence intervals on the parameters in the DM model.
The first step of the analysis is the evaluation of upper limits on the possible gamma-ray signal events. This calculation is
performed by properly choosing, for each source, a signal and a background region (see Sect.s \ref{sec:ana_meth} and \ref{sec:anadwarfs}) and following a Bayesian approach to evaluate upper limits on the signal counts. In this way no models are required for the source and for the background. Moreover, the effects of systematic uncertainties can be easily taken into account by integrating over a nuisance parameter (either the J-factor or the effective area) the posterior probability distributions (Sect. \ref{sec:SistErr}). Finally, the upper limits on the photon counts can then be converted into upper limits on $\Phi^{PP}(E)$, and consequently on $\langle \sigma v \rangle$, once a DM model has been implemented. In the present analysis we used a sample of gamma-ray data collected by the Fermi LAT during its first 3 years of operation in survey mode. The data were analyzed using the most recent LAT IRFs (P7SOURCE\_V6 and P7CLEAN\_V6). Candidate photons converting in both the front and back part of the instrument in the energy range from $562 \units{MeV}$ to $562 \units{GeV}$ were used for the analysis. Upper limits on $\langle \sigma v \rangle$ as a function of $m_{\chi}$ were obtained from the analysis of individual dSph galaxies (Sect. \ref{sec:anadwarfs}) and from their combined analysis, as well as from the analysis of the Milky Way Halo (Sect. \ref{sec:halo}).

\section{The instrument and the data}
\label{sec:data}

The LAT is a pair-conversion gamma-ray telescope designed to measure gamma rays in the energy range from $20\units{MeV}$ to more than $300\units{GeV}$. In this paper a brief description of the LAT is given, while full details can be found in~\cite{Atwood2009}.

The LAT is composed of a $4 \times 4$ array of $16$ identical towers designed to convert incident gamma-rays into $e^{+} e^{-}$ pairs, and to determine their arrival directions and energies. Each tower hosts a tracker module and a calorimeter module. Each tracker module consists of $18$ x-y planes of silicon-strip detectors, interleaved with tungsten converter foils, for a total on-axis thickness equivalent to $1.5$ radiation lengths (r.l.). Each calorimeter module, $8.6$ r.l. on-axis thick, hosts $96$ CsI(Tl) crystals,
hodoscopically arranged in $8$ perpendicular layers. The instrument is surrounded by a segmented anti-coincidence detector that tags the
majority of the charged-particle background. 

A sample of gamma-ray data collected by the Fermi LAT during its first three years of operation in survey mode was used for this analysis, overlapping substantially with the data used for the second LAT source catalog~\cite{2FGL}. The Pass7 IRFs~\cite{Pass7} event selection cuts (for SOURCE and CLEAN event classes), with candidate photons converting in both the front and back parts of the instrument, were used. To avoid contamination from the bright limb of the Earth, data taken during any time period when the angular separation of a cone of $10\degrees$ angular radius centered on the source direction with respect to the Zenith direction exceeded $105\degrees$ were discarded, as well as data taken during any time period when the LAT rocked to an angle exceeding $52\degrees$. The data taken during time periods when the source was observed with an off-axis angle larger than $66.4\degrees$ were also discarded.

We performed the spectral analysis using the internal LAT Collaboration software package \textit{FermiUnfolding}~\cite{unf0, unf1, unf2}, which enables gamma-ray spectra to be reconstructed without assuming any model for the sources or the background. The data analysis was performed selecting gamma rays with energies from $562\units{MeV}$ to $562\units{GeV}$. The energy interval was divided into $12$ bins, equally spaced on a logarithmic scale ($4$ bins per decade). We emphasize that, to take energy dispersion into account, in the unfolding approach there is a distinction between the observed photon energies and the true ones. The relationship between observed and true energy is expressed in terms of a smearing matrix, which represents the IRF and is evaluated by means of a full Monte Carlo simulation~\cite{Atwood2009}.

\section{Analysis methods}
\label{sec:ana_meth}

\subsection{Study of individual sources}
\label{sec:individualanalysis}

For each individual source a signal region and a background region were defined. The signal region, in which gamma rays emitted from the source are expected, was defined as a cone of a given angular radius, centered on the nominal position of the source. On the other hand, the background region was usually defined as an annulus centered on the source position and external to the signal region. To rule out possible contaminants in the background evaluation, when defining the background regions all the sources in the 2FGL catalog~\cite{2FGL} were masked. The values of the angular radii adopted in this analysis to define the signal and background regions, as well as for masking the 2FGL catalog sources, are given in \S~\ref{sec:anadwarfs} and in \S~\ref{sec:halo}.

Since the possible gamma-ray signal is expected to be faint, in each energy energy bin we set upper limits on the signal counts. The evaluation of the upper limits was performed following the Bayesian approach illustrated in Ref.~\cite{lopmaz}. Following the notation of Ref.~\cite{lopmaz}, we indicate with $n$ and $m$ the number of photons detected in a given energy bin in the signal and background regions, respectively (in the following, to keep the notation simple, we will suppress the energy dependence of these variables).  We assume that the probabilities of measuring the pair ($n$,$m$) are both Poissonian with expectation values $s+cb$ and $b$, respectively, where $s$ is the expectation value of the signal counts (in the signal region), $b$ is the expectation value of the background counts (in the background region) and $c$ is defined as:

\begin{equation}
c = \frac{\Delta \Omega_{s}}{\Delta \Omega_{b}} 
\label{eq:cdef}
\end{equation} 
where $\Delta \Omega_{s,b}$ are the solid angles of the signal and of
the background regions respectively. In principle the definition of $c$ in the
previous equation should include the livetime ratio $T_{s}/T_{b}$,
where $T_{s,b}$ are respectively the livetimes of the signal and of
the background regions. However, since the data selection cuts
illustrated in \S\ref{sec:data} are performed on a cone of $10\degrees$
angular radius centered on the source, and since the outer radius of
the background annulus used for the present analyses is always less than
$10\degrees$ (see \S\ref{sec:anadwarfs}), the livetime ratio
$T_{s}/T_{b}$ is always equal to $1$.

The posterior probability density function (PDF) of the signal counts $s$ was calculated assuming a uniform prior PDF for both $s$ and $b$ and is given by~\cite{lopmaz}:

\begin{equation}
p(s | n, m) = \sum_{k=0}^{n} a_{k} s^{k} \textrm{e}^{-s}
\label{eq:snglsignalpdf}
\end{equation}
with the coefficients $a_{k}$ defined as:

\begin{equation}
a_{k} = \frac{1}{\mathcal{N}} \frac{\Gamma(m + n - k + 1)}{\Gamma(k+1)\Gamma(n - k + 1)} 
\left( \frac{c}{c+1} \right)^{n - k} 
\label{eq:coeffpdf}
\end{equation}
where $\mathcal{N}$ is a normalization constant.

In case of the absence of a background ($c=0$, $m=0$), the posterior PDF on the signal reduces to~\cite{lopmaz}:

\begin{equation}
p(s | n) = 
\frac{s^n \textrm{e}^{-s}}{\Gamma(n + 1)}. 
\label{eq:nobkgsnglsignalpdf}
\end{equation}

The upper limit on the signal counts $s_u$ at the confidence level (or credibility level, CL) $1-\alpha$ was evaluated by numerically solving the integral equation:

\begin{equation}
\int_{0}^{s_u} p(s | n, m) ds = 1-\alpha .
\end{equation}

The upper limits on the signal counts were finally converted into upper limits on the flux by means of the unfolding procedure described in~\cite{unf0, unf1, unf2}. The smearing matrix associated with each sky direction was built by taking into account the pointing history recorded by the LAT~\cite{unf0} and was evaluated from the Monte Carlo simulation of the LAT.

The measured upper limits on the flux were then converted into upper limits on $\langle \sigma v \rangle$. From Eqs.~\ref{eq:PPfactor} and~\ref{eq:DMflux2} it follows that:

\begin{equation}
\langle \sigma v \rangle = \frac{1}{J(\Delta \Omega)} 
\Phi_{\gamma}(E,\Delta \Omega) \times
\frac{8 \pi m^{2}_{\chi}}{\sum_{f} B_{f} N_{f}(E, m_{\chi})}.
\end{equation}
For each value of $m_{\chi}$ the conversion of the limits on the gamma-ray flux into limits on $\langle \sigma v \rangle$ was performed by requiring that the flux predicted from the model must not exceed the measured photon flux in any energy bin. The expected gamma-ray flux from the DM annihilation channels was evaluated as a function of energy using the DMFIT package~\cite{dmfit} based on DarkSUSY~\cite{darksusy}, as implemented in the LAT Science Tools~\cite{ST}. For large DM masses (around or above $1 \units{TeV}$), the radiation of soft electroweak bosons leads to additional gamma rays in the energy range of relevance for the present analysis 
(see e.g.~\cite{bell,ciaf}). This emission mechanism is not included in the DMFIT package. Therefore the present analysis provides conservative upper limits on $\langle \sigma v \rangle$. 

\subsection{Stacking analysis}
\label{sec:stacking}

According to Eq.~\ref{eq:DMflux2}, the particle physics factor is independent of the source under investigation. This feature suggests the possibility of combining the data from all individual sources in order to improve the constraints on the DM models. 

Once the individual sky directions were analyzed, a stacking analysis was performed. In this case the counts from the signal and background regions corresponding to each source were added, and the upper limits on the signal were evaluated following the same procedure as for individual sources.

In order to implement the same analysis procedure as for individual sources, in the stacking analysis the ratio between the signal and background regions was defined as: 

\begin{equation}
c = \frac{\sum_{i} \Delta \Omega_{si} T_{si}}{\sum_{i} \Delta \Omega_{bi} T_{bi}}
\label{eq:cdef0}
\end{equation}
where $\Delta \Omega_{si,bi}$ and $T_{si,bi}$ are respectively the
solid angles and the livetimes of the signal and background regions 
of the $i$-th source ($T_{si}=T_{bi}$ according to the discussion in 
\S\ref{sec:individualanalysis}).  

We note that a more detailed statistical
analysis (see the discussion in Appendix~\ref{app:cdef}) shows that the coefficient 
$c$ should be defined as: 
\begin{equation}
c = \frac{\sum_i c_{i} ( m_{i} + 1)}{\sum_i (m_{i} + 1)}
\label{eq:cdef1}
\end{equation} 
where $c_{i}$ is the coefficient defined in Eq.~\ref{eq:cdef} for the
$i$-th source and $m_{i}$ are the counts in the background region of
the $i$-th source.

If the coefficient $c$ is defined as in Eq.~\ref{eq:cdef0}, its value
will depend only on the extensions (solid angles) of the signal 
and background regions and on the livetimes of the stacked sources. 
On the other hand, if $c$ is defined as in Eq.~\ref{eq:cdef1}, its
value will also depend on the data (counts in the individual
background regions). In the stacking analysis of the dSph galaxies 
we evaluated the coefficients $c$ using both the definitions in Eq.~\ref{eq:cdef0} 
and Eq.~\ref{eq:cdef1}. We found that the differences in the values of $c$
obtained implementing the two different definitions were negligible in
all the energy bins.

The different exposures of the individual sources were also taken into account in the evaluation of the smearing matrix \cite{unf0}, which was performed by stacking the pointing histories of all the sources. This procedure is equivalent to stacking the events from each sky direction on top of one another and then analyzing the resulting image (as an example see the last panels in Figs.~\ref{fig:countmaps} and \ref{fig:countdist} for the case of the dSph analysis). 

Since the sources may be seen by the instrument with different exposures, the $J$-factor value used in the stacking analysis was defined as the average of the $J$-factors of individual sources, each one weighted with its exposure in the whole energy interval under investigation. In principle, different $J$-factors should be determined for each energy bin, with each one evaluated taking into account the exposures in the corresponding bin. We performed this calculation in the case of the dSph galaxies, and the differences between the $J$-factors evaluated using the exposures in individual energy bins with respect to the $J$-factor evaluated using the overall exposure were less than $1\%$. Since these differences are small, we decided to use the same $J$-factor for the whole energy interval, evaluated using the overall exposure. In this way we also avoided introducing an energy dependence of the $J$-factor that may seem unphysical since, according to Eq.~\ref{eq:jfactor}, the $J$-factor is determined only by the DM density profile.

The upper limits on $\langle \sigma v \rangle$ were evaluated in the same way as for individual sources.

\subsection{Composite analysis}
\label{sec:composite}

In the previous analysis the events from all the sources were stacked. This is equivalent to considering the set of all the sources as a single source with a $J$-factor given by the average value weighted with the exposures of all the sources. In the stacking method all the sources are treated in the same way, and photons from a source with a small $J$-factor are considered as likely to originate from DM as photons from a source with a large $J$-factor. However, in the absence of a clear gamma-ray signal,
i.e., if the counts in the signal region $n$ are compatible with the expected background ($n \approx c~m$), a source with a higher $J$-factor will yield a lower upper limit with respect to a source with a lower $J$-factor. The ``DM sensitivity'' of each source is therefore determined by its $J$-factor. 

To account for the different sensitivities of each observation we developed a composite analysis approach that combines the results from all the sources taking into account the individual $J$-factors. Unlike the approaches discussed in sections~\ref{sec:individualanalysis} and~\ref{sec:stacking}, for simplicity, in this approach we do not treat the energy resolution. Since the $68\%$ containment of the energy resolution of the LAT in the energy range chosen for the present analysis is less than $15\%$~\cite{Pass7}, we expect that neglecting the energy dispersion in the evaluation of the flux could yield a similar uncertainty. Indeed we verified, in the case of dSph galaxies, that the differences between the fluxes reconstructed either neglecting or taking into account the energy dispersion are of the order of a few percent in the whole energy range of the analysis.  

Indicating with $s_i$ the expected signal counts from the $i$-th source in the energy interval $[E,E+\Delta E]$ (again, for simplicity, we will suppress the energy dependence of these variables), it is possible to define the random variable $u$ as:

\begin{equation}
u = \eta_{i} s_{i} 
\end{equation}
with the factor $\eta_{i}$ defined as:

\begin{equation}
\eta_{i} = \frac{1}{J_{i} \mathcal{E}_{i}(E) \Delta E }
\end{equation}
where $J_{i}$ is the $J$-factor of the $i$-th source and $\mathcal{E}_{i}(E)$ is its exposure in the energy bin $[E,E+\Delta E]$, which is given by:

\begin{equation}
\mathcal{E}_{i}(E) = \int dt f_{LT}(t) A_{i}(E,t) 
\label{eq:exp}
\end{equation}
where $A_{i}(E,t)$ is the effective area and $f_{LT}(t)$ is the livetime fraction. The dependence on $t$ in Eq.~\ref{eq:exp} indicates that the aspect angles (off-axis and azimuthal angles in the instrument frame) corresponding to the given sky direction (source) are changing with time.

Since $u$ is equal to $(1/J)\Phi_{\gamma}(E, \Delta \Omega)$, and hence to $\Phi^{PP}(E)$, it is independent of the particular source under investigation. 

A set of PDFs for the random variable $u$ can be evaluated starting from the data of each source. Indicating with $n_{i}$ and $m_{i}$ the counts in the signal and background regions of the $i$-th source, the PDF for $s_{i}$ is given by Eq.~\ref{eq:snglsignalpdf}, which can be rewritten explicitly indicating the source index as: 
 
\begin{equation}
p_{i}(s_{i} | n_{i}, m_{i}) = \sum_{k=0}^{n_{i}} a_{ik} s_{i}^{k} \textrm{e}^{-s_{i}}
\label{eq:snglsignalpdf2}
\end{equation}
with the coefficients $a_{ik}$ defined as in Eq.~\ref{eq:coeffpdf}.

The $i$-th PDF for the variable $u$ can be derived from Eq.~\ref{eq:snglsignalpdf2}, and is given by:

\begin{equation}
p_{i} (u|n_{i},m_{i}) = \textrm{e}^{-u/\eta_{i}} \sum_{k=0}^{n_{i}} b_{ik} u^{k} 
\end{equation}
with the coefficients $b_{ik}$ defined as:

\begin{equation}
b_{ik} = \frac{a_{ik}}{\eta_i^{k+1}}.
\end{equation}

To combine a set of $N$ sources we build the likelihood function:

\begin{equation}
\label{eq:like1}
\mathcal{L}(u | n_{1}, m_{1}; n_{2}, m_{2}; \ldots ; n_{N}, m_{N}) =
\prod_{i=1}^{N} p_{i} (u | n_{i}, m_{i} ).
\end{equation} 
Expanding the calculations in the previous equation, the final expression of the likelihood function is given by:

\begin{equation}
\label{eq:like2}
\mathcal{L}(u) = \textrm{e}^{-u/\eta} \sum_{k=0}^{n_{max}} f_{k} u^{k}
\end{equation}
where $\eta$ and the set of coefficients $f_{k}$ are defined as follows:

\begin{equation}
1 / \eta = \sum_{i=1}^{N} \frac{1}{\eta_{i}}
\label{eq:jtot}
\end{equation}

\begin{equation}
f_{k} = \sum_{\substack{k_1, k_2, \ldots k_N \\ k_1 + k_2 + \ldots k_N = k}} \prod_{i=1}^{N} b_{ik_{i}}.
\end{equation}
In the summation of Eq.~\ref{eq:like2} the maximum value of $k$ yielding a non-zero coefficient $f_k$ is $n_{max}=n_1+n_2+\ldots+n_{N}$.

In case of the absence of a background the expression of the likelihood function becomes simpler. Starting from Eq.~\ref{eq:nobkgsnglsignalpdf}, it is straightforward to show that the expression of the likelihood function is given by:

\begin{equation}
\label{eq:like3}
\mathcal{L}(u) = \textrm{e}^{-u/\eta} u^{n} \prod_{i=1}^{N} \frac{1}{\eta_{i}^{n_{i}+1} \Gamma(n_{i}+1)}
\end{equation}
where 

\begin{equation}
n = \sum_{i=1}^{N} n_{i}.
\end{equation}

The likelihood function obtained from Eq.~\ref{eq:like2} (or Eq.~\ref{eq:like3}) is not normalized because, having assumed that $u=s_{i}/\eta_{i}$ is independent of the source under investigation, the measurements $(n_{i}, m_{i})$ are not independent of each other. To ensure normalization, the function $\mathcal{L}(u)$ must be multiplied by a constant $\mathcal{A}$, which in the general case of Eq.~\ref{eq:like2} is given by:

\begin{equation}
\mathcal{A} = \cfrac{1}{\sum \limits_{k=0}^{n_{max}}\cfrac{f_k \Gamma(k+1)}{\eta^{k+1}}}.
\end{equation}

Once the likelihood function is normalized, the upper limit $u^{*}$ at the CL $1-\alpha$ can be evaluated by numerically solving the equation:

\begin{equation}
\int_{0}^{u^{*}} \mathcal{A} \mathcal{L}(u)du = 1-\alpha.
\label{eq:uldef}
\end{equation}

\subsection{Systematic uncertainties}
\label{sec:SistErr}

Systematic uncertainties on the $J$-factor as well as on the effective area can be taken into account in the above procedures introducing a nuisance parameter in the definition of the random variable $u$. In the following we will illustrate the calculations to take into account the systematic uncertainties on the $J$-factors; the mathematical formalism used in the calculations to take into account systematic uncertainties on the effective area is similar so we do not present it here. 

Similar to the approach in \S\ref{sec:composite}, it is possible to define the random variable $u$ as:

\begin{equation}
u = \rho_{i} \frac{s_{i}}{J_{i}}
\end{equation}
where the the factor $\rho_{i}$ is defined as: 

\begin{equation}
\rho_{i} = \frac{1}{\mathcal{E}_{i}(E) \Delta E}. 
\end{equation}
Unlike in \S\ref{sec:composite}, in this case the dependence of $u$ on $J_{i}$ is written explicitly in order to take fluctuations in $J_{i}$ into account.

The posterior PDF for $u$ can be obtained starting from the joint PDF $p_{i}(s_{i},J_{i})$ for $s_{i}$ and $J_{i}$ as:

\begin{equation}
p_{i}(u) = \frac{1}{\rho_{i}} \int J_{i} ~ p_{i}\left( J_{i} u/ \rho_{i}, J_{i}\right) ~ dJ_{i}.
\end{equation}

Since $s_{i}$ and $J_{i}$ are independent random variables, their joint PDF can be factorized, and the previous equation rewritten as:

\begin{equation}
p_{i}(u) = \frac{1}{\rho_{i}} \int J_{i} ~ p_{i}(J_{i} u/\rho_{i}) ~ p_{i}(J_{i}) ~ dJ_{i}.
\end{equation}
where the PDF $p_{i}(s_{i})$ is given by Eq.~\ref{eq:snglsignalpdf2}.

To make the calculation simpler, for the $J$-factors a uniform PDF in the range $[J_{i1}, J_{i2}]$ is assumed, i.e. $p_{i}(J_{i})=1/\Delta J_{i}$. Introducing these PDFs in the previous equation, the posterior PDF for $u$ is given by:

\begin{equation}
p_{i}(u) = \frac{1}{\rho_{i} \Delta J_{i}} \sum_{k=0}^{n} a_k \int_{J_{i1}}^{J_{i2}} J ~ \left( \frac{uJ}{\rho_{i}} \right)^k \textrm{e}^{-uJ/\rho_{i}} ~ dJ.
\label{eq:sysJ}
\end{equation}

The upper limit $u^{*}$ at the CL $1-\alpha$ is evaluated by numerically solving the integral equation:

\begin{equation}
\int_{0}^{u^{*}} p_{i}(u) du = 1-\alpha
\end{equation}
where $p_{i}(u)$ is given by Eq.~\ref{eq:sysJ}. 

A similar approach is implemented to evaluate the effects of the
systematic uncertainties on the effective area of the instrument. In
this case the effective area is treated as a uniformly distributed
random variable, while the $J$-factor is assumed to be known. 

For the stacking analysis, the same procedure was implemented as for the analysis of individual sources. In this case the PDF for $u$ was obtained starting from the cumulative signal and background counts, as in \S\ref{sec:stacking}. 

In the case of the composite analysis, the likelihood function was built by multiplying all the individual PDFs $p_{i}(u)$ in Eq.~\ref{eq:sysJ}, and then the upper limits on $u$ were evaluated as discussed in \S\ref{sec:composite}.

\section{Analysis of the Dwarf Spheroidal Galaxies}
\label{sec:anadwarfs}

This analysis was performed using P7SOURCE\_V6 class events. For
each source, events within a cone of $10\degrees$ angular radius
centered on the nominal position of the source were selected. Again we note that, because of the selection cuts described in
\S\ref{sec:data}, all sky directions within $10\degrees$ from the
source will have the same live time. 

The positions of the dSph galaxies considered in the present analysis 
and the corresponding values of their $J$-factors are reported in Tab.~\ref{tab:dwarfs}, which is taken
from Ref.~\cite{lat_DMdwarf_paper}. These dSph galaxies are not included in 
the second catalog of the Fermi LAT~\cite{2FGL}, i.e., they are not detected 
in the gamma-ray energy band above $100 \units{MeV}$. 
In our analysis we assumed that the $J$-factor distribution for each dSph is well described by a log-normal function (see Ref.~\cite{lat_DMdwarf_paper} for more details), with average value and standard deviation of $\log_{10}J$ reported in Tab.~\ref{tab:dwarfs}. The half-light radii of the dSph galaxies used to compute the $J$-factors are less than or close to $0.5\degrees$. The average values $\langle J_{i} \rangle$ were calculated from the log-normal distributions as:

\begin{equation}
\langle J_{i} \rangle = \exp \left( \mu_{i} + \frac{1}{2} \sigma_{i}^{2} \right)  
\end{equation}
where $\mu_{i}$ and $\sigma_{i}^{2}$ are the average value and the variance of the distributions of $\ln J_{i}$, which can be calculated multiplying the values reported in Tab.~\ref{tab:dwarfs} by $\ln10$.

\begin{table}[ht]
{\small
\begin{tabular}{||l||c|c|c||} \hline \hline
Name & Galactic  & Galactic & $log_{10}(J)$ \\ 
     & longitude & latitude & ($\units{GeV^{2}cm^{-5} sr}$) \\ \hline
Bootes I 	& $358.08\degrees$ 	& $69.62\degrees$ 	& $17.7 \pm 0.34$ \\
Carina   	& $260.11\degrees$ 	& $-22.2\degrees$ 	& $18.0 \pm 0.13$ \\
Coma Berenices 	& $241.9\degrees$	& $83.6\degrees$	& $19.0 \pm 0.37$ \\
Draco		& $86.37\degrees$	& $34.72\degrees$	& $18.8 \pm 0.13$ \\
Fornax		& $237.10\degrees$	& $-65.7\degrees$	& $17.7 \pm 0.23$ \\
Sculptor	& $287.15\degrees$	& $-83.16\degrees$	& $18.4 \pm 0.13$ \\
Segue I		& $220.48\degrees$	& $50.42\degrees$	& $19.6 \pm 0.53$ \\
Sextans		& $243.4\degrees$	& $42.2\degrees$	& $17.8 \pm 0.23$ \\
Ursa Major II	& $152.46\degrees$ 	& $37.44\degrees$	& $19.6 \pm 0.40$ \\
Ursa Minor	& $104.95\degrees$	& $44.80\degrees$	& $18.5 \pm 0.18$ \\ \hline \hline
\end{tabular}
\caption{List of the dSph galaxies used in this analysis. The $J$-factors are assumed to be distributed according to a log-normal distribution with $\langle \log_{10} J \rangle$ and $\sigma_{\log_{10}J}$ given here. The half-light radii of the dSph galaxies used to compute the $J$-factors are less than or close to $0.5\degrees$~\cite{lat_DMdwarf_paper}. }
\protect{\label{tab:dwarfs}}
}
\end{table}

\begin{figure}[hb]
\includegraphics[width=0.42\textwidth]{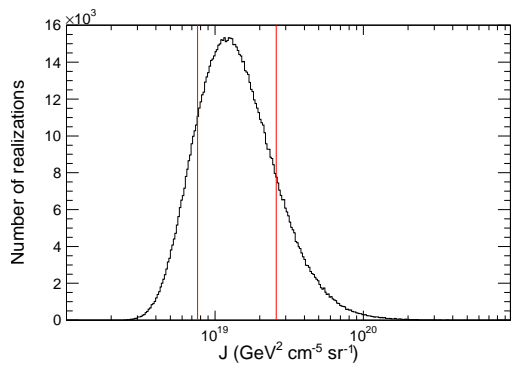}
\caption{Distribution of $10^{6}$ $J$-factor values for the stacked
sources. Each realization is obtained by sampling the $10$ log-normal
distributions of the $J$-factors of individual sources and evaluating
the average value weighted by the exposures. The red lines
correspond to the $16\%$ and $84\%$ quantiles of the distribution.}
\label{fig:jdist}
\end{figure}

\begin{figure*}
\includegraphics[width=0.29\textwidth]{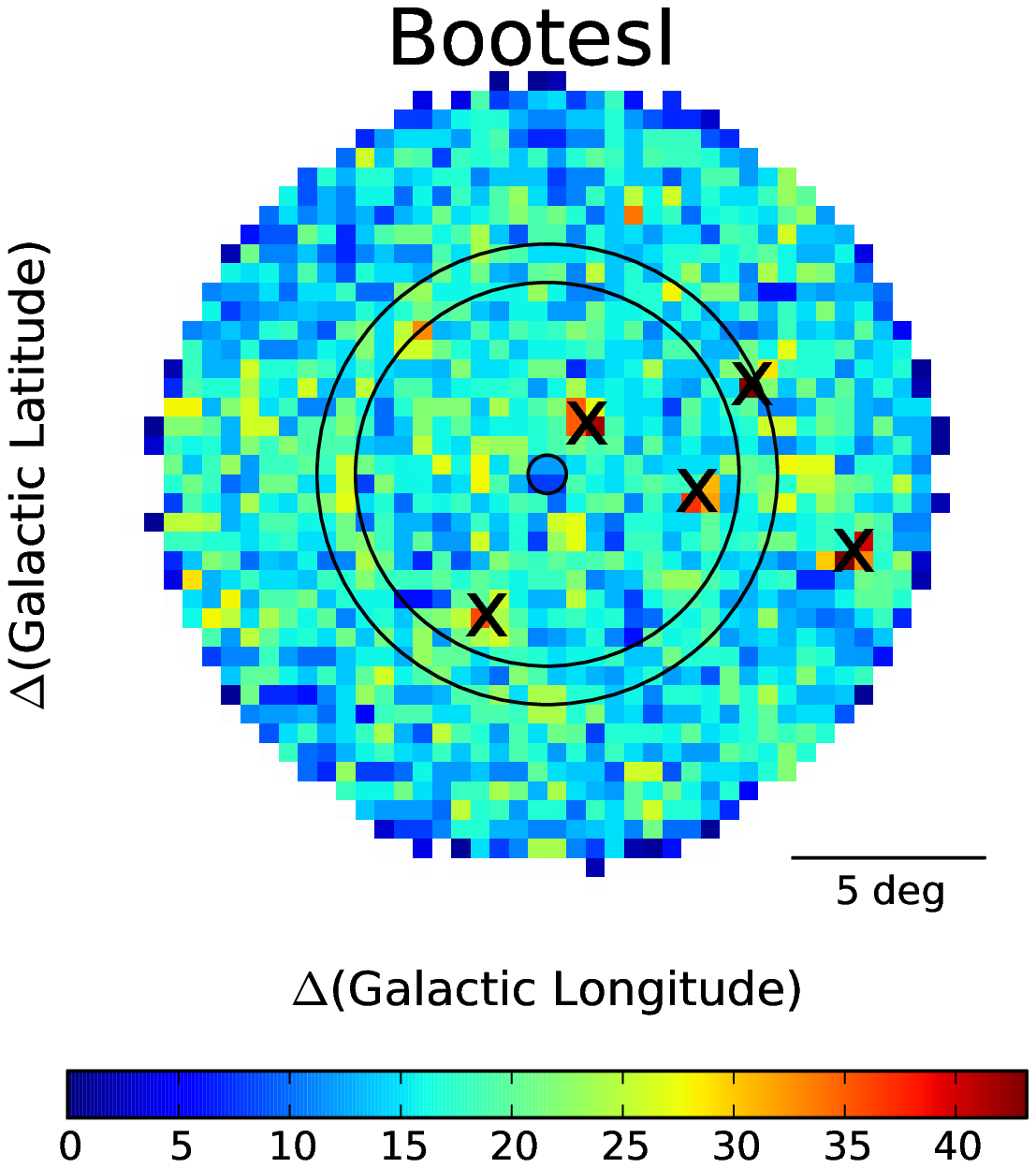}
\includegraphics[width=0.29\textwidth]{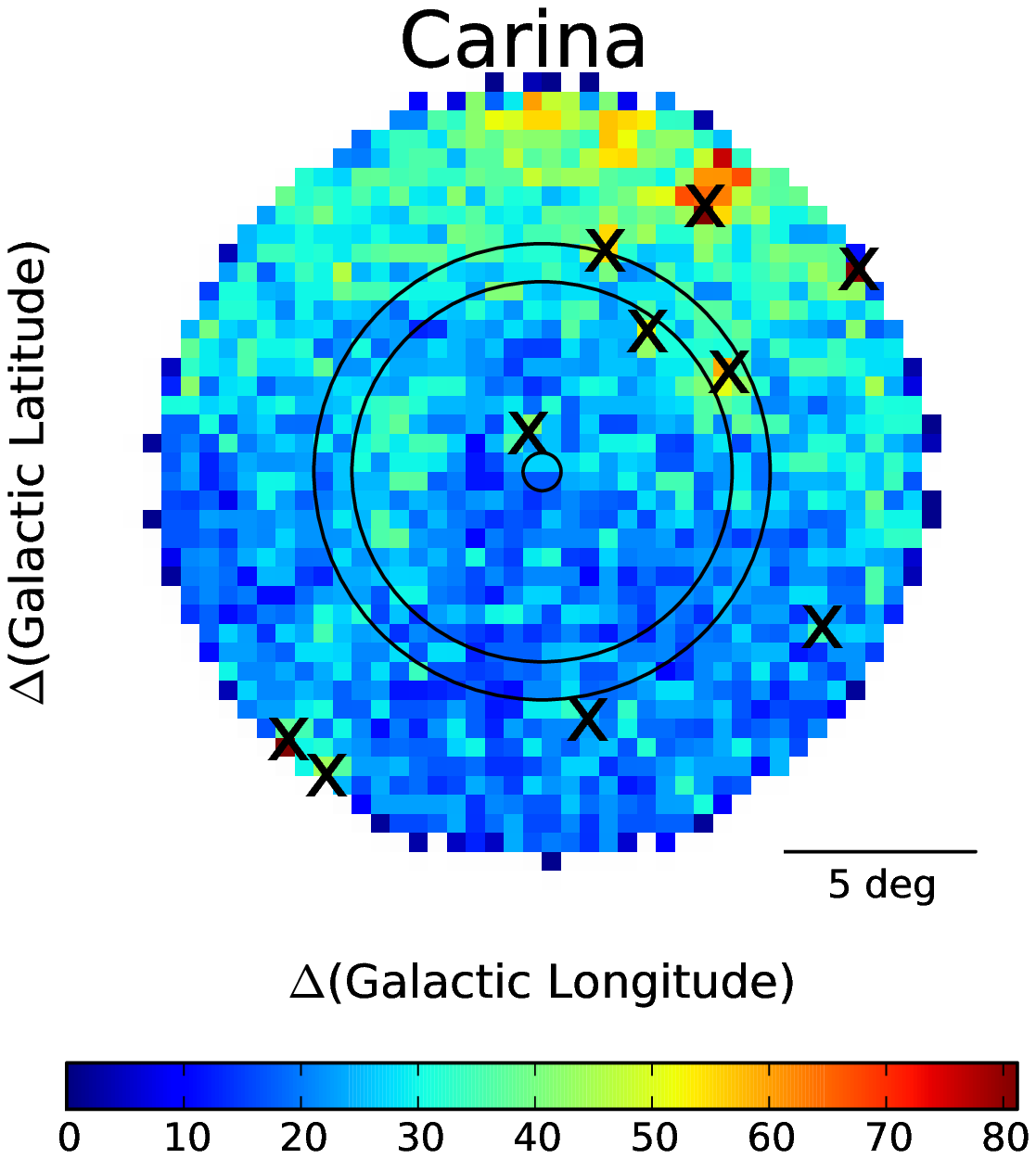}
\includegraphics[width=0.29\textwidth]{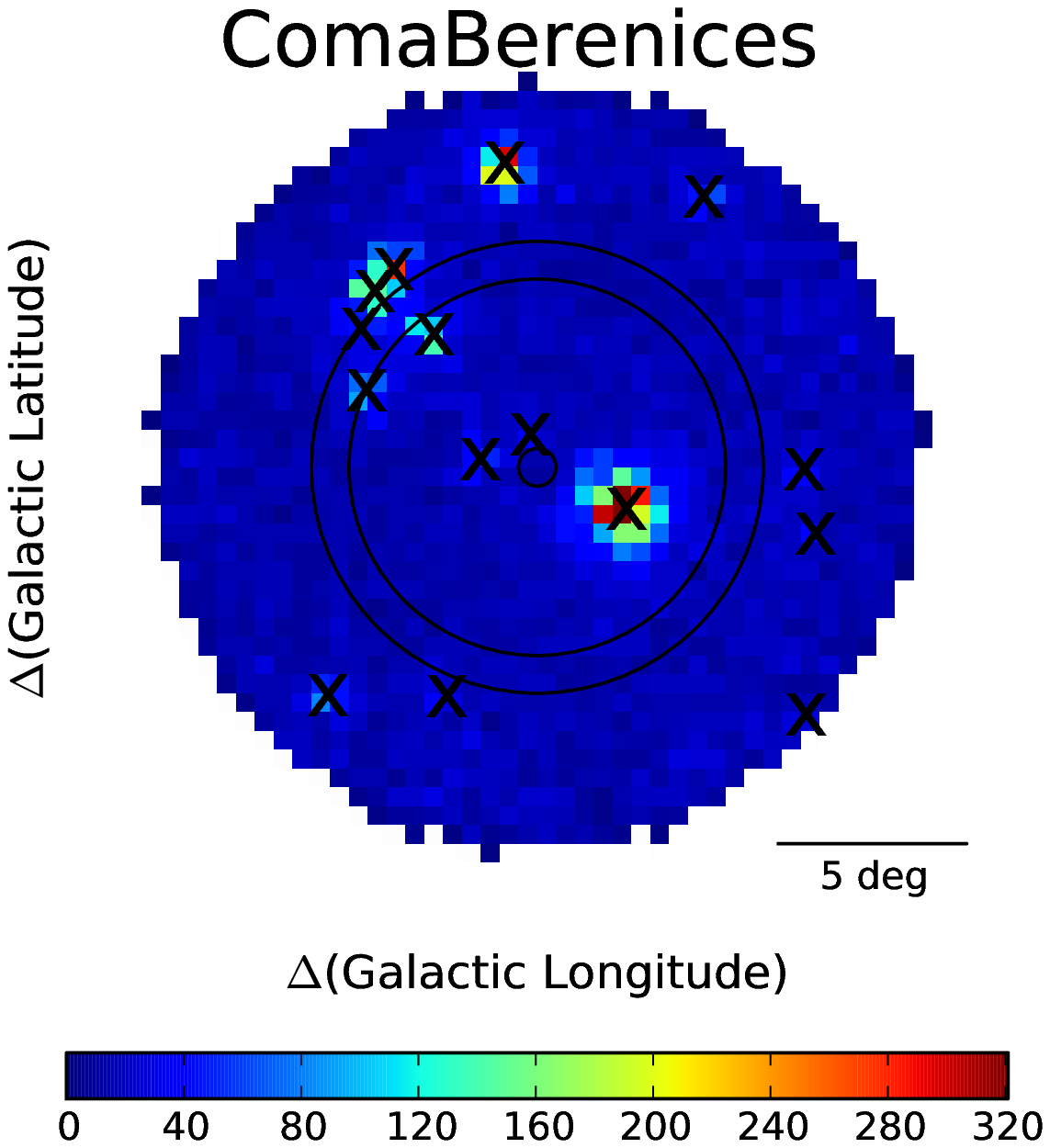}
\includegraphics[width=0.29\textwidth]{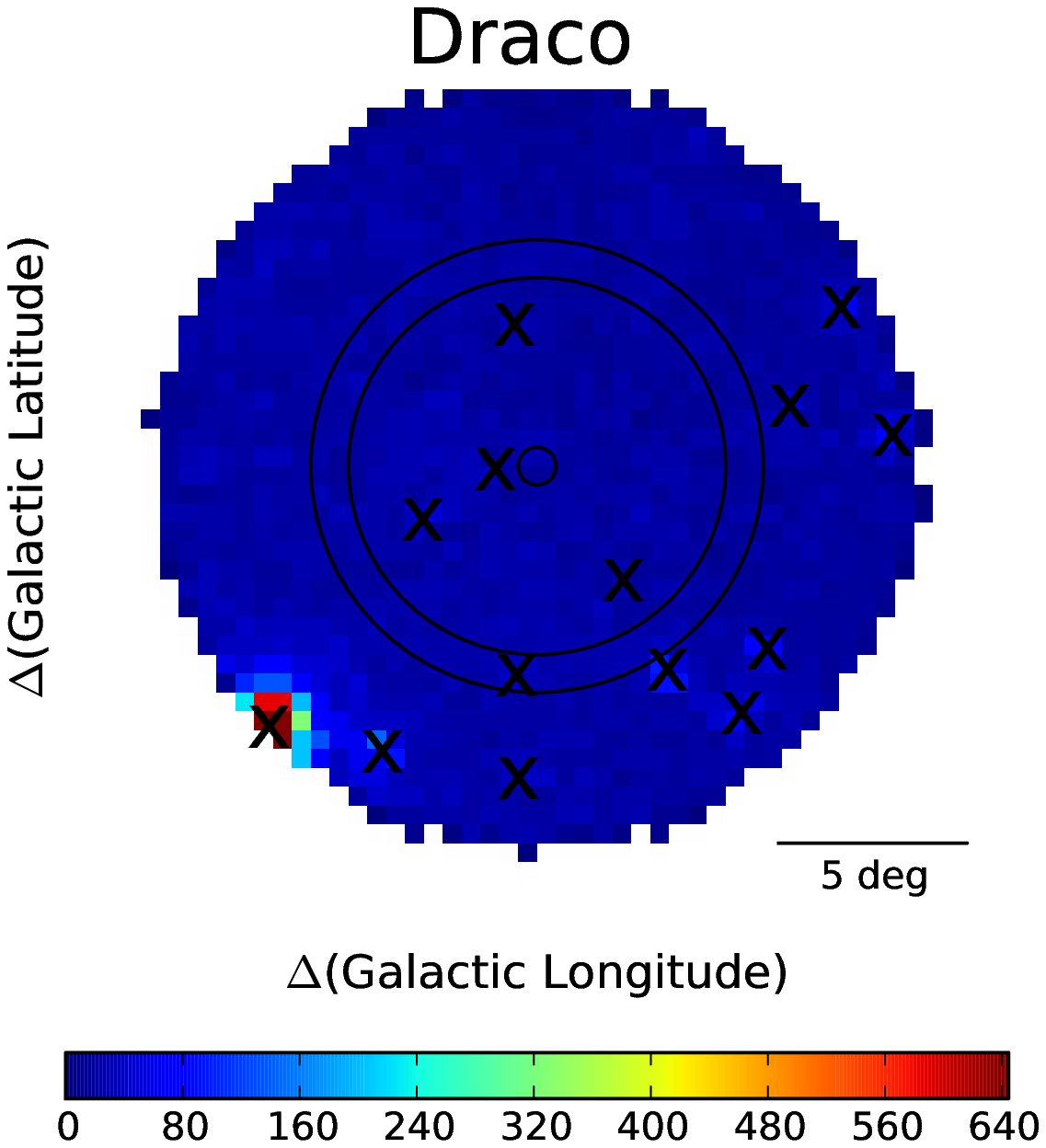}
\includegraphics[width=0.29\textwidth]{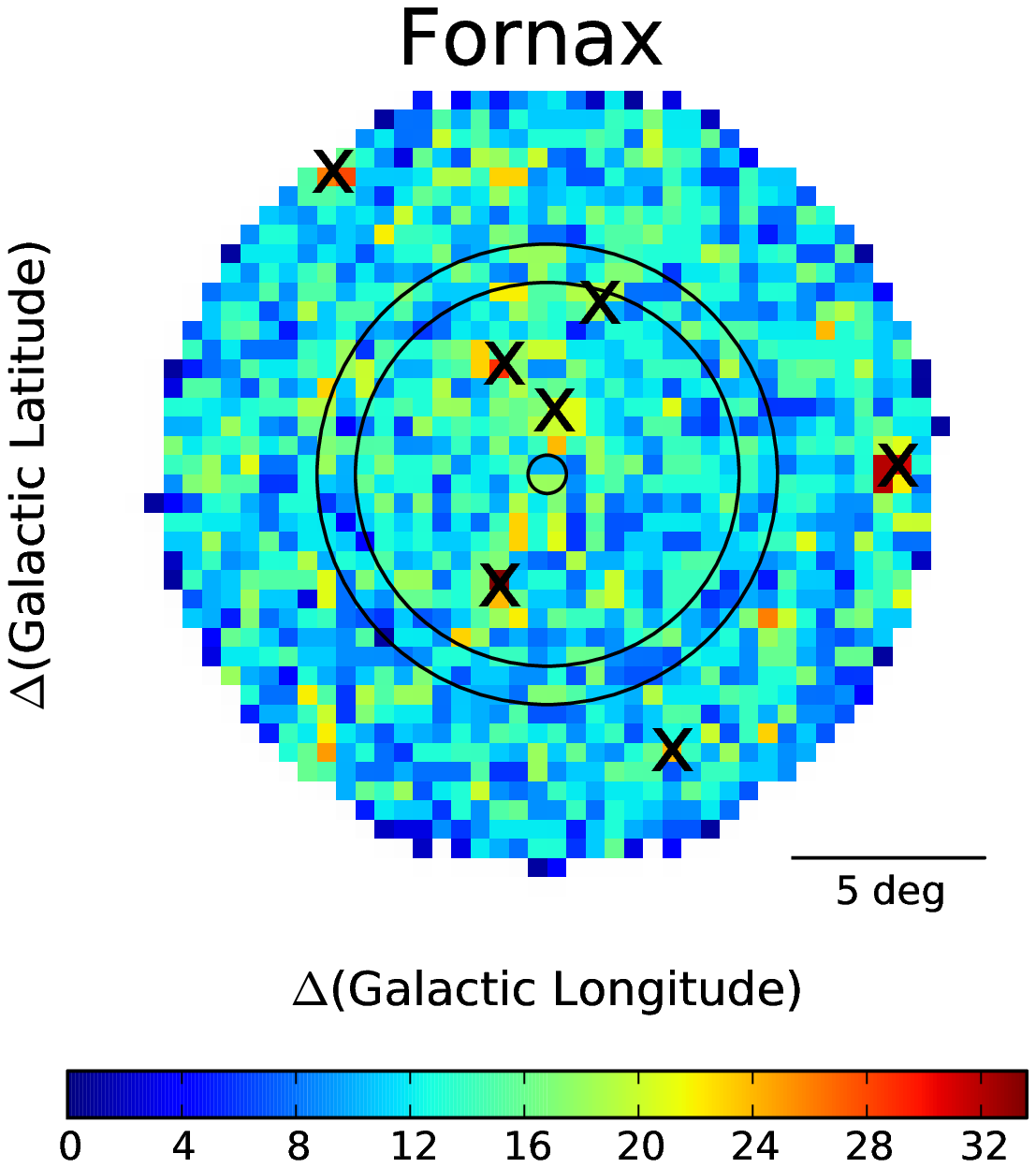}
\includegraphics[width=0.29\textwidth]{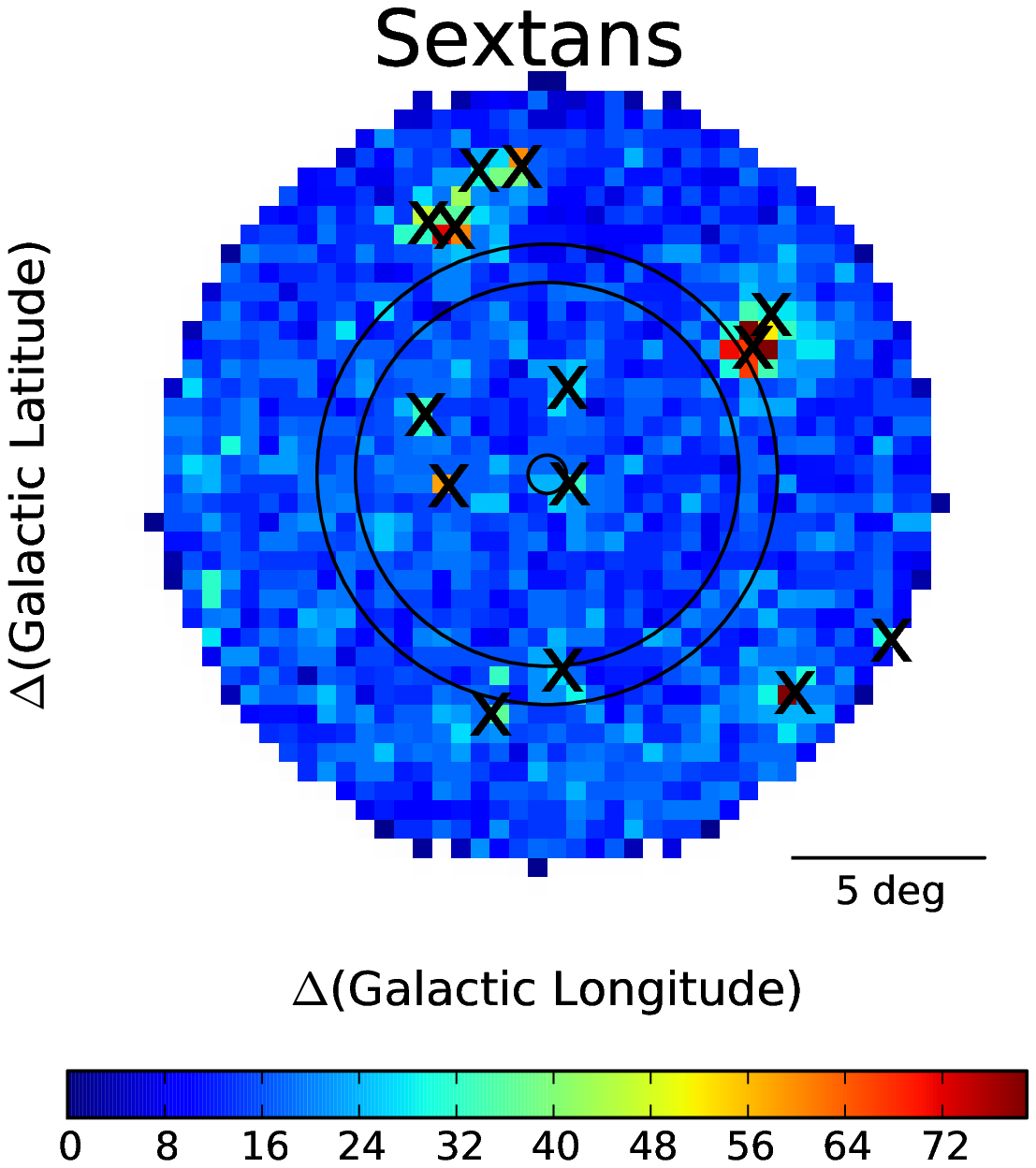}
\includegraphics[width=0.29\textwidth]{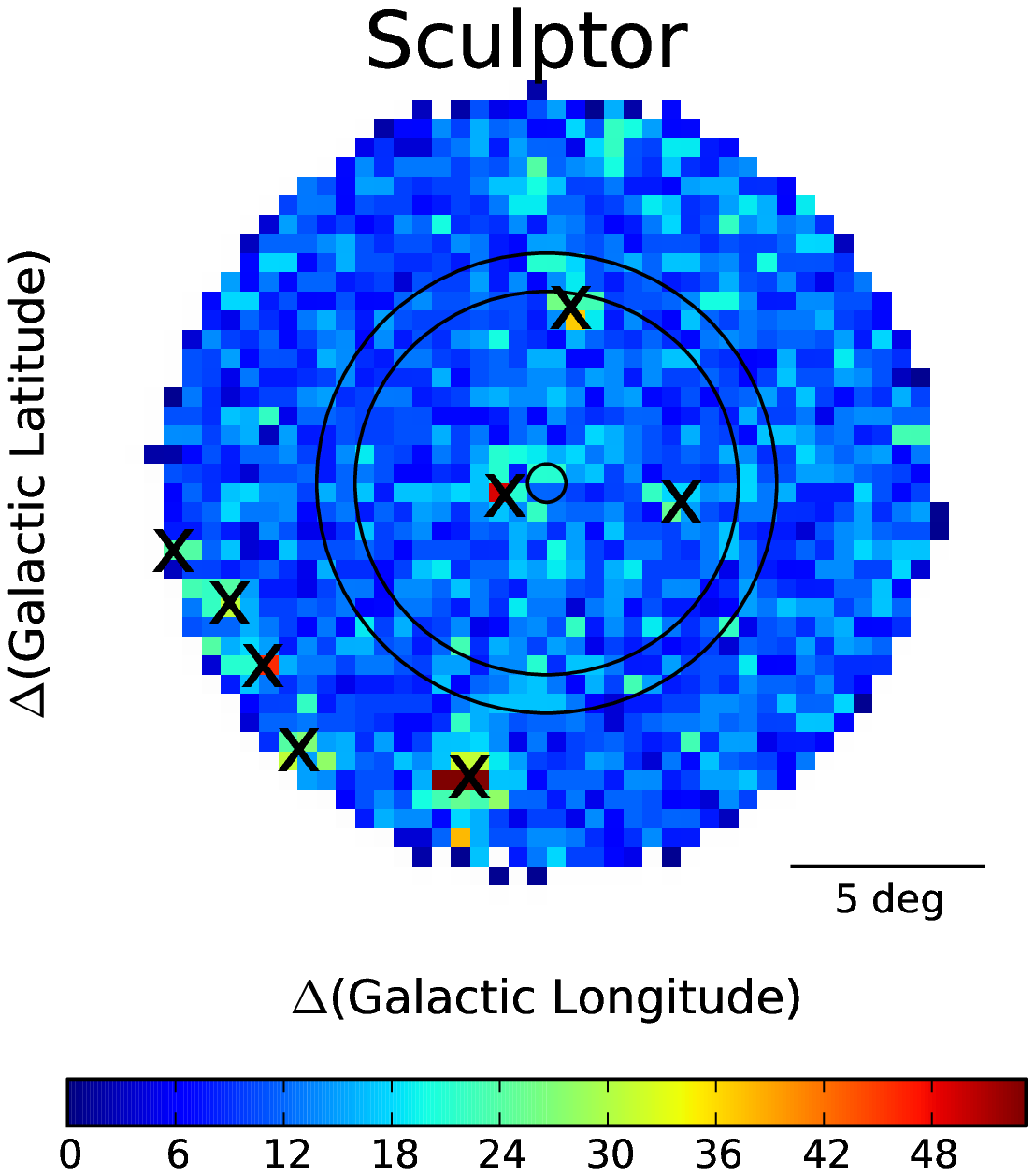}
\includegraphics[width=0.29\textwidth]{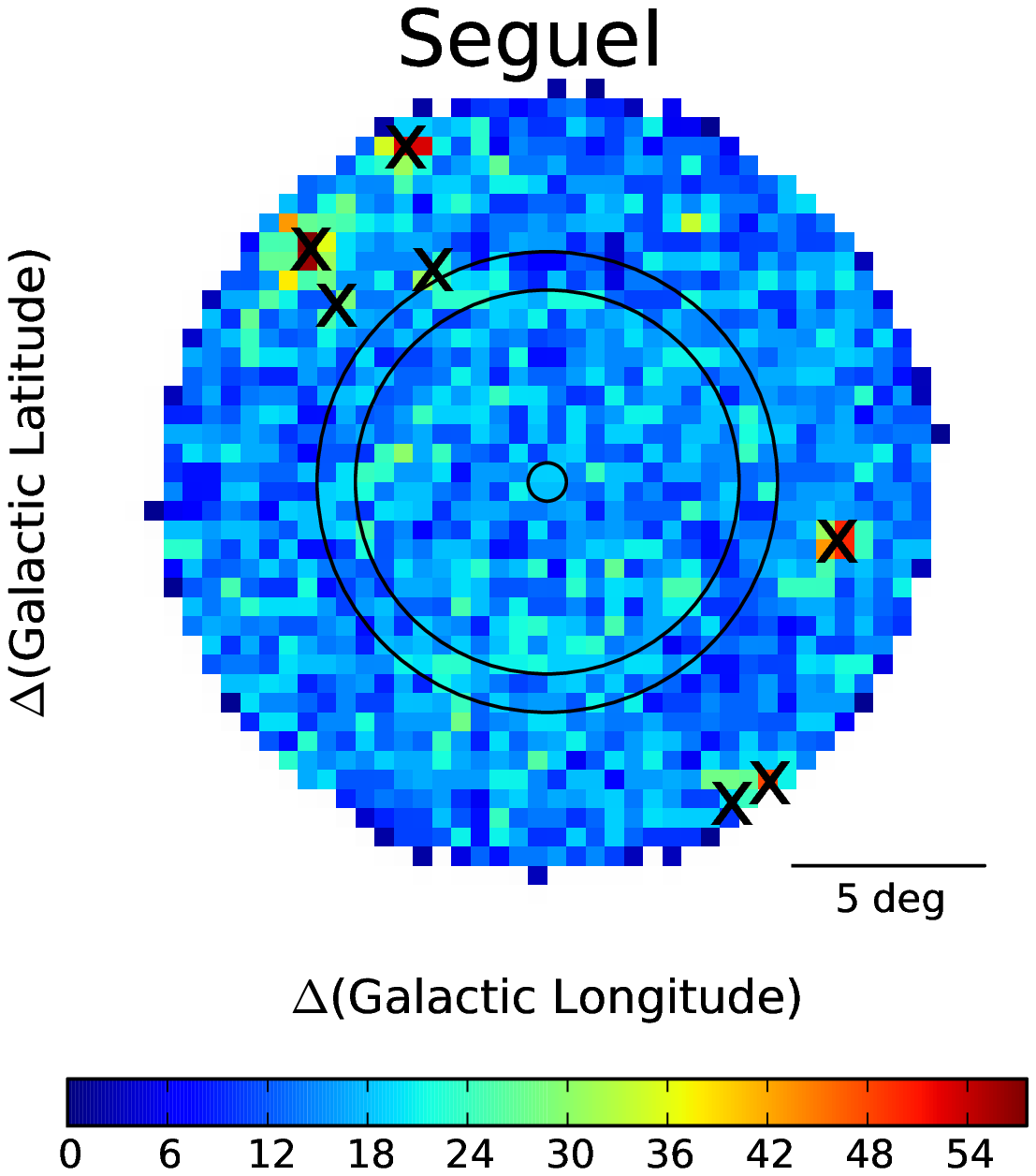}
\includegraphics[width=0.29\textwidth]{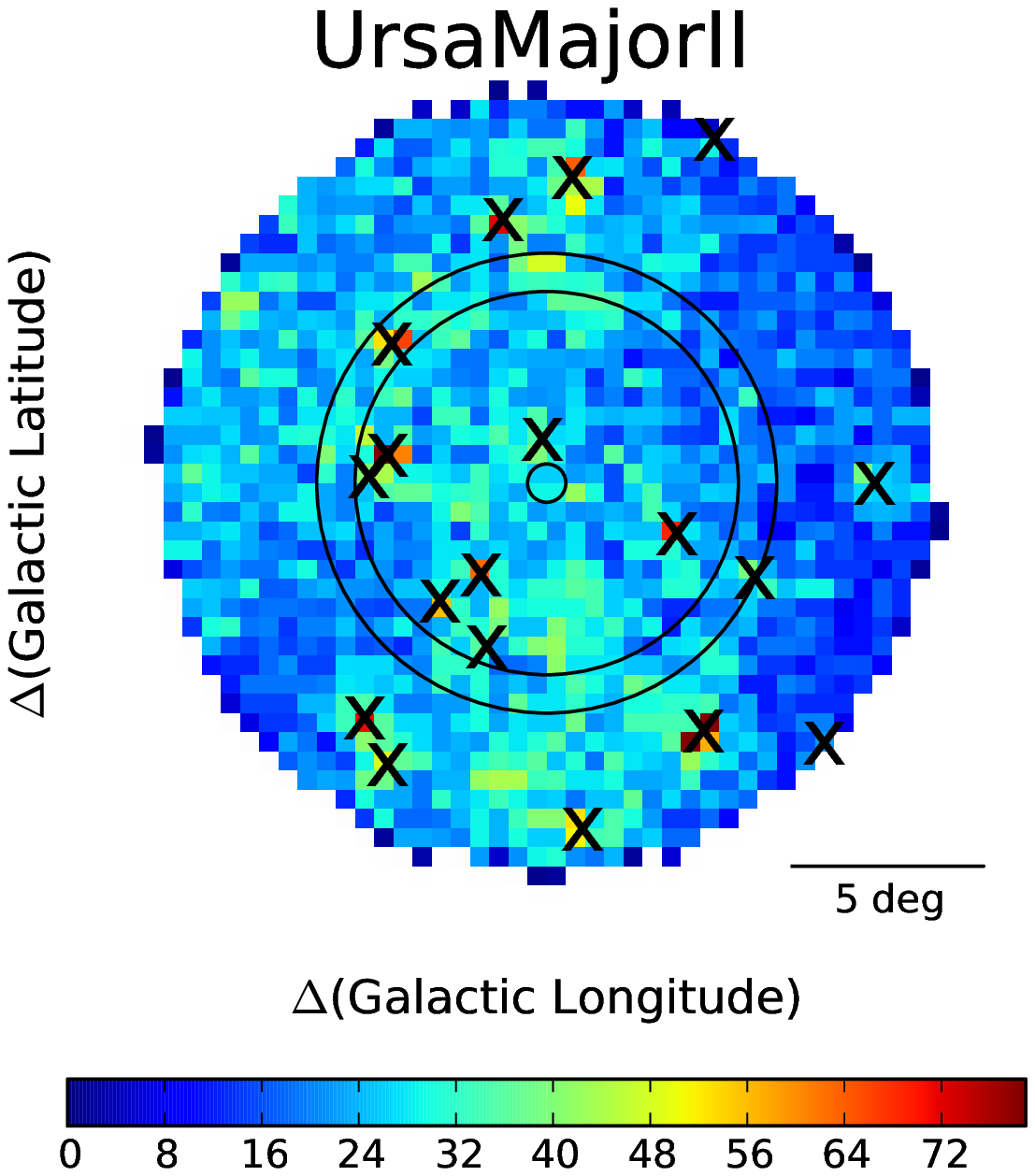}
\includegraphics[width=0.29\textwidth]{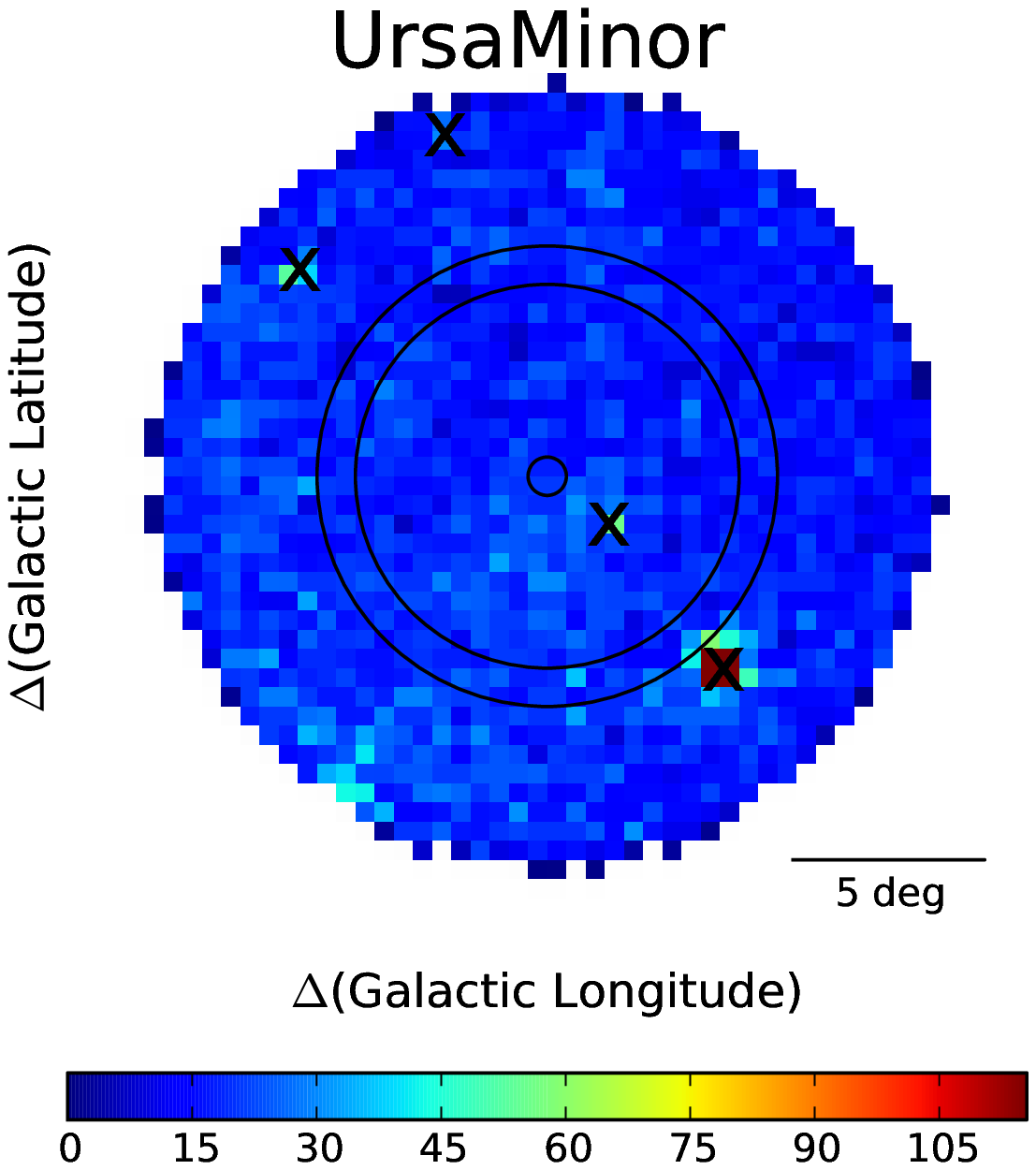}
\includegraphics[width=0.29\textwidth]{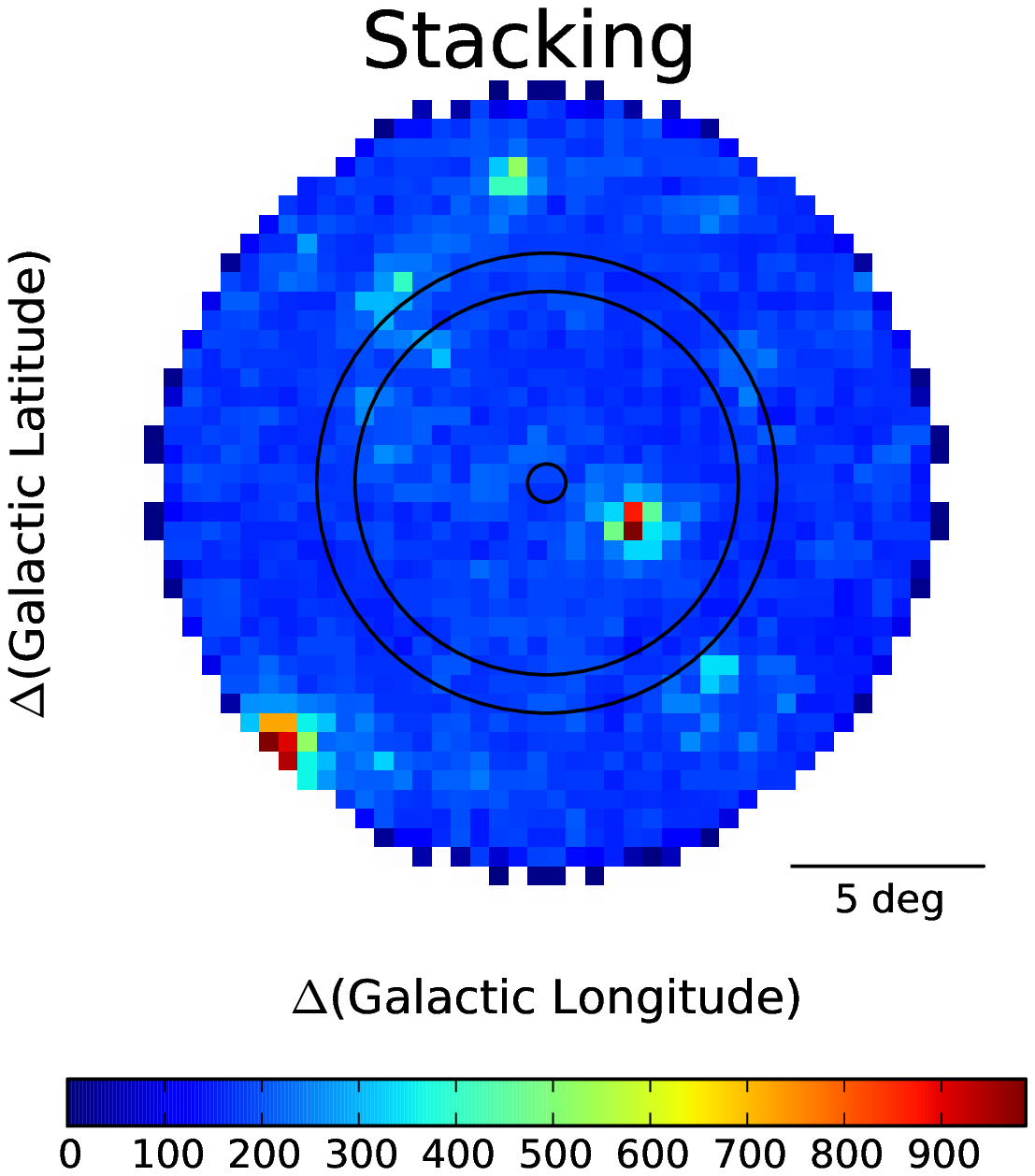}
\caption{Photon count maps in the observed energy range from
$562\units{MeV}$ to $562\units{GeV}$ for the dSph galaxies considered
in this analysis. The black circles indicate the cones of angular
radii of $0.5\degrees$, $5\degrees$ and $6\degrees$, representing the
boundaries of the signal and background regions. The sources in the
2FGL Catalog are indicated with crosses. Each map is centered on the
position of the corresponding source. The map in the bottom right
panel was obtained by stacking the data from all the dSph galaxies.}
\label{fig:countmaps}
\end{figure*}

\begin{figure*}
\includegraphics[width=0.3\textwidth]{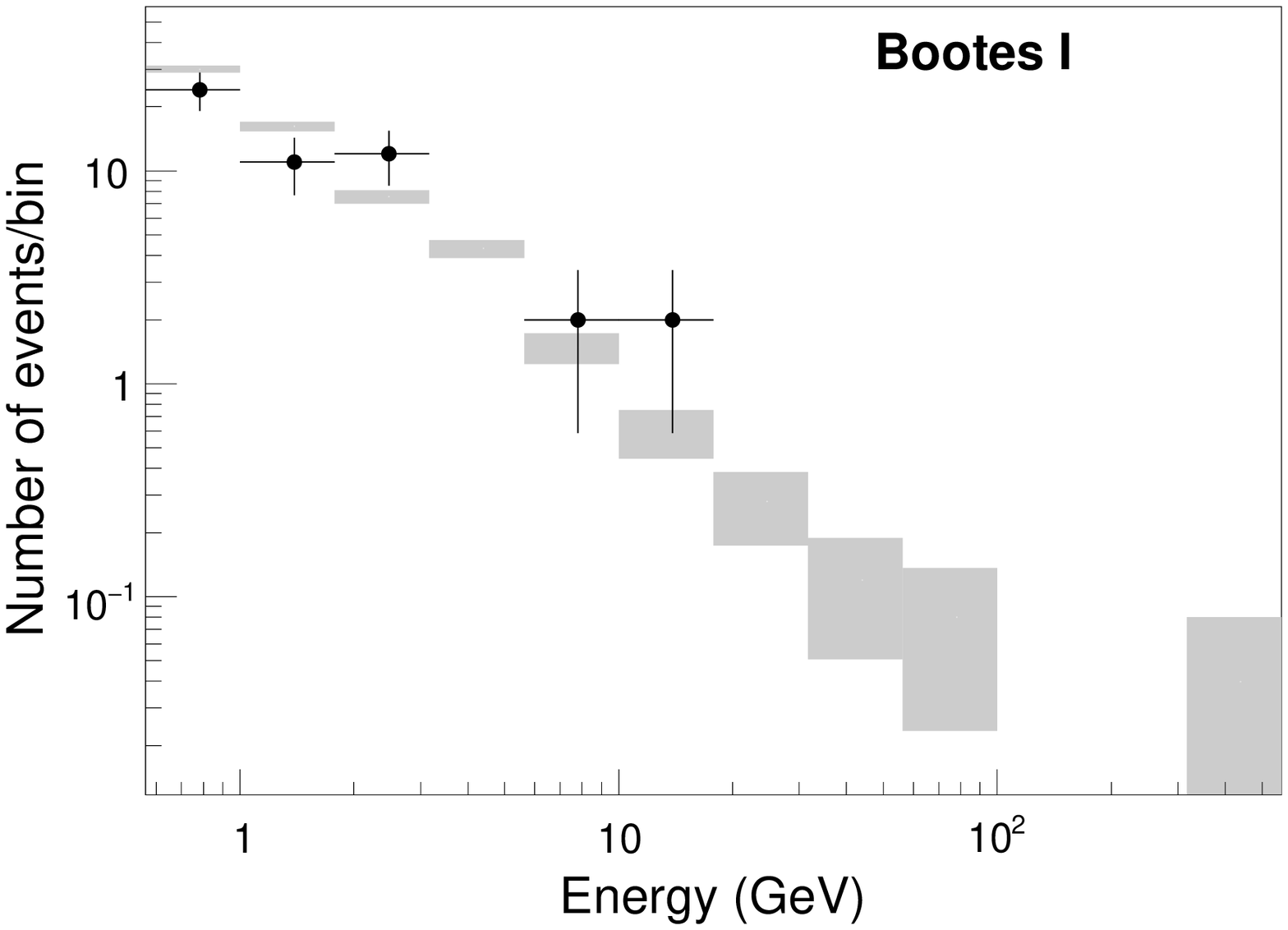}
\includegraphics[width=0.3\textwidth]{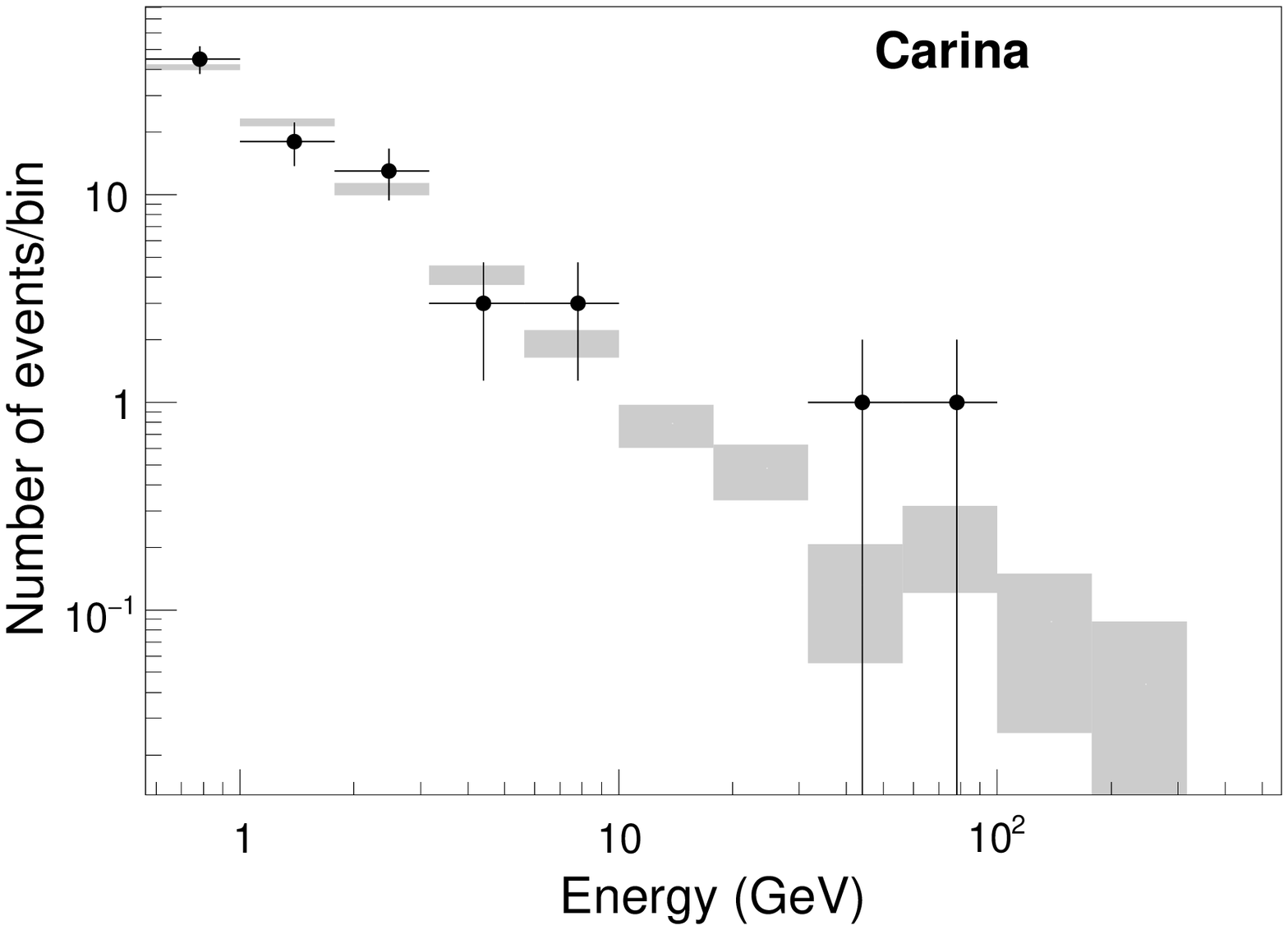}
\includegraphics[width=0.3\textwidth]{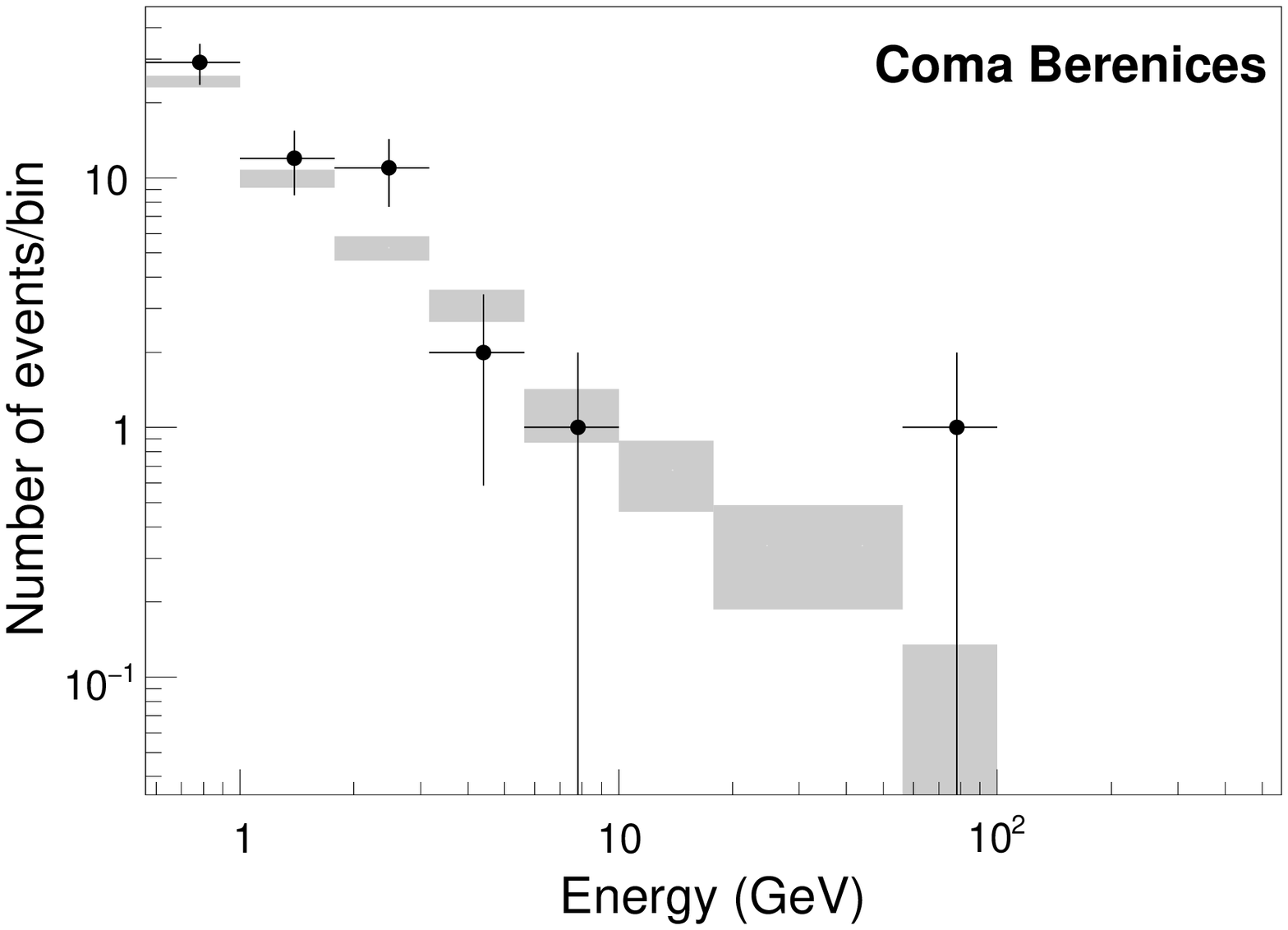}
\includegraphics[width=0.3\textwidth]{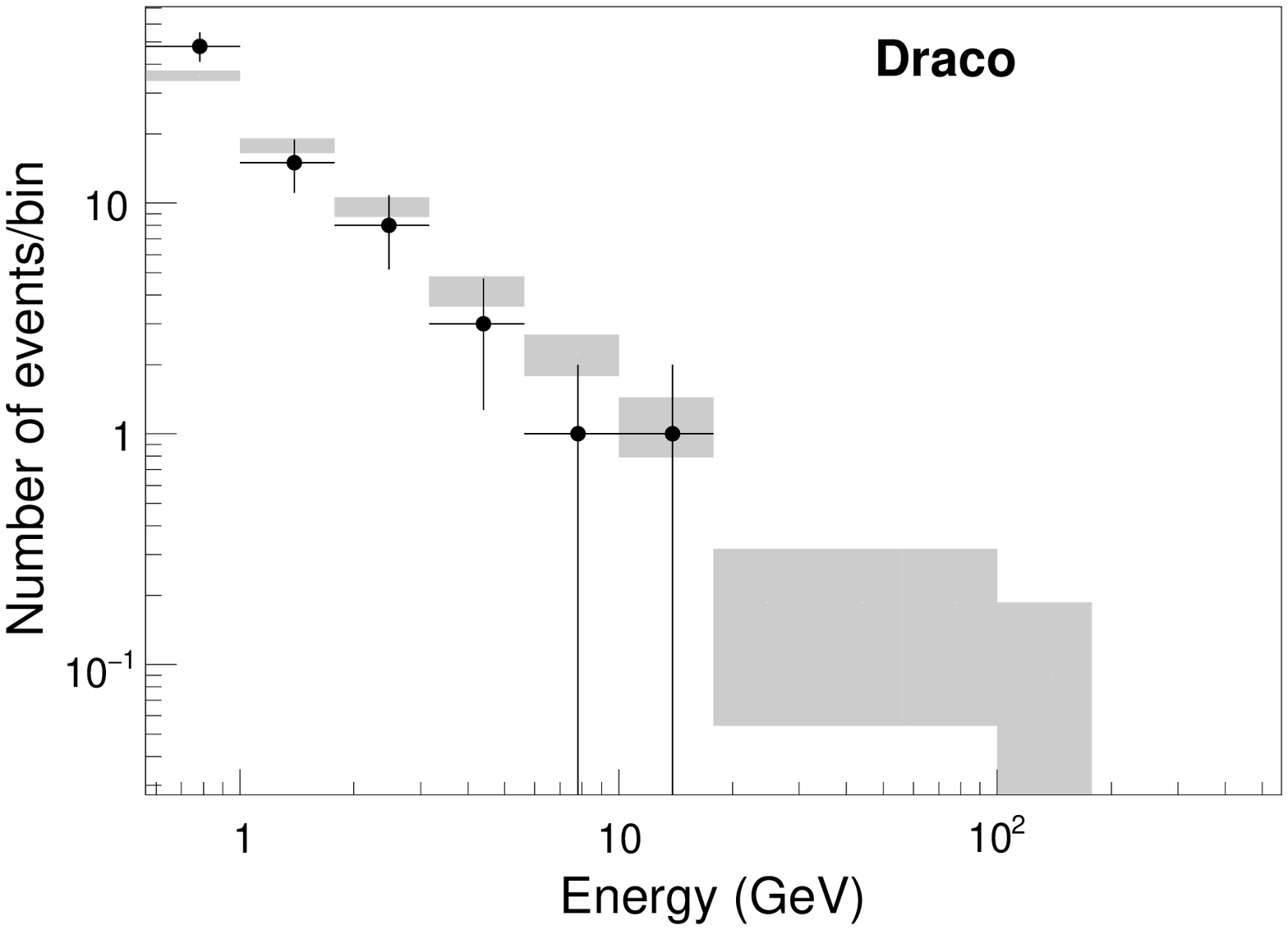}
\includegraphics[width=0.3\textwidth]{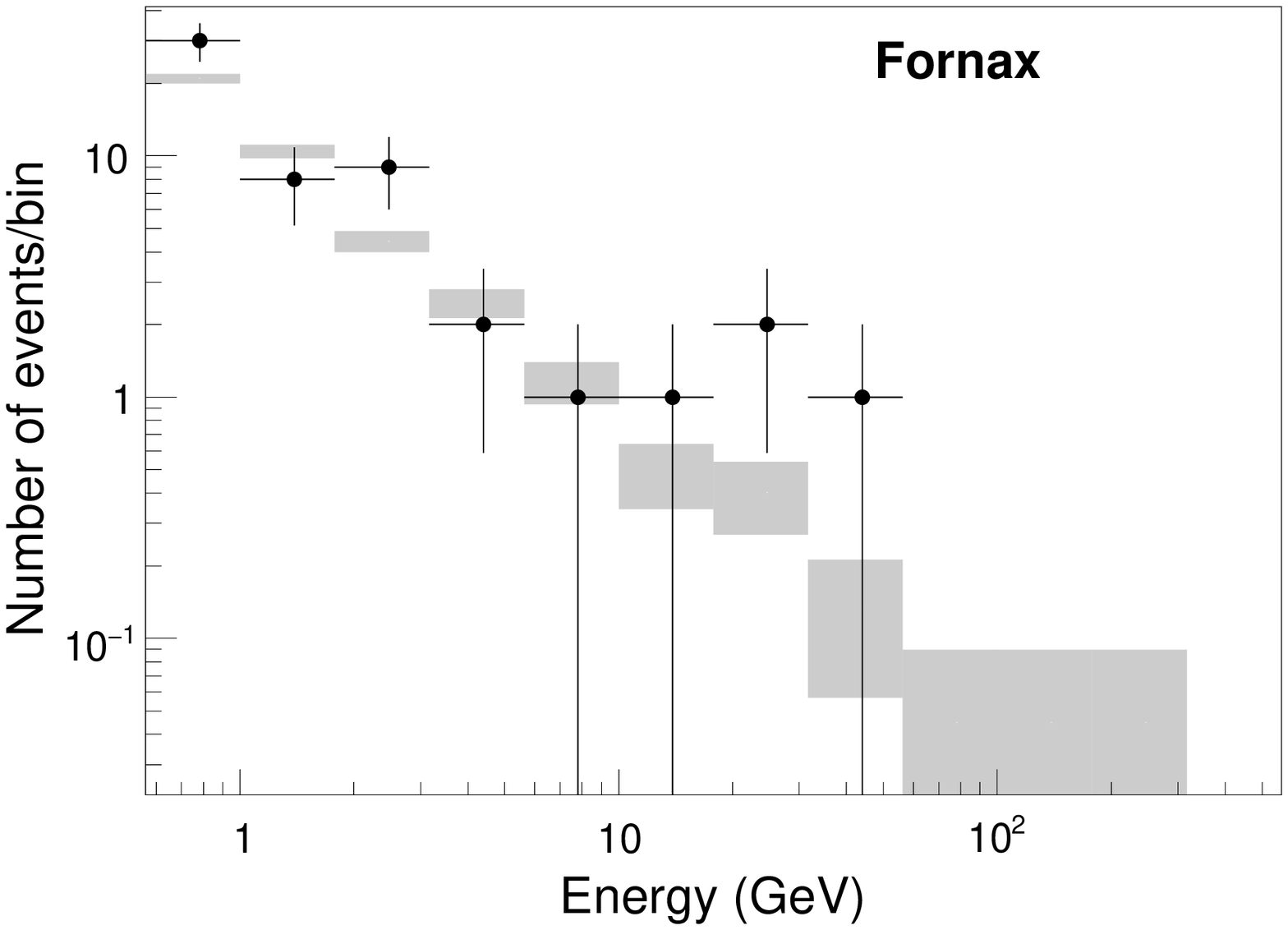}
\includegraphics[width=0.3\textwidth]{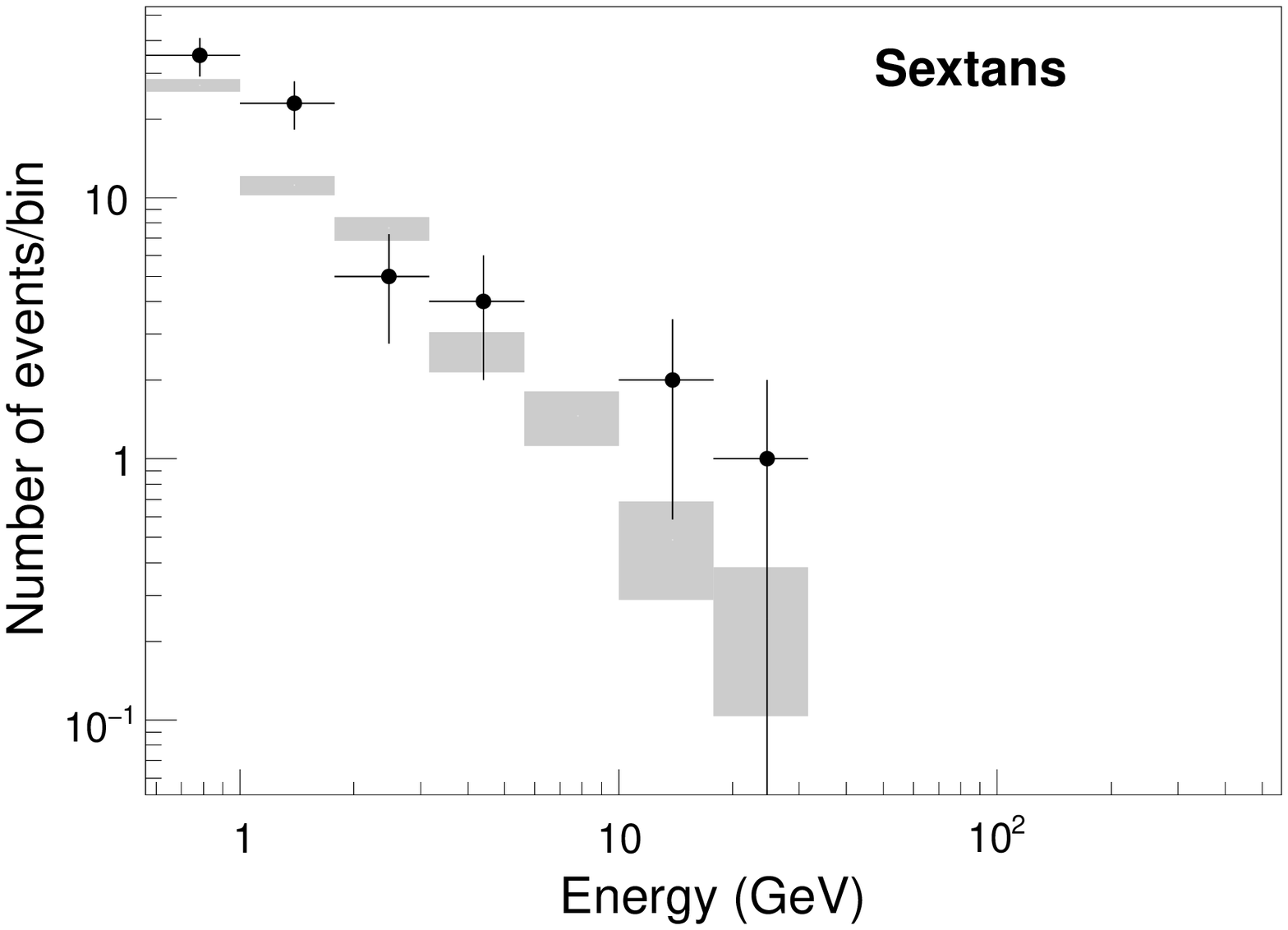}
\includegraphics[width=0.3\textwidth]{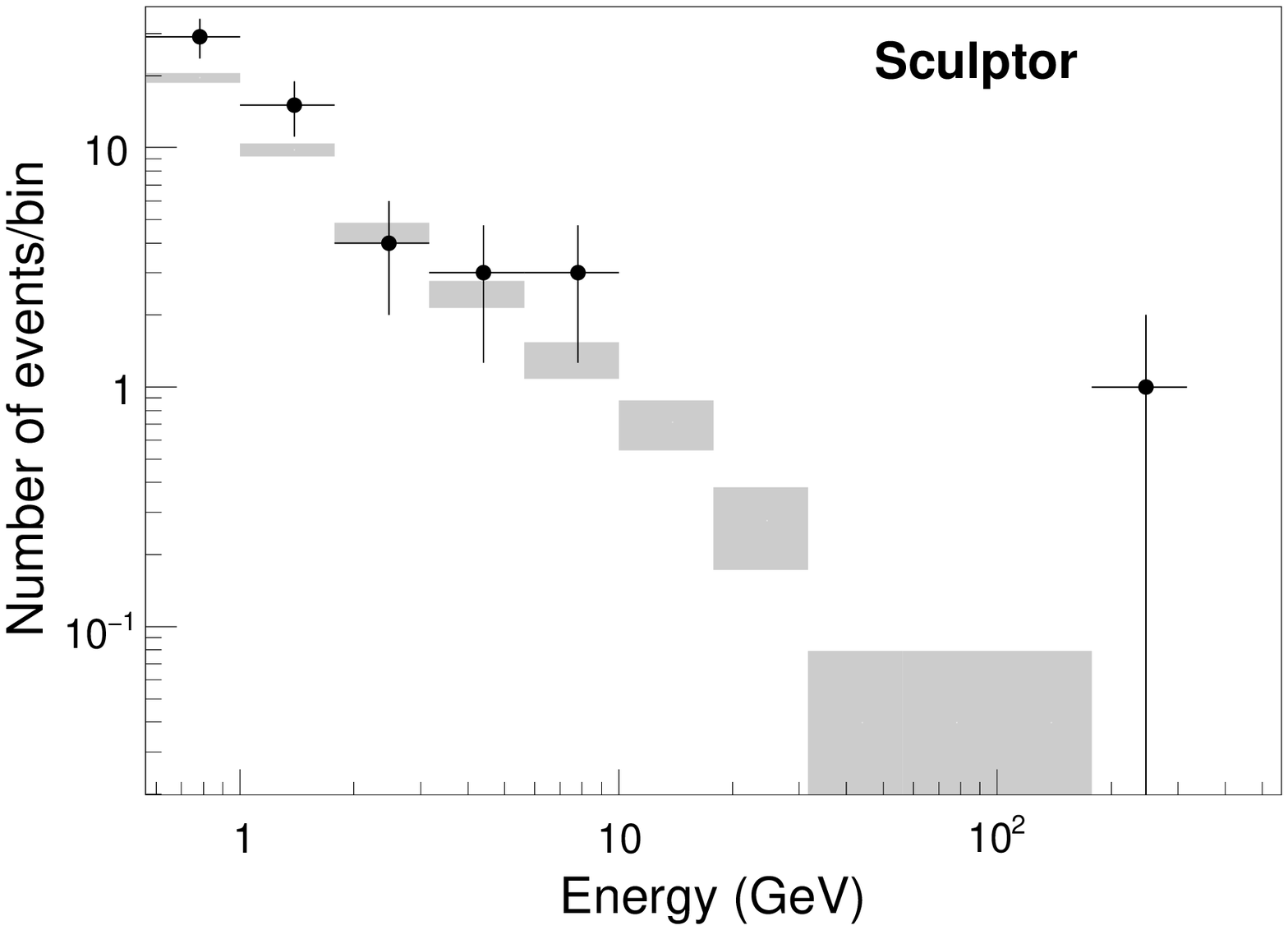}
\includegraphics[width=0.3\textwidth]{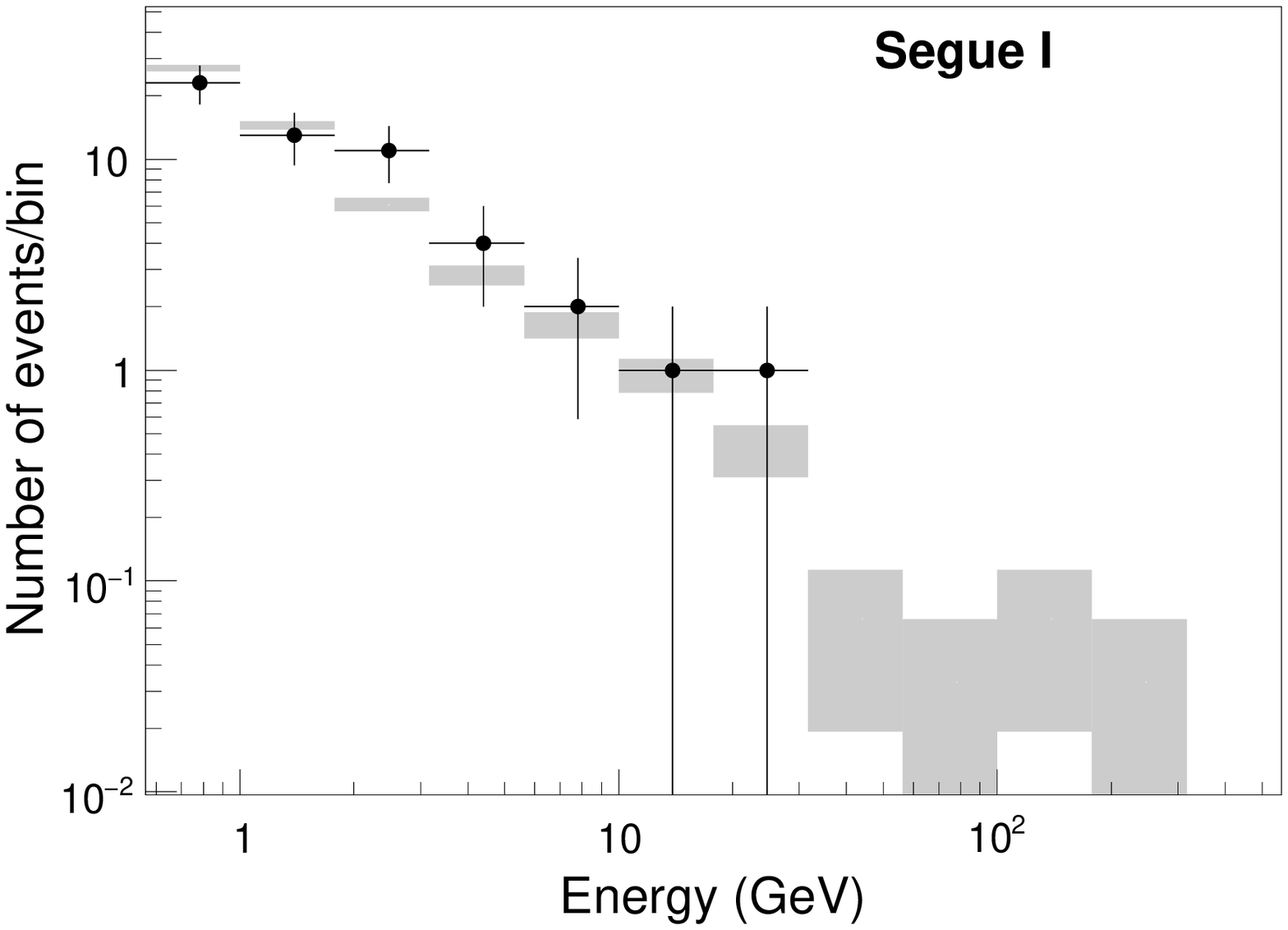}
\includegraphics[width=0.3\textwidth]{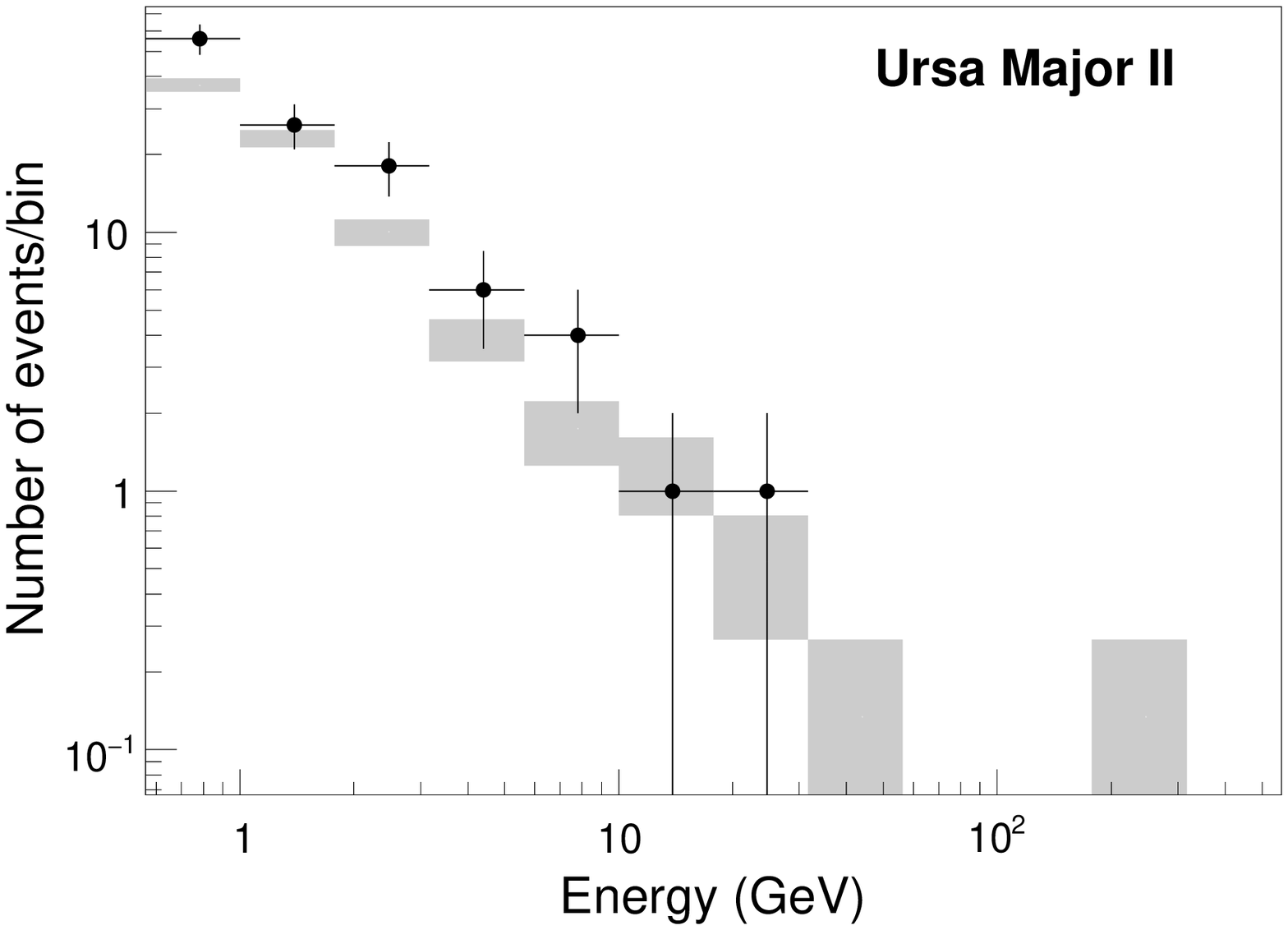}
\includegraphics[width=0.3\textwidth]{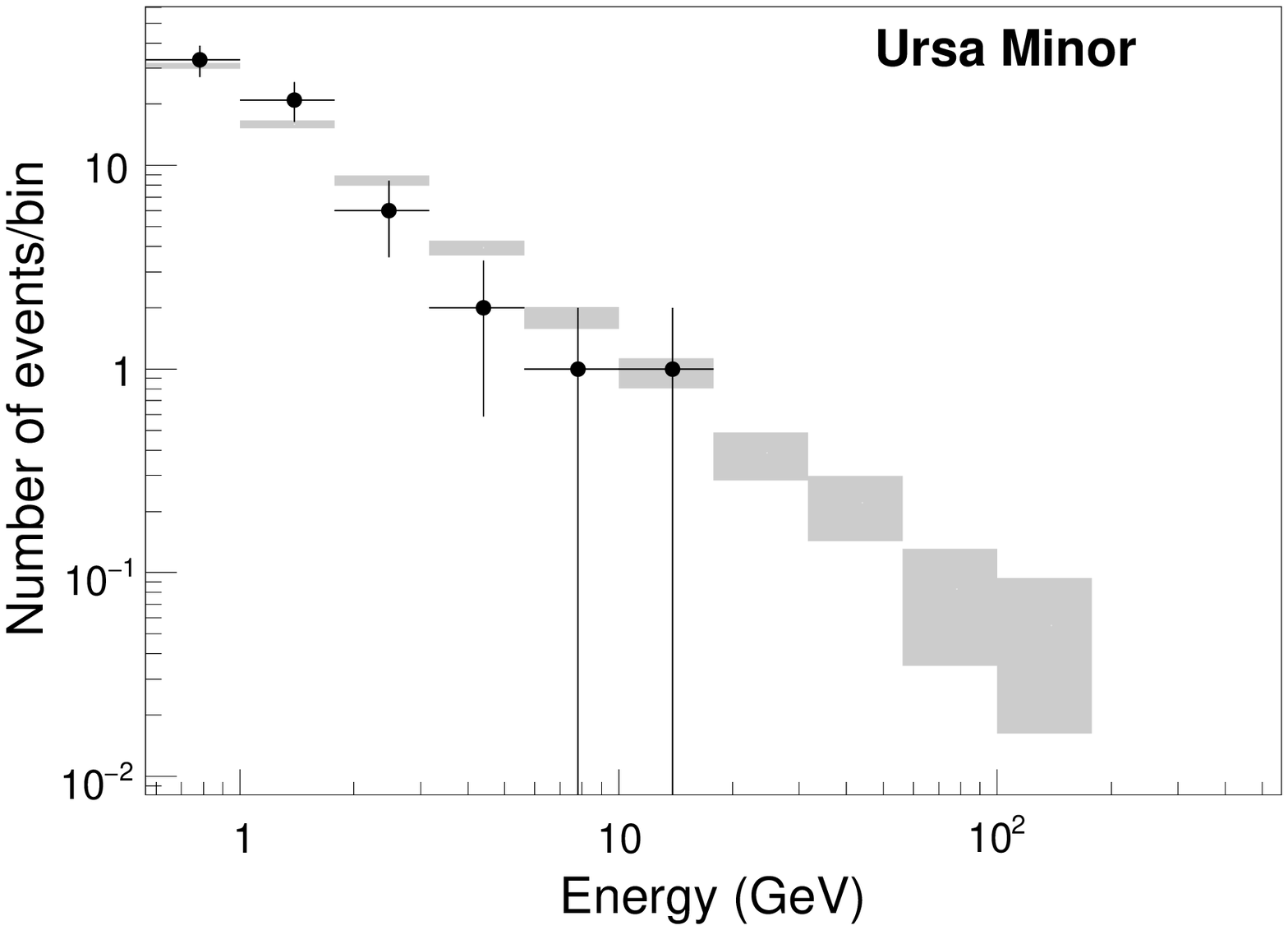}
\includegraphics[width=0.3\textwidth]{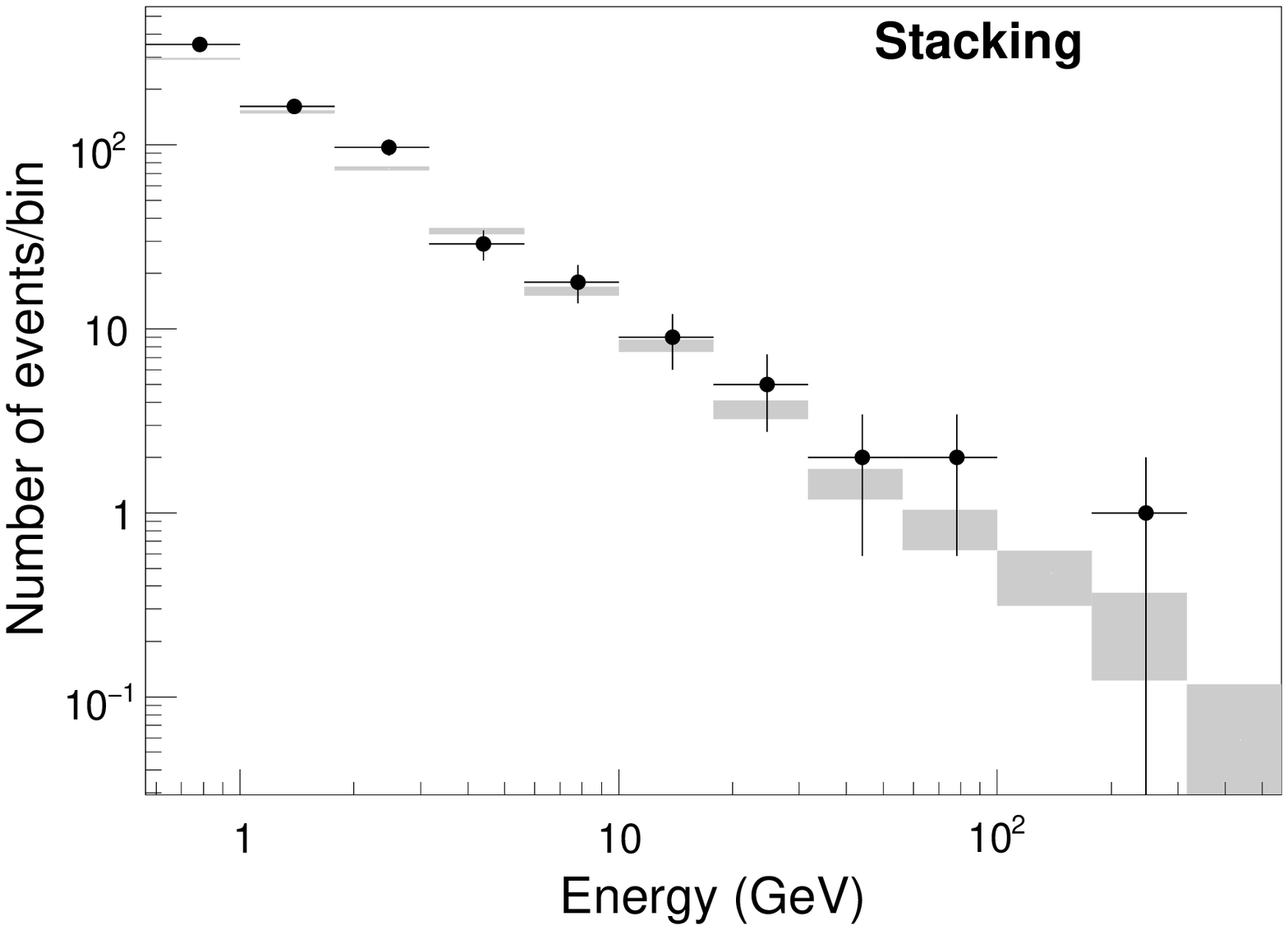}
\caption{Count distributions in the observed energy range from
$562\units{MeV}$ to $562\units{GeV}$  for the dSph galaxies considered
in this analysis. The black points represent the counts in the signal
region; the grey areas represent the equivalent number of background
counts (i.e., the counts in the background region scaled by the
coefficient $c$) with their errors. The bottom right panel shows the distribution obtained by
stacking all the dSph galaxies. }
\label{fig:countdist}
\end{figure*}

When evaluating the effects of systematic uncertainties on the
$J$-factor, we followed the procedure discussed in
\S\ref{sec:SistErr}. The fluctuations of $J_i$ were described by a 
uniform PDF in the interval $[J_{i1},J_{i2}]$, with $J_{i1}$ and
$J_{i2}$ corresponding to the $16\%$ and $84\%$ quantiles of 
the log-normal distribution. 

In the case of the stacking analysis, as discussed in
\S\ref{sec:stacking}, the $J$-factor was evaluated as the weighted
average of the $J$-factors of individual dSph galaxies with the
exposures. The distribution of the $J$-factors of the stacked sources
was built sampling a large set of events ($10^{6}$) from the $J$-factor
distributions of individual sources, and is shown in
Fig.~\ref{fig:jdist}. The average value of the $J$-factor for the
stacked sources is \mbox{$\langle J \rangle = 1.75 \cdot 10^{19}
\units{GeV^{2}cm^{-5}sr}$}. As in the case of individual sources, to
study the systematic uncertainties on $J$ we used a uniform PDF in the
interval $[J_{1}, J_{2}]$, with $J_{1}$ and $J_{2}$ corresponding to
the $16\%$ and $84\%$ quantiles of the resulting $J$-factor distribution.

For each dSph galaxy the signal region was defined as a cone of angular radius $\Delta \theta = 0.5\degrees$ centered on the source position. The value of $\Delta \theta$ is the same as the one used to evaluate the $J$-factor, and is consistent with the Point Spread Function (PSF) of the instrument, the $68\%$ containment radius of which is smaller than $1\degrees$ in the energy range above $1\units{GeV}$~\cite{Pass7}. The background region was defined as an annulus centered on the source position, with an inner radius of $5\degrees$ and an outer radius of $6\degrees$. In order to prevent contamination of the background sample from photons emitted by other astrophysical sources, all the sources in the 2FGL Catalog~\cite{2FGL} were masked. Using the HEALPix~\cite{healpix} pixelization scheme with $N_{\rm side}=256$, the sky was divided into $786432$ equal area pixels, each covering a solid angle of $1.6 \cdot 10^{-5} \units{sr}$. The background region was composed of all the pixels in the annulus, excluding those at an angular distance less than $3\degrees$ from any point source and those at an angular distance less than $3\degrees$ plus the angular size of the semi-major axis from any extended source.
The solid angle of the background region, $\Delta \Omega_{bi}$, was then evaluated by adding the solid angles corresponding to all the unmasked pixels in the annulus.

Fig.~\ref{fig:countmaps} shows the photon count maps with energy
greater than $562\units{MeV}$ for the $10$ dSph galaxies considered in
the present analysis. A qualitative inspection of the count maps shows no evidence of a gamma-ray signal from any dSph galaxy. On the other hand, from Fig.~\ref{fig:countmaps}, bright gamma-ray sources close to some dSph galaxies are evident. However, as mentioned above, these sources are not considered when evaluating the background because of the masking procedure.  

Photons emitted by possible gamma-ray point sources lying close to a dSph galaxy might be detected in the signal region. These photons will not be accounted for in the background and, therefore, they might be confused with a DM annihilation signal. The result is that the upper limits on the DM signal will be higher and so, in this sense, this analysis is conservative.

Fig.~\ref{fig:countdist} shows the signal and background count
distributions for all the dSph galaxies that were analyzed. The
background counts have been scaled taking into account the solid angle
ratio between the signal and background regions, according to
Eq.~\ref{eq:cdef} (Eq.~\ref{eq:cdef0} in the case of the stacking
analysis). In all cases no evidence is observed of a net signal excess with respect to the background in any energy bin. 

\begin{figure}[hb]
\includegraphics[width=0.5\textwidth]{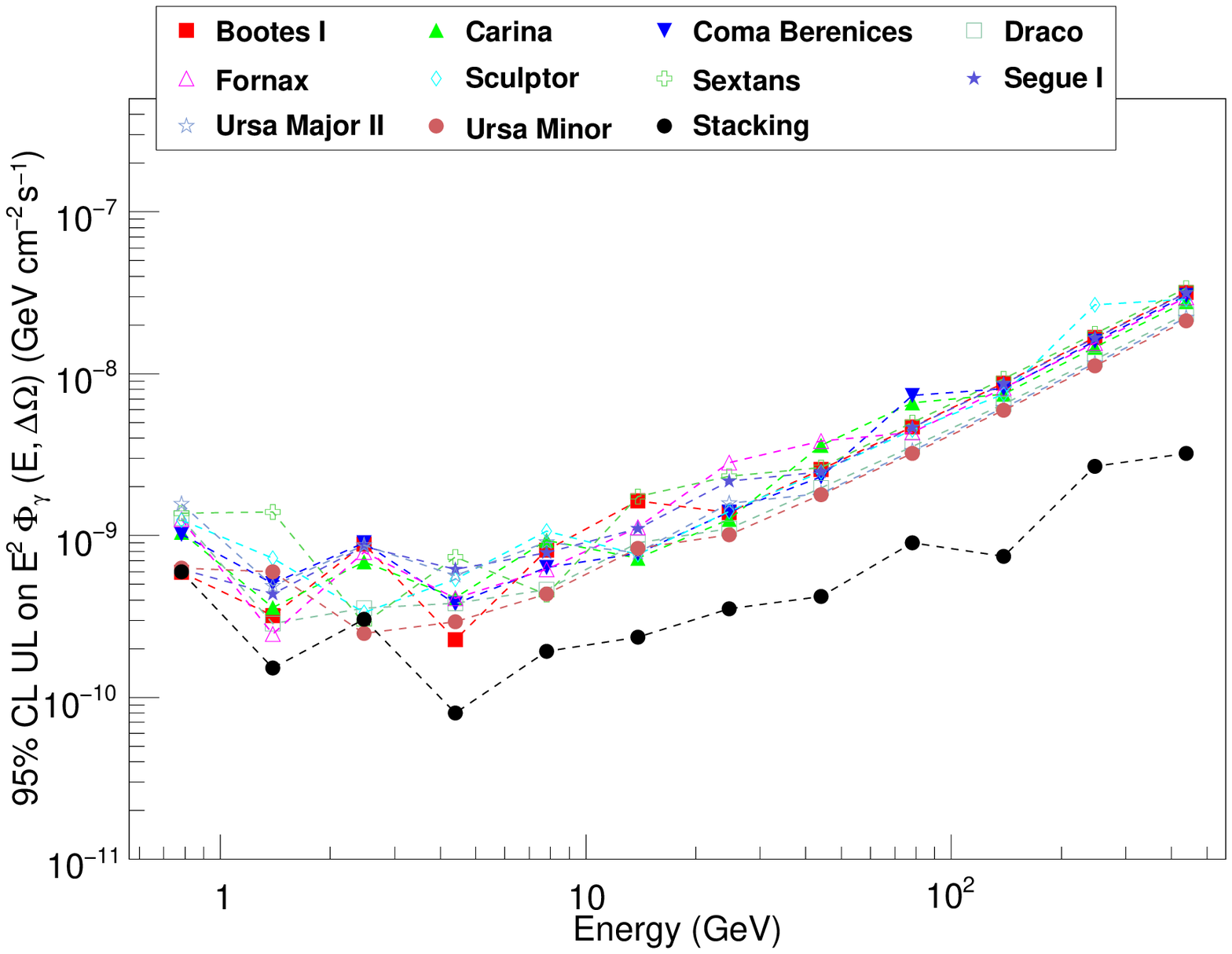}
\includegraphics[width=0.5\textwidth]{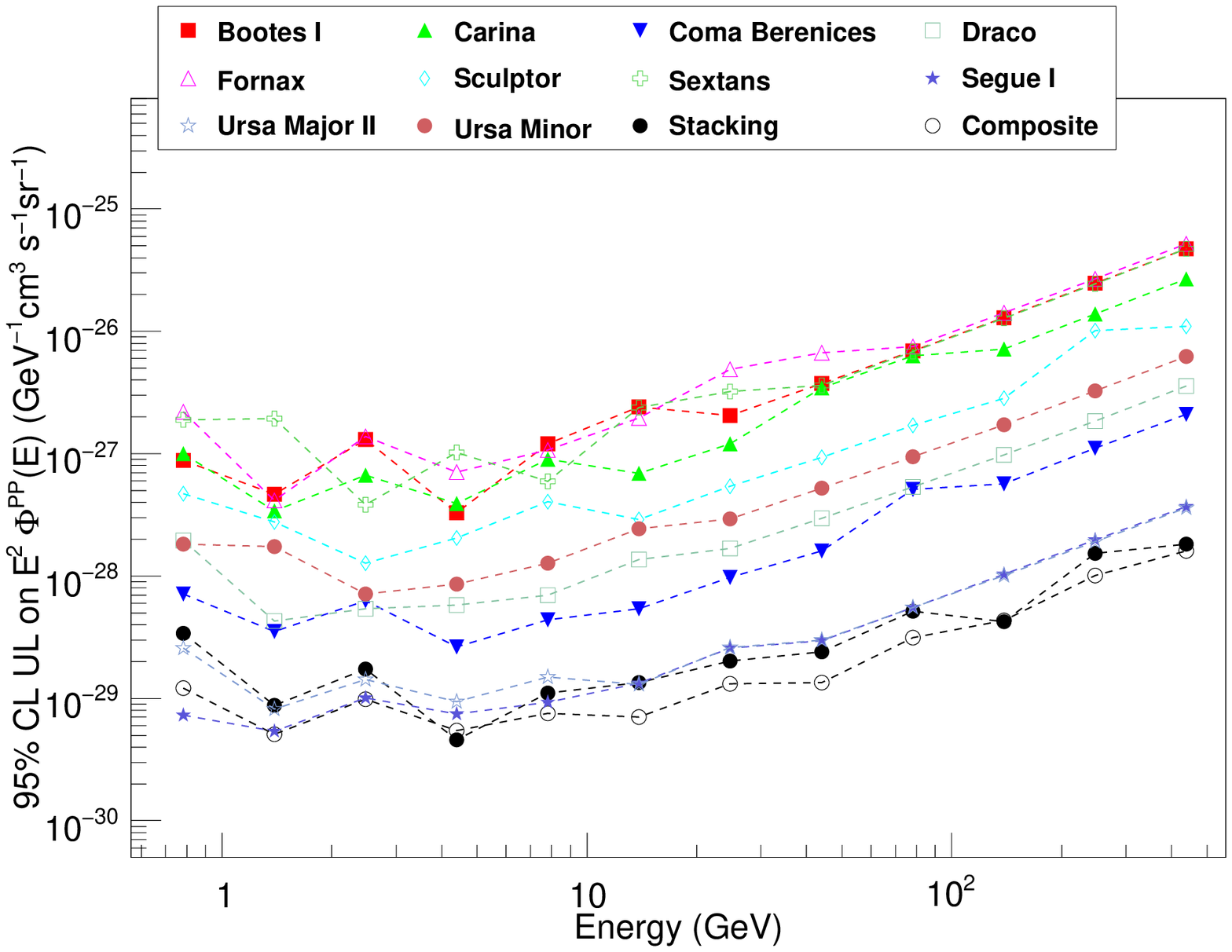}
\caption{Top panel: Upper limits at $95\%$ CL on the
gamma-ray flux as function of energy. Bottom panel: Upper limits
at $95\%$ CL on $\Phi^{PP}(E)$ as function of energy. The colored symbols correspond to the results obtained from the individual dSph galaxies. The open black circles indicate the results obtained from the composite analysis, while the filled black circles indicate the results obtained from the stacking analysis of all the dSph galaxies.}
\label{fig:ULFluxOverJ}
\end{figure}

\begin{figure*}
\includegraphics[width=0.48\textwidth]{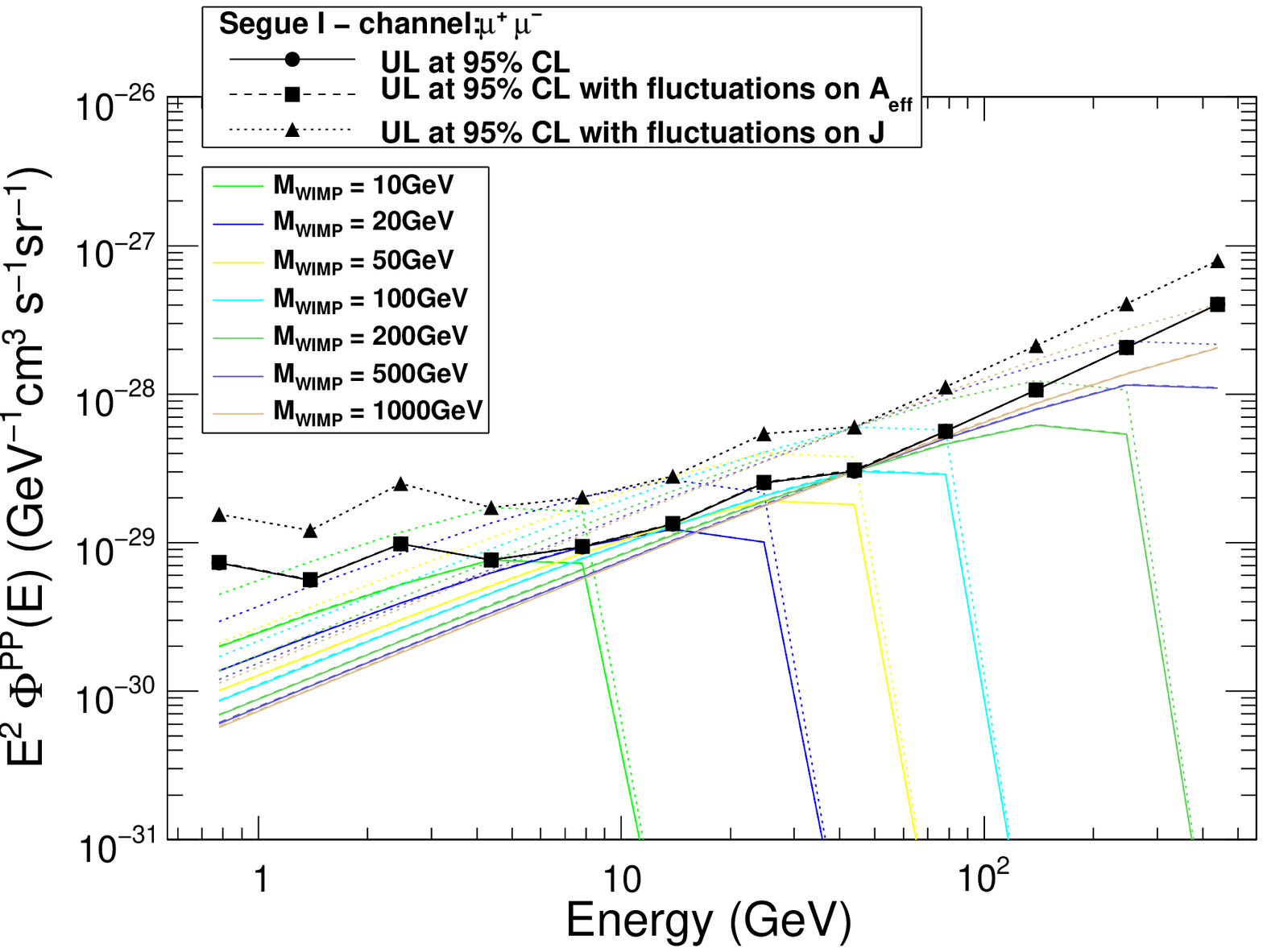}
\includegraphics[width=0.48\textwidth]{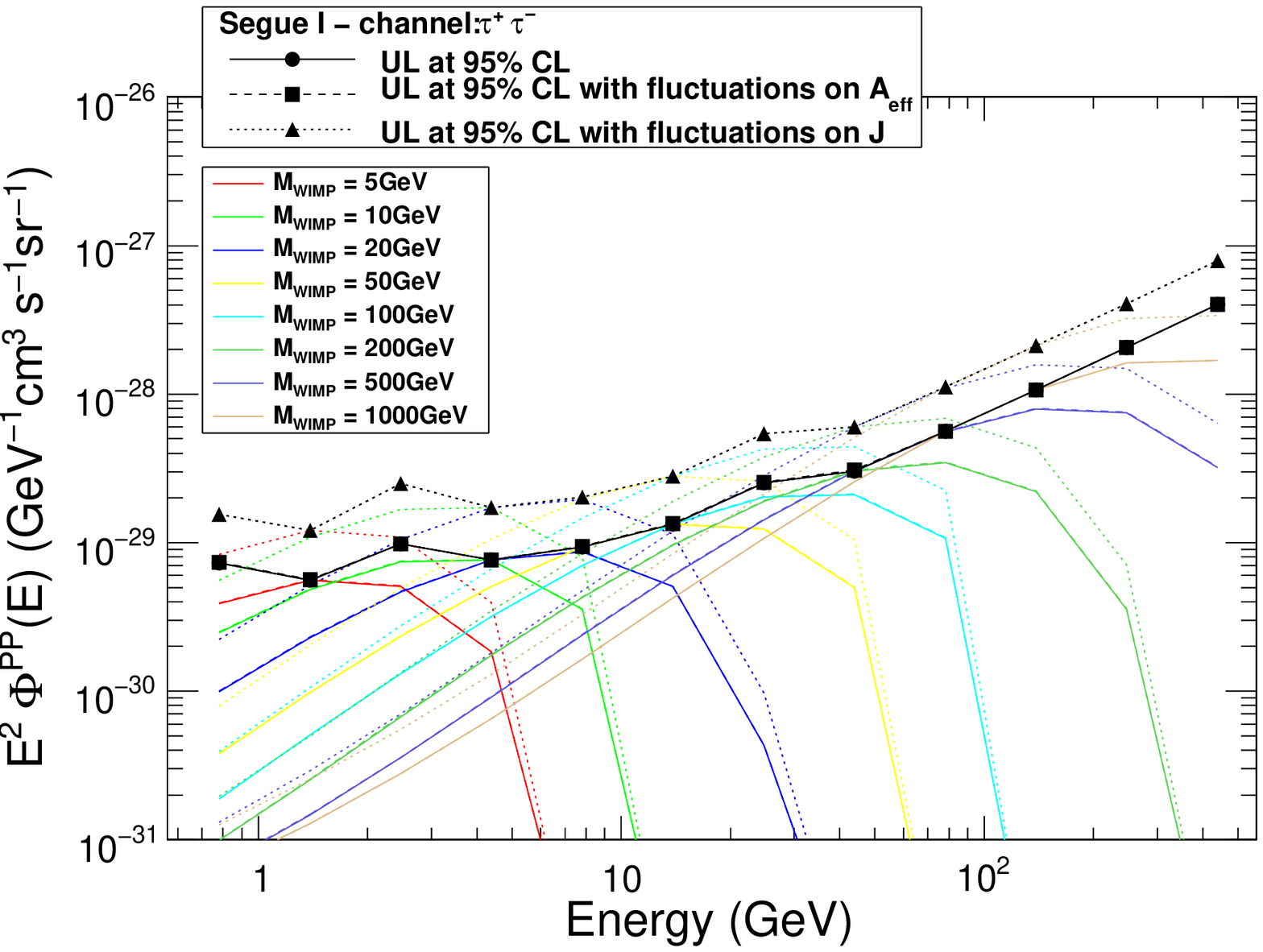}
\includegraphics[width=0.48\textwidth]{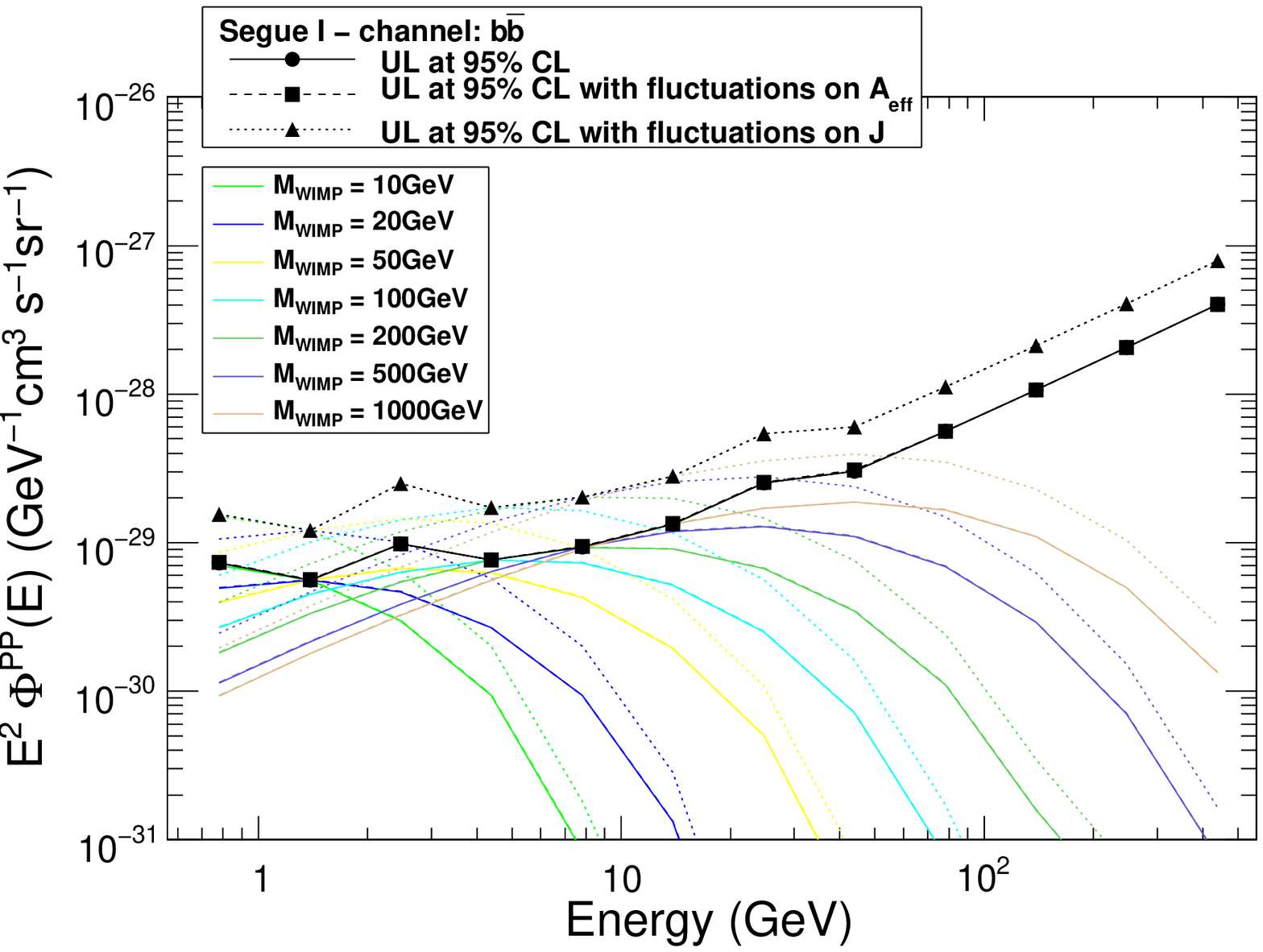}
\includegraphics[width=0.48\textwidth]{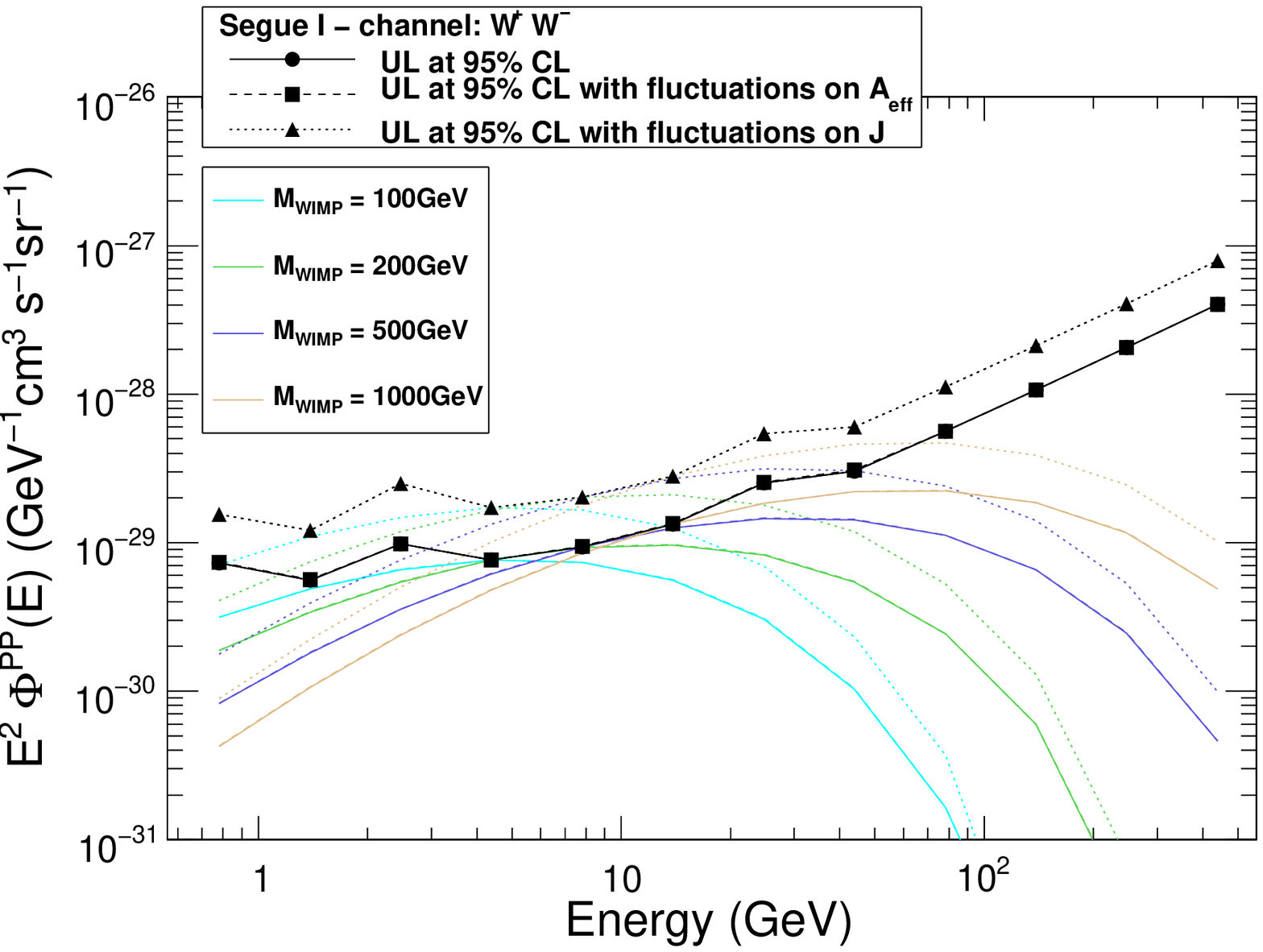}
\caption{Evaluation of the upper limits on $\langle \sigma v \rangle$
as a function of the true energy for several WIMP mass values in the
case of Segue I. The blacks lines with full circles correspond to the upper limits at $95\%$ CL on $\Phi^{PP}(E)$. The colored lines, each corresponding to a specified WIMP mass, indicate the maximum allowed values of $\Phi^{PP}(E)$ that do not exceed the measured upper limits. The four panels refer to WIMP annihilation into $\mu^{+} \mu^{-}$, $\tau^{+} \tau^{-}$, $b \bar{b}$ and $W^{+}W^{-}$ (as labeled). The dashed lines with filled squares and the dotted lines with filled triangles indicate the upper limits evaluated taking into account the systematic uncertainties on the effective area and on the J-factors, respectively. The effects of the systematic uncertainties on the effective area are negligible (dashed lines and filled squares are almost coincident with continuous lines and filled circles).}
\label{fig:ulexample}
\end{figure*}

\begin{figure*}
\includegraphics[width=0.32\textwidth]{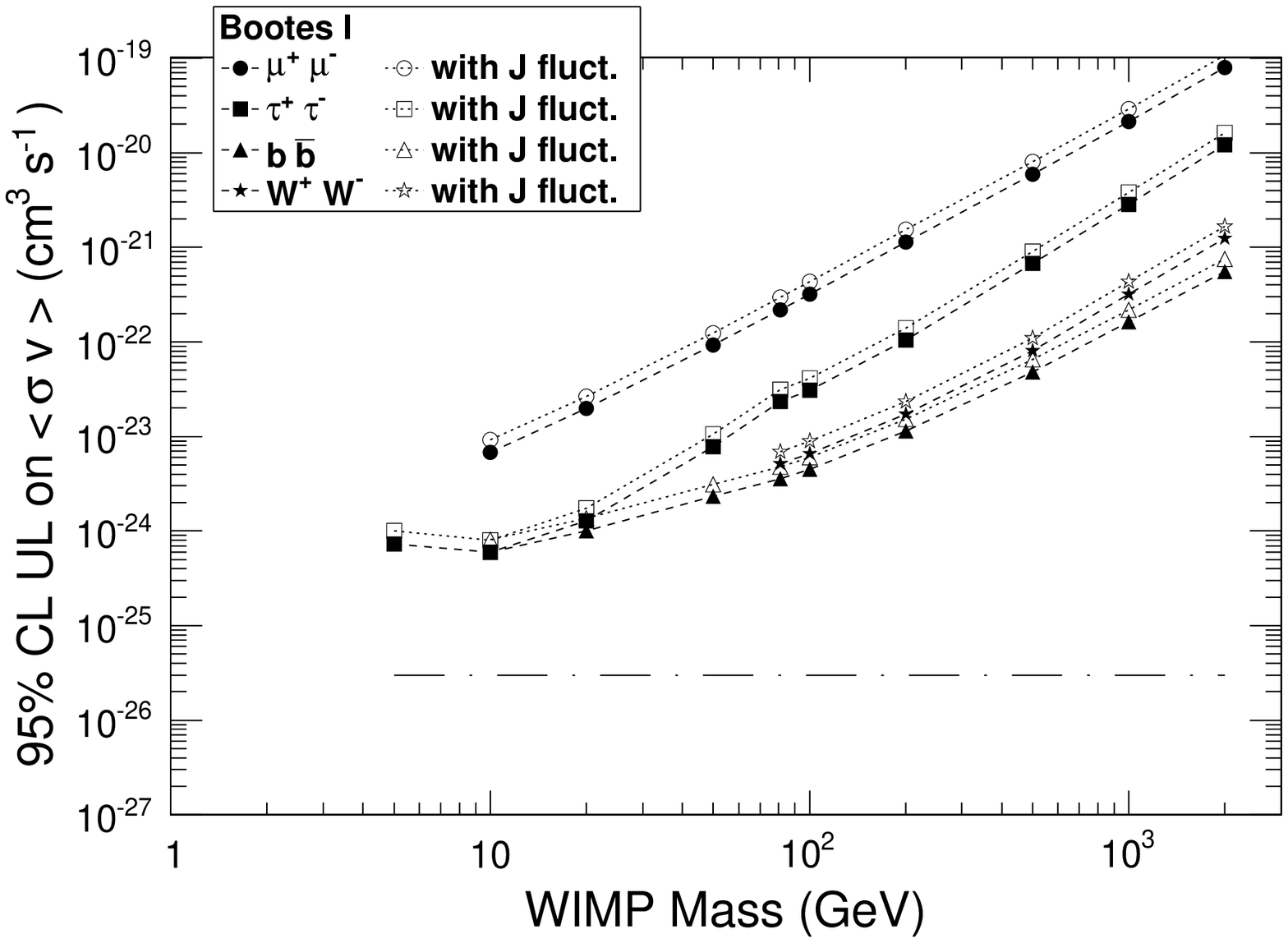}
\includegraphics[width=0.32\textwidth]{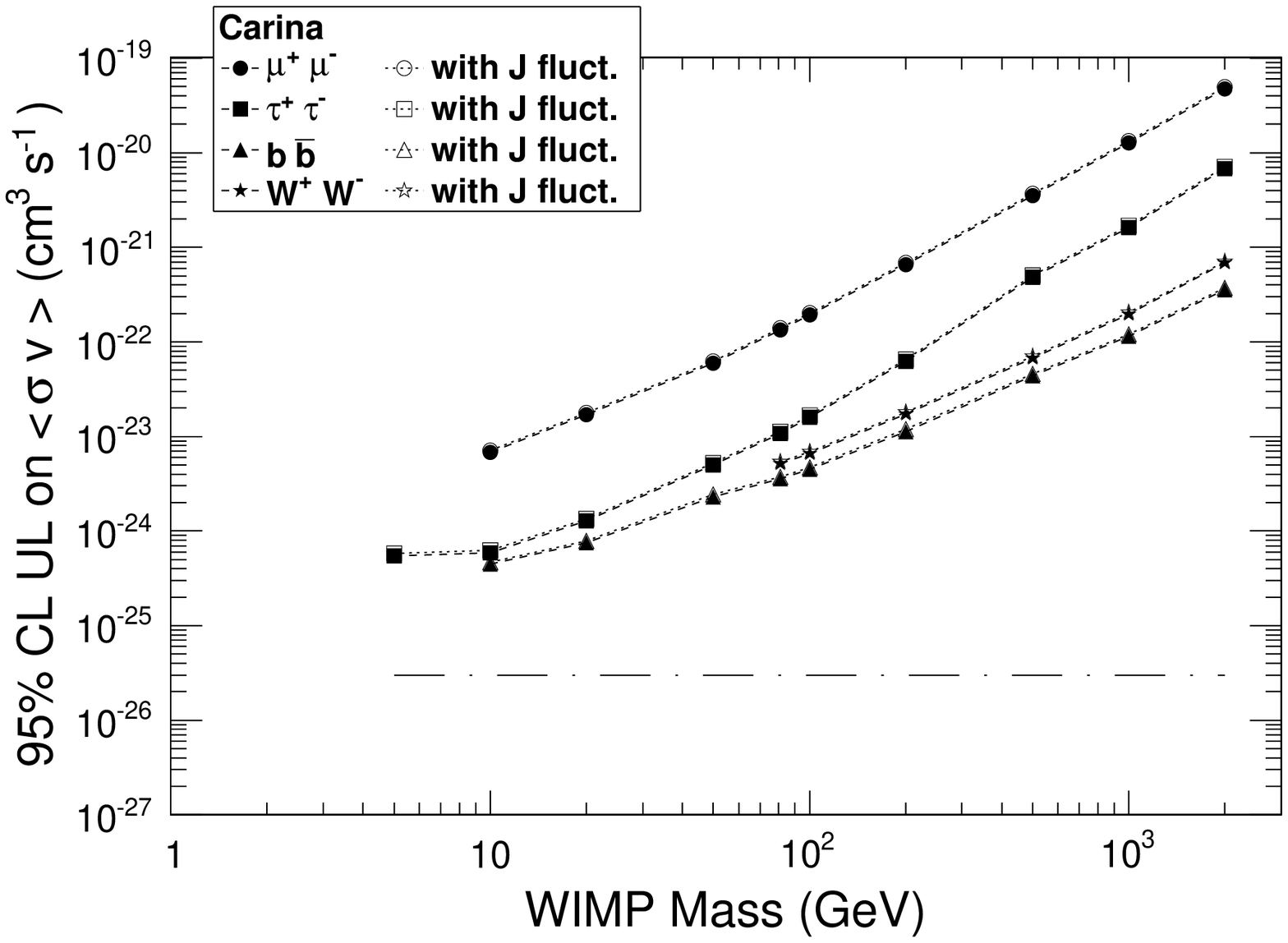}
\includegraphics[width=0.32\textwidth]{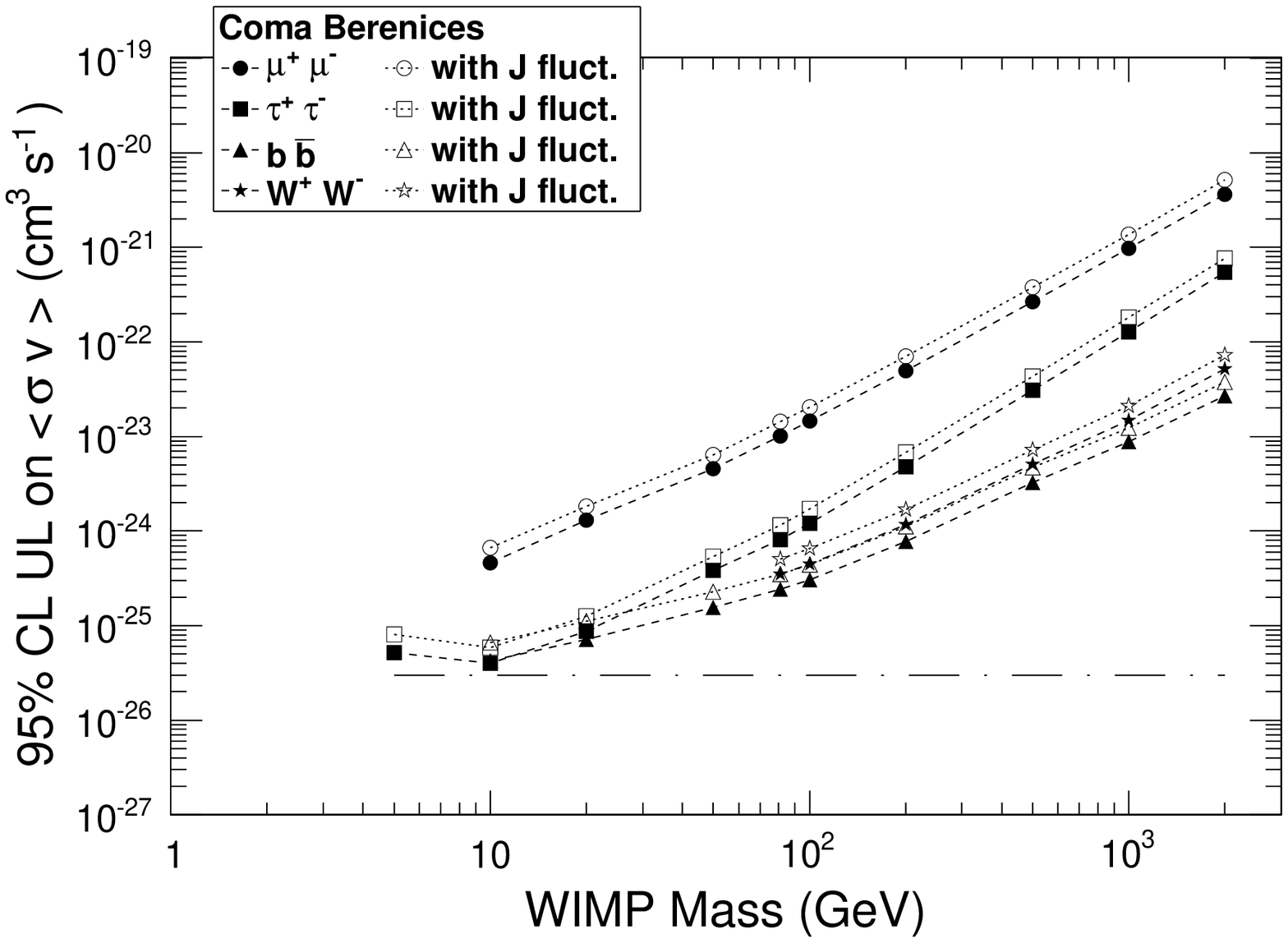}
\includegraphics[width=0.32\textwidth]{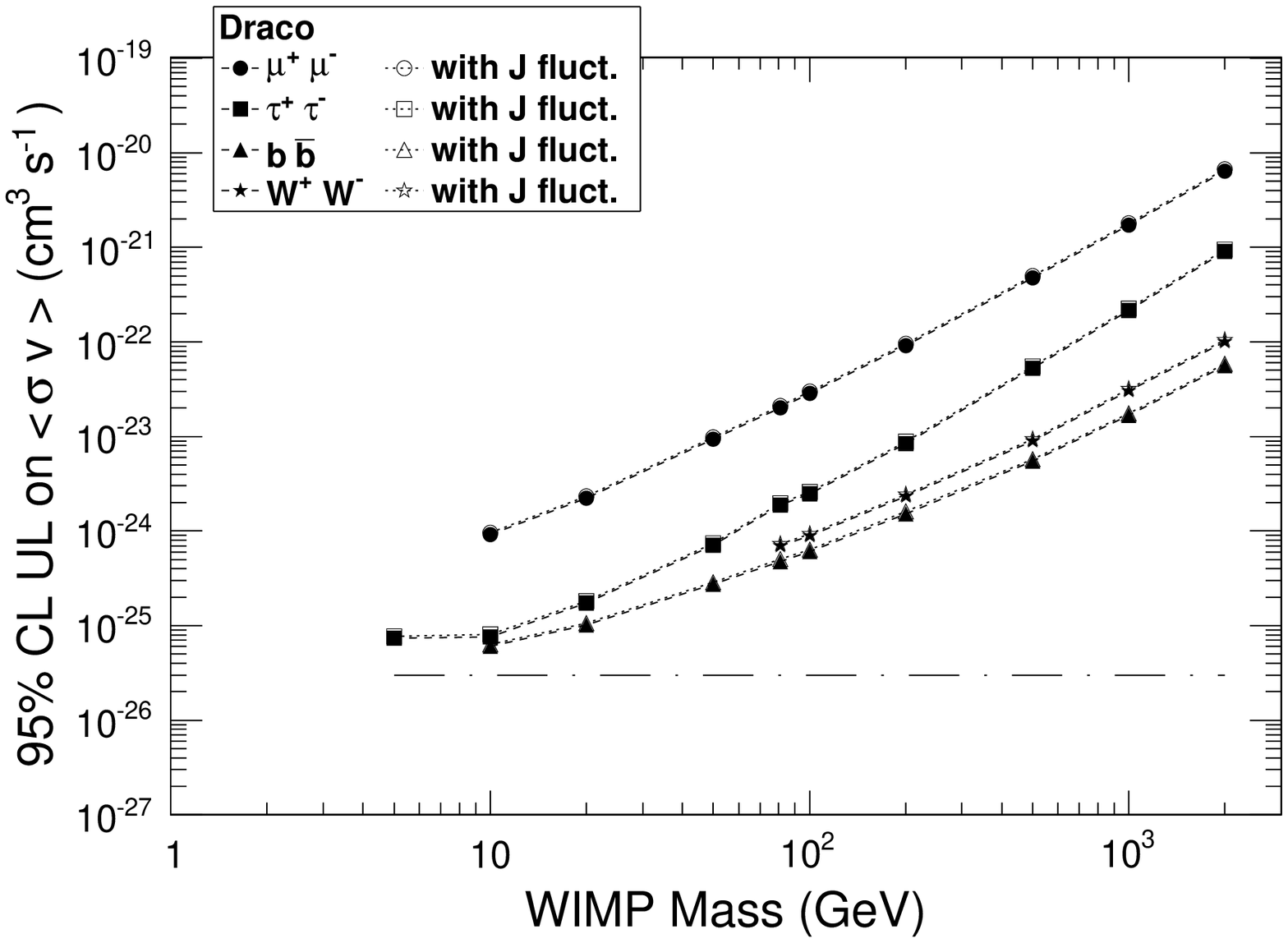}
\includegraphics[width=0.32\textwidth]{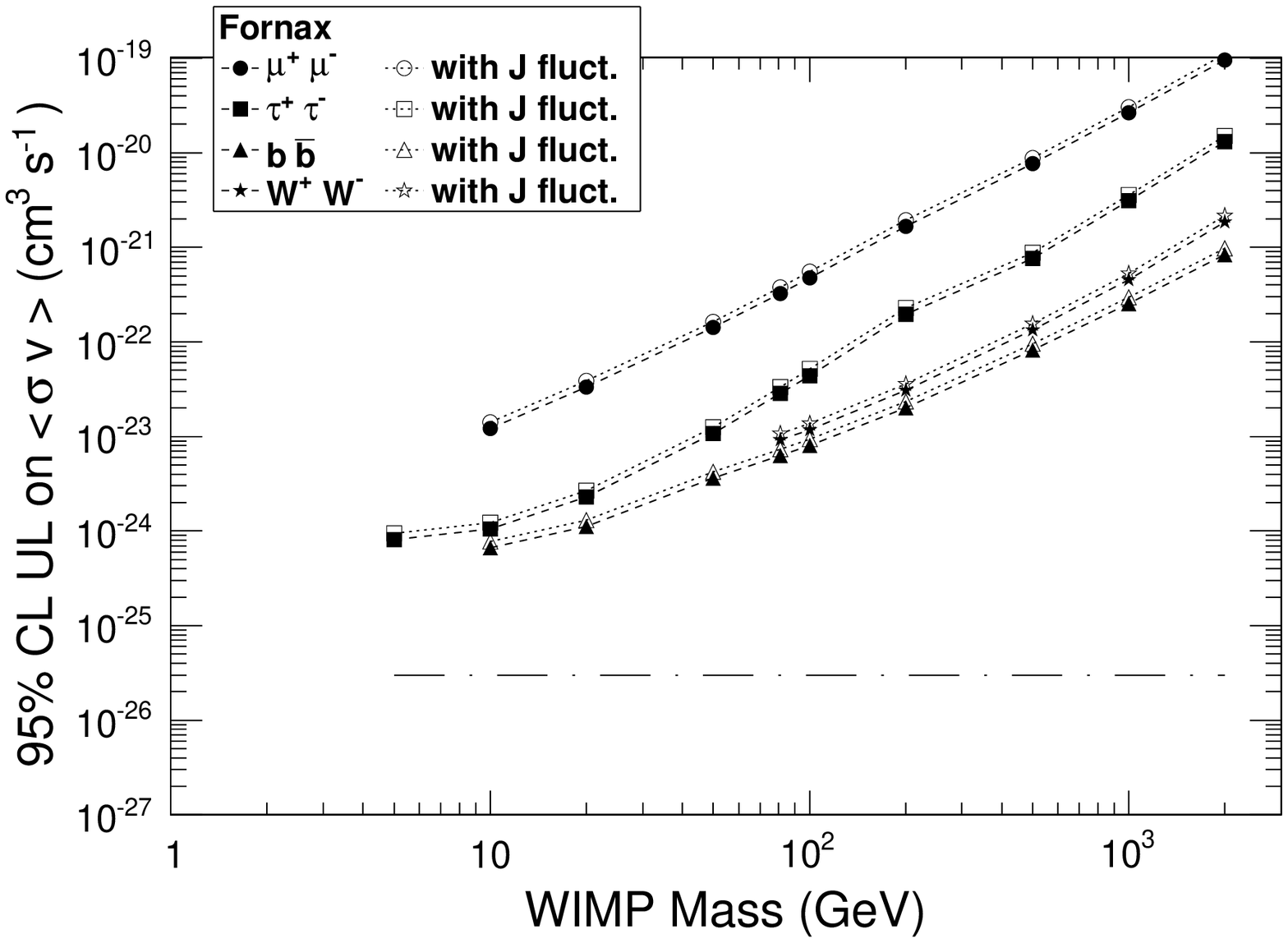}
\includegraphics[width=0.32\textwidth]{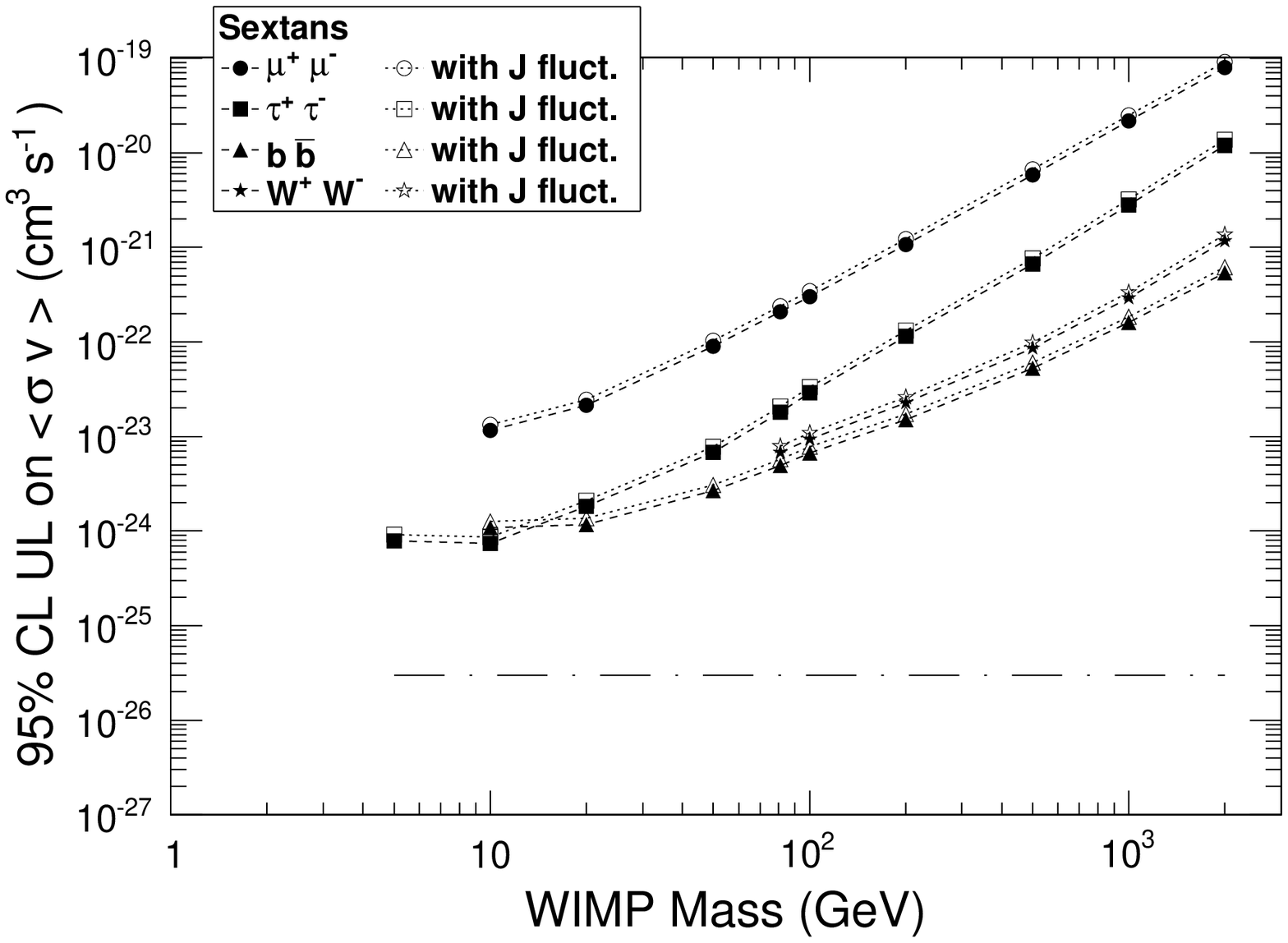}
\includegraphics[width=0.32\textwidth]{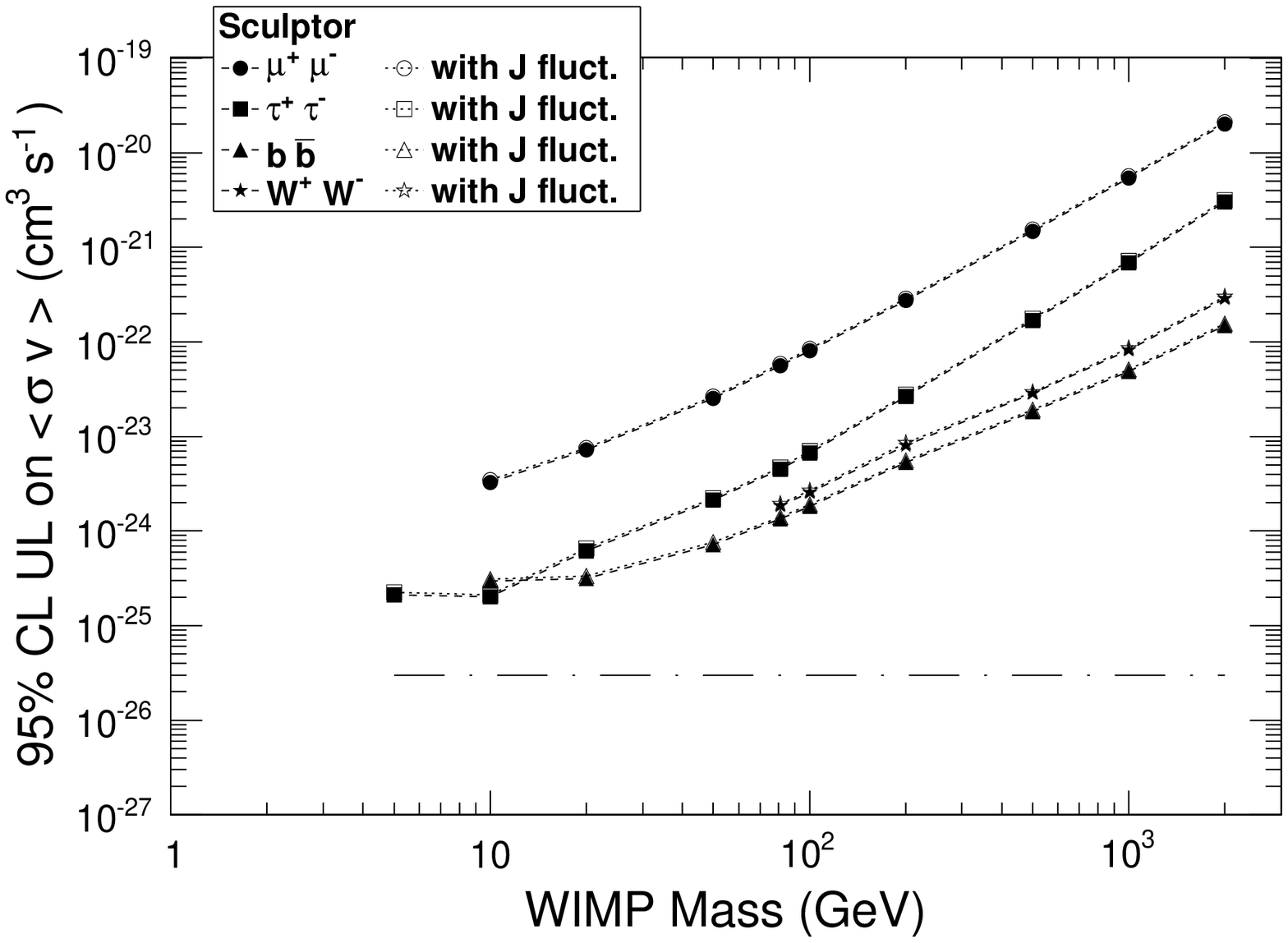}
\includegraphics[width=0.32\textwidth]{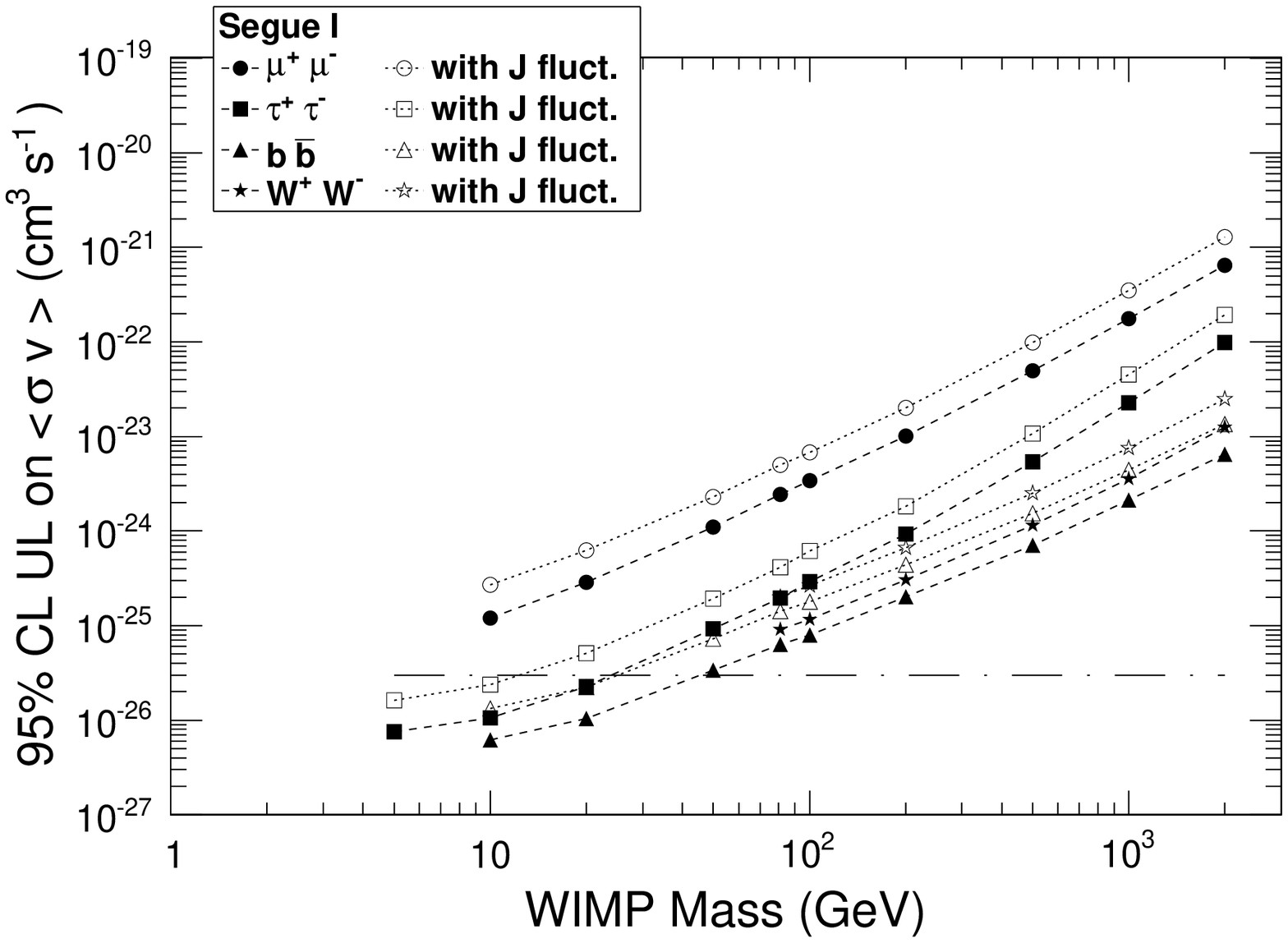}
\includegraphics[width=0.32\textwidth]{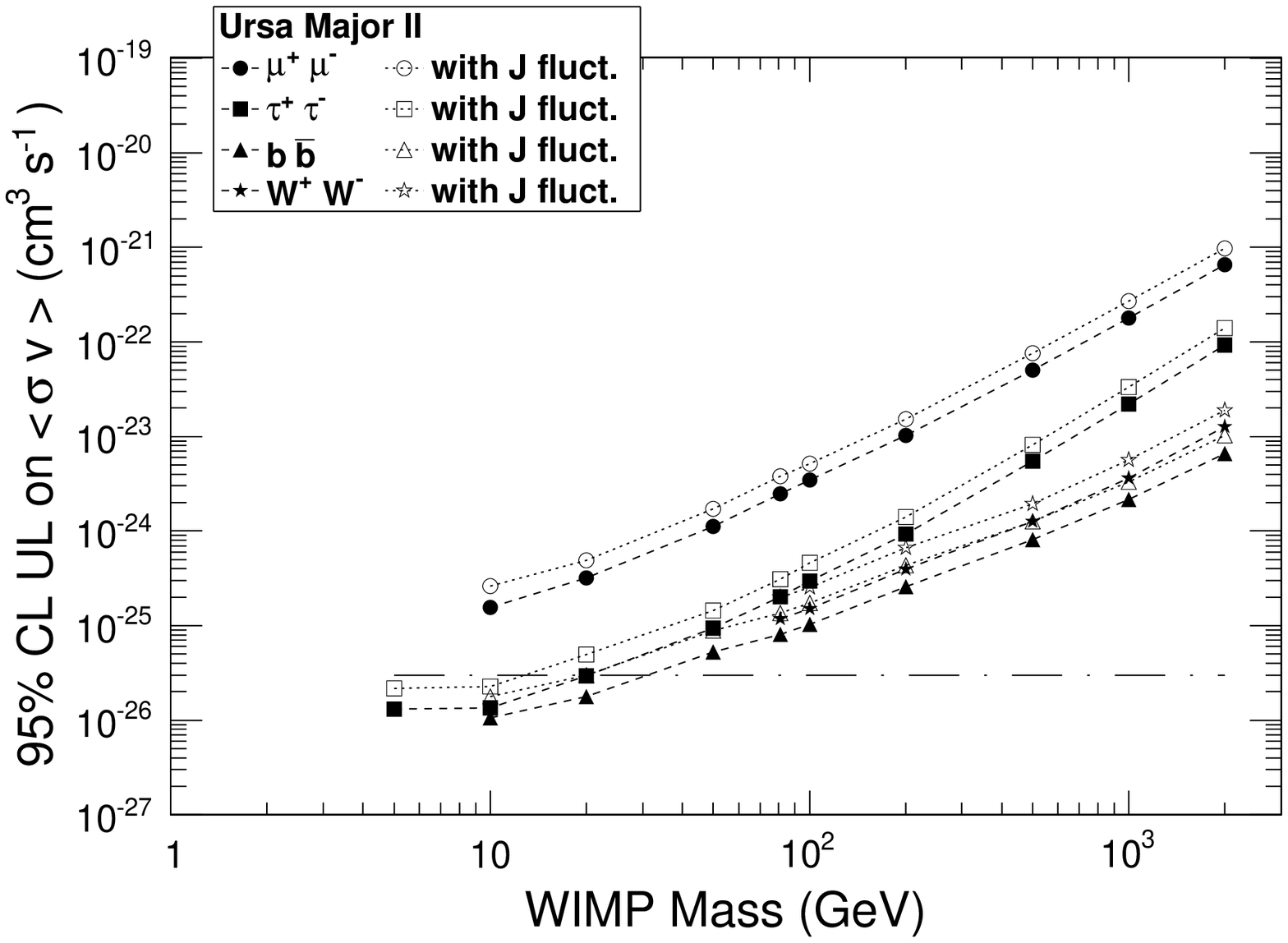}
\includegraphics[width=0.32\textwidth]{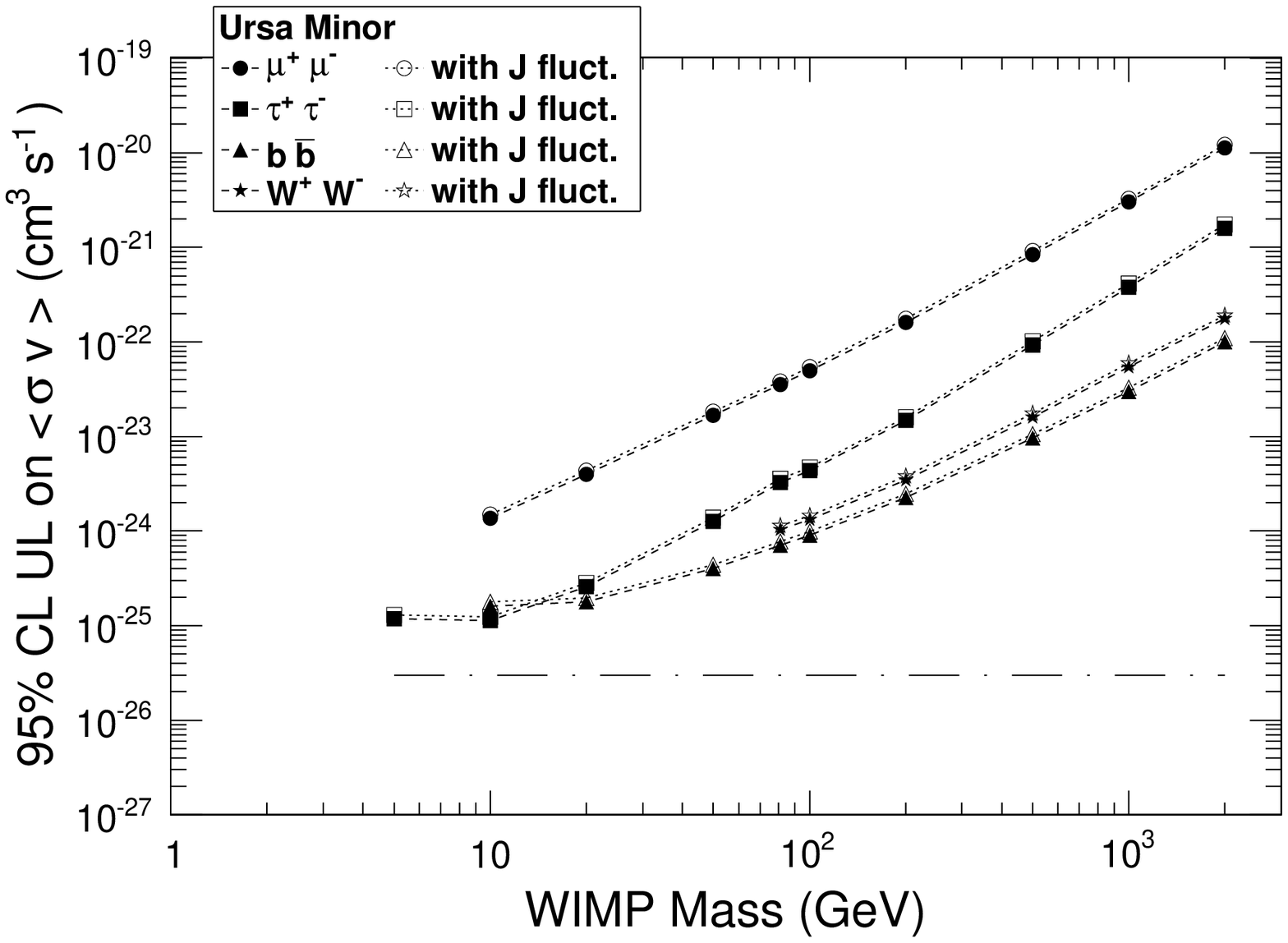}
\includegraphics[width=0.32\textwidth]{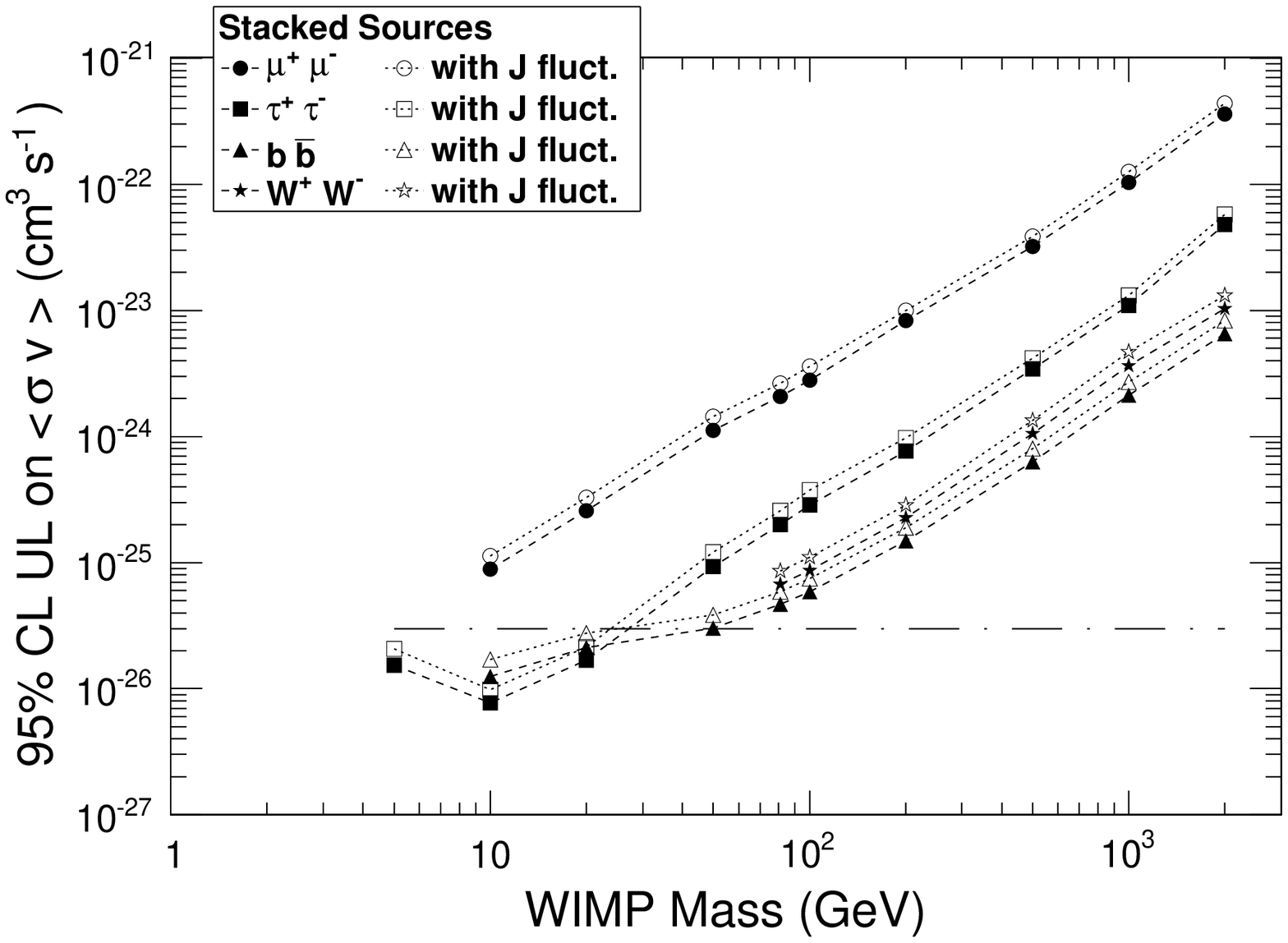}
\includegraphics[width=0.32\textwidth]{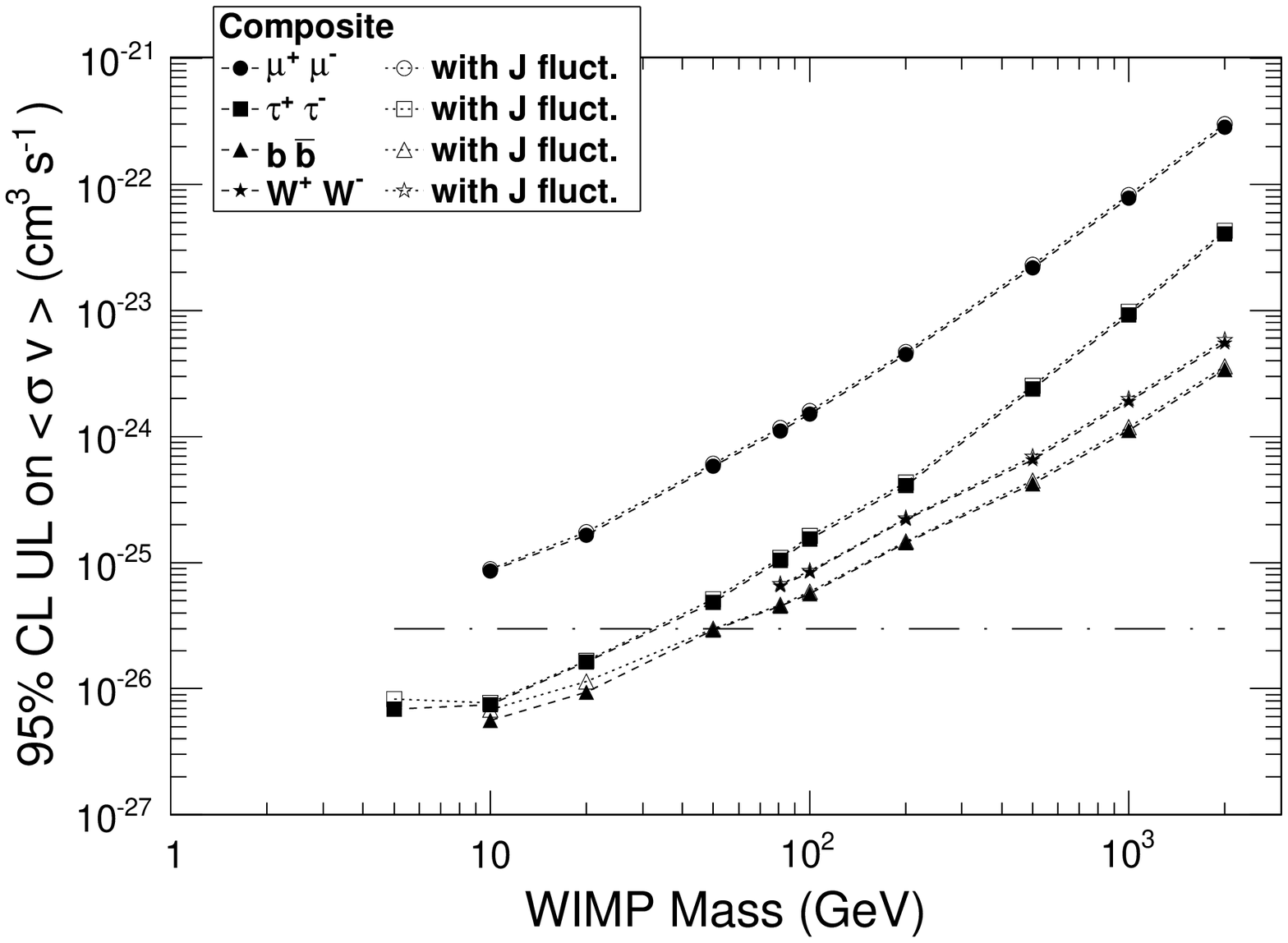}
\caption{Upper limits at $95\%$ CL on $\langle \sigma v \rangle$ as function of the WIMP mass for the annihilation channels $\mu^{+} \mu^{-}$, $\tau^{+} \tau^{-}$, $b \bar{b}$ and $W^{+} W^{-}$. The plots show the upper limits obtained from the analysis of individual dSph galaxies and from the stacking and composite analyses. The continuous lines indicate the upper limits obtained neglecting the systematic uncertainties, while the dotted lines indicate the upper limits obtained including the uncertainties on the $J$-factors. The long dashed line corresponds to the canonical value of the annihilation cross section of $3\times10^{-26} \units{cm^3 s^{-1}}$ in the thermal relic WIMP scenario.}
\label{fig:sigmavlimits}
\end{figure*}

Fig.~\ref{fig:ULFluxOverJ} shows the upper limits at $95\%$ CL on the flux (top panel) and on $\Phi^{PP}(E)$ (bottom panel) as function of the energy, for each of the dSph galaxies considered in this analysis. The more constraining limits are those obtained from the dSph galaxies with the highest $J$-factors. 

Once the individual sources were analyzed, the stacking analysis and the composite analysis were implemented following the procedures described in \S\ref{sec:stacking} and in \S\ref{sec:composite}. As mentioned above, for the stacking analysis the counts from the signal and background regions corresponding to each source were added, and the upper limits on the signal were evaluated following the same procedure as for individual sources. This approach is equivalent to stacking the events from all the dSph galaxies 
and then analyzing the single image obtained from the superposition of all the individual images (see the last plots in Figs.~\ref{fig:countmaps} and \ref{fig:countdist}). 

It is worth noting that in the high-energy bands, where the counts in the signal and background regions are both null (i.e. $n = m = 0$), as shown in Fig.~\ref{fig:countdist}, the upper limit evaluated on the signal counts is always constant (i.e., at $95\%$ CL the upper limits on $s$ corresponds to about $3$ counts). Hence, when performing the stacking analysis, the upper limit on the flux (see top panel of Fig.~\ref{fig:ULFluxOverJ}) will decrease linearly with the number of stacked sources. In other words, in the stacking analysis the observations of different sources are added and the result is expected to be equivalent to a single observation of an individual source with a total live time corresponding to the sum of the live times of each observation. On the other hand, in the low-energy band, where the counts in the signal region are roughly equal to those in the background region (i.e. $n \approx c~m$), the upper limit on the signal counts $s$ is roughly proportional to the square root of the observed events~\cite{lopmaz}. In this case, since the total live time will be roughly proportional to the number of stacked sources, the upper limit on the stacked flux will improve with the square root of the number of stacked sources. 

In the bottom panel of Fig.~\ref{fig:ULFluxOverJ} the upper limits on
$\Phi_{PP}(E)$ are shown for all the candidate sources as well as for
the stacking and composite analyses. As shown in
Tab.~\ref{tab:dwarfs}, the J-factors of the $10$ dSph galaxies studied
in the present analysis are distributed in an interval that spans two
orders of magnitude. As a consequence, since the upper limits on the
photon fluxes are roughly similar, the upper limits on $\Phi^{PP}(E)$
will span two orders of magnitude. The stacking analysis
improves the upper limits on $\Phi^{PP}(E)$ by a factor that ranges
from a few to about $10$ with respect to those obtained from the
analysis of individual dSph galaxies. When considering the quantity
$\Phi^{PP}(E)$, since the $J$-factor used in the stacking analysis is
evaluated as the average of the $J$-factors of individual sources
weighted  with their exposures, the result is an improvement of a factor of a few
 with respect to the upper limits obtained from the analysis of
the source with the highest $J$-factor.

The results from the composite analysis are in general more
constraining than those from the stacking analysis. This could
be due to the fact that the random variable used to evaluate the upper
limits in the composite analysis is $\propto \prod_{i} s_{i}/J_{i}$,
while the random variable used in the stacking analysis is $\propto
(\sum_{i} s_{i}) / \langle J \rangle$. This means that the ``effective
$J$-factor'' for the composite analysis could in principle be different
from that for the stacking analysis. 

The measured upper limits on $\Phi^{PP}(E)$ were converted
into upper limits on $\langle \sigma v \rangle$ following the
procedure described in \S\ref{sec:individualanalysis}. 
Fig.~\ref{fig:ulexample} shows an example of this calculation in the
case of Segue I for the annihilation channels $\mu^{+} \mu^{-}$, $\tau^{+}
\tau^{-}$, $b \bar{b}$ and $W^{+} W^{-}$. The upper limits were also
evaluated taking into account separately the uncertainties on the
effective area and on the $J$-factors. To describe the systematic
uncertainties on the effective area we assumed a uniform PDF centered
on the average value $A(E)$ in each energy bin with fluctuations
of $\pm10\%$. These uncertainties have a negligible effect on the
upper limits. As discussed above, the effects on the upper limits due
to the systematic uncertainties on the J-factor were evaluated
assuming for $J$ a uniform PDF in a range corresponding to the $68\%$
area of the actual $J$-factor distribution. As shown in
Fig.~\ref{fig:ulexample}, the effects of the uncertainties on the
$J$-factor are not negligible and, depending on the source under
investigation, the upper limits on $\Phi^{PP}(E)$ can increase by up to a factor of a few. 

Fig.~\ref{fig:sigmavlimits} shows the upper limits at $95\%$ CL on $\langle \sigma v \rangle$ obtained from the analysis of individual dSph galaxies, from the stacking analysis and from the composite analysis, as a function of the WIMP mass for the annihilation channels $\mu^{+} \mu^{-}$, $\tau^{+} \tau^{-}$, $b \bar{b}$ and $W^{+} W^{-}$. The upper limits obtained by taking into account the effects of the uncertainties on the J-factors are also shown. 

\section{Analysis of the Milky Way Halo}
\label{sec:halo}

The study of the Milky Way halo is quite complex because its gamma-ray emission has to be disentangled from that of known gamma-ray sources. However, a possible approach to the study of the Milky Way halo is that of selecting a set of sky directions that are well-separated from known gamma-ray sources.

\begin{figure}[t]
\begin{center}
\includegraphics[width=0.48\textwidth]{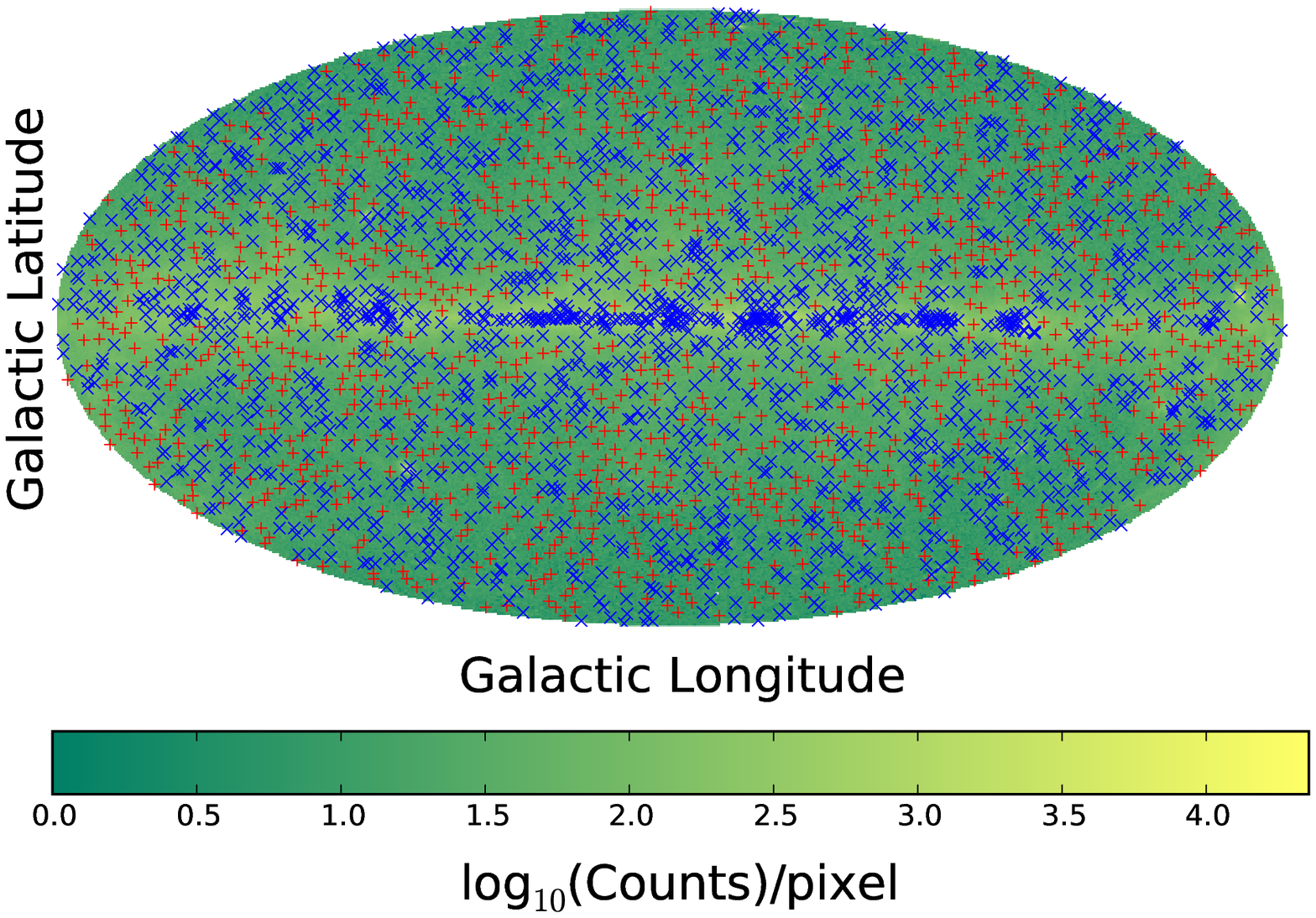}
\end{center}
\caption{
Count map used for the analysis of the Milky Way Halo. 
The map was built in the Galactic reference frame, using the HEALPix
pixelization scheme with $N_{side}=128$ ($196608$ pixels, each
covering a solid angle of $6.4 \cdot 10^{-5}\units{sr}$), and is displayed in the
Aitoff projection. The Galactic
Center is in the middle of the map. The $1000$ random directions are 
indicated with the red markers; the sources in the 2FGL Catalog
are indicated with the blue markers.
}
\label{fig:maprandom}
\end{figure}

For this analysis a set of $1000$ random directions was generated in the sky, each direction located at an angular distance of at least $3\degrees$ from all the $1873$ point sources and at least $3\degrees$ plus twice the size of the semi-major axis from all the $11$ extended sources in the 2FGL Catalog~\cite{2FGL}. The random positions are illustrated in Fig.~\ref{fig:maprandom} in Galactic coordinates. Since many gamma-ray sources are concentrated in the region of the Galactic plane, we also decided to perform a separate analysis selecting only random directions at an angular distance larger than $10\degrees$ from the Galactic plane (i.e. all directions with Galactic latitude $|\beta|>10\degrees$). A subset of the initial sample, consisting of $866$ random directions, was used for this analysis.  

The analysis of the Milky Way Halo was performed selecting P7CLEAN\_V6 class events in order to guarantee optimal rejection of the charged particle background. As in the case of the dSph galaxies, discussed in \S\ref{sec:anadwarfs}, the data analysis was performed selecting gamma rays with energies from $562\units{MeV}$ to $562\units{GeV}$, with the energy interval being divided into $12$ bins, equally spaced on a logarithmic scale. 

In the case of the Galactic halo analysis, we hypothesized an extreme
scenario in which all the detected photons originate from DM
annihilation. In this analysis the upper limits on $\Phi^{PP}(E)$, and
consequently those on $\langle \sigma v \rangle$, were evaluated assuming 
an absence of background events. For each random direction the 
signal region was defined as a cone of angular radius $\Delta \theta =
1\degrees$ centered on it. The analysis was then performed 
stacking the data from all the random directions, without 
background subtraction (i.e. $c=0$ and $m=0$). 
This strategy relies on the assumption that a possible DM-induced 
gamma-ray flux must be lower than the total observed flux.
The upper 
limits obtained from this analysis therefore will be conservative. 

\begin{figure}[h]
\includegraphics[width=0.48\textwidth]{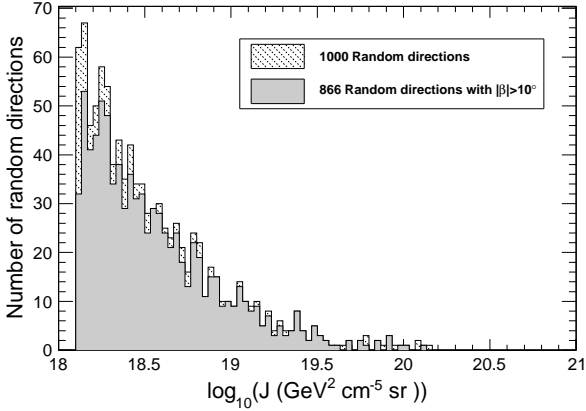}
\caption{Distribution of the $J$-factors evaluated for the random directions in the Galactic halo. The dashed filled region shows the J-factors of all the $1000$ directions; the grey filled region shows the J-factors for the $866$ sky directions that are separated more than $10 \degrees$ from the Galactic plane.}
\label{fig:HaloJdist}
\end{figure}

\begin{figure}[hb]
\includegraphics[width=0.48\textwidth]{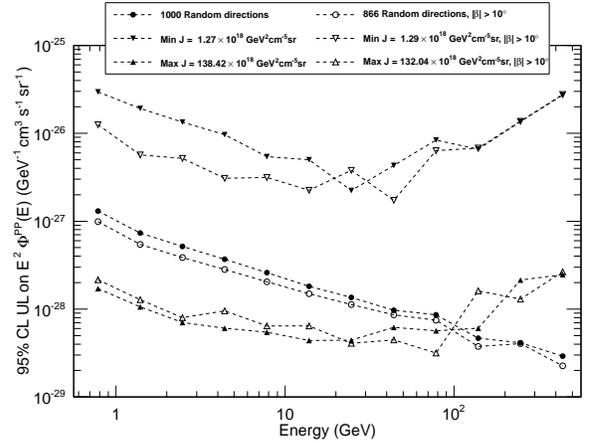}
\caption{Comparison of the $95\%$ CL upper limits on
$\Phi^{PP}(E)$ as function of the energy for all of the $1000$ random
directions and for the $866$ directions outside the Galactic plane 
($|\beta| > 10\degrees$). The upper limits obtained from the
directions corresponding to the highest and lowest $J$-factors are also
shown. The lowest $J$-factor among the $1000$ sky directions is
associated with the direction (in Galactic coordinates) 
$(\lambda, \beta) = (183.616\degrees, -0.189\degrees)$, while the
highest $J$-factor is associated with the direction 
$(356.796\degrees, 9.23\degrees)$. The lowest $J$-factor among the $866$
sky directions outside the Galactic plane is associated with the 
direction $(179.968\degrees,-11.5651\degrees)$, while the highest $J$-factor
 is associated with the direction $(0.621474\degrees, 10.0793\degrees)$.}
\label{fig:4ul_comparison}
\end{figure}

For any given random direction (hereafter we will refer to random
directions as sources) the $J$-factor was evaluated using
Eq.~\ref{eq:jfactor}. Since the signal region is a narrow cone of 
$1\degrees$ angular radius, Eq.~\ref{eq:jfactor} reduces to:

\begin{equation}
J \approx \Delta \Omega \int \rho^{2}(l(\psi)) dl.
\label{eq:jfactorrandom}
\end{equation}
where $\Delta \Omega \approx 9.6\cdot 10^{-4} \units{sr}$ is the solid
angle corresponding to the signal region. In the previous equation we
explicitly wrote the dependence of the DM density on the angle $\psi$, which
represents the angular separation of the source from the Galactic Center. 

In performing the calculations we assumed that the detector is 
located at the Sun's position and we used the Galactic reference
frame. The angle $\psi$ can then be calculated 
from the Galactic longitude and latitude ($\lambda$, $\beta$) using
the following relation:
 
\begin{equation}
\cos \psi = \cos \lambda \cos \beta.
\end{equation}

For the DM density we assumed a Navarro-Frenk-White (NFW) profile~\cite{nfw}:

\begin{equation}
\label{eq:nfw}
\rho(r) = \frac{\rho_{0}}
{\left( r/r_{s} \right) \left( 1 + r/r_{s} \right)^{2}}
\end{equation}
where $\rho_{0}=0.3 \units{GeV/cm^3}$ and $r_s = 20\units{kpc}$. The
coordinate $r$ in Eq.~\ref{eq:nfw} represents the distance 
from the Galactic Center, and is given by:

\begin{equation}
r = \sqrt{l^{2} + R_{0}^{2} - 2lR_{0}\cos\psi}
\end{equation}
where $R_{0}=8.5\units{kpc}$ is the distance of the Galactic Center from the
Sun and $l$ is the distance of the line element $dl$ from the Sun.

Fig.~\ref{fig:HaloJdist} shows the distribution of the $J$-factors evaluated for the sky directions shown in Fig.~\ref{fig:maprandom}. As expected, the directions with the highest J-factors are near the Galactic Center; on the other hand, the directions with the lowest $J$-factors are near the Galactic Anti-center.

\begin{figure*}
\includegraphics[width=0.48\textwidth]{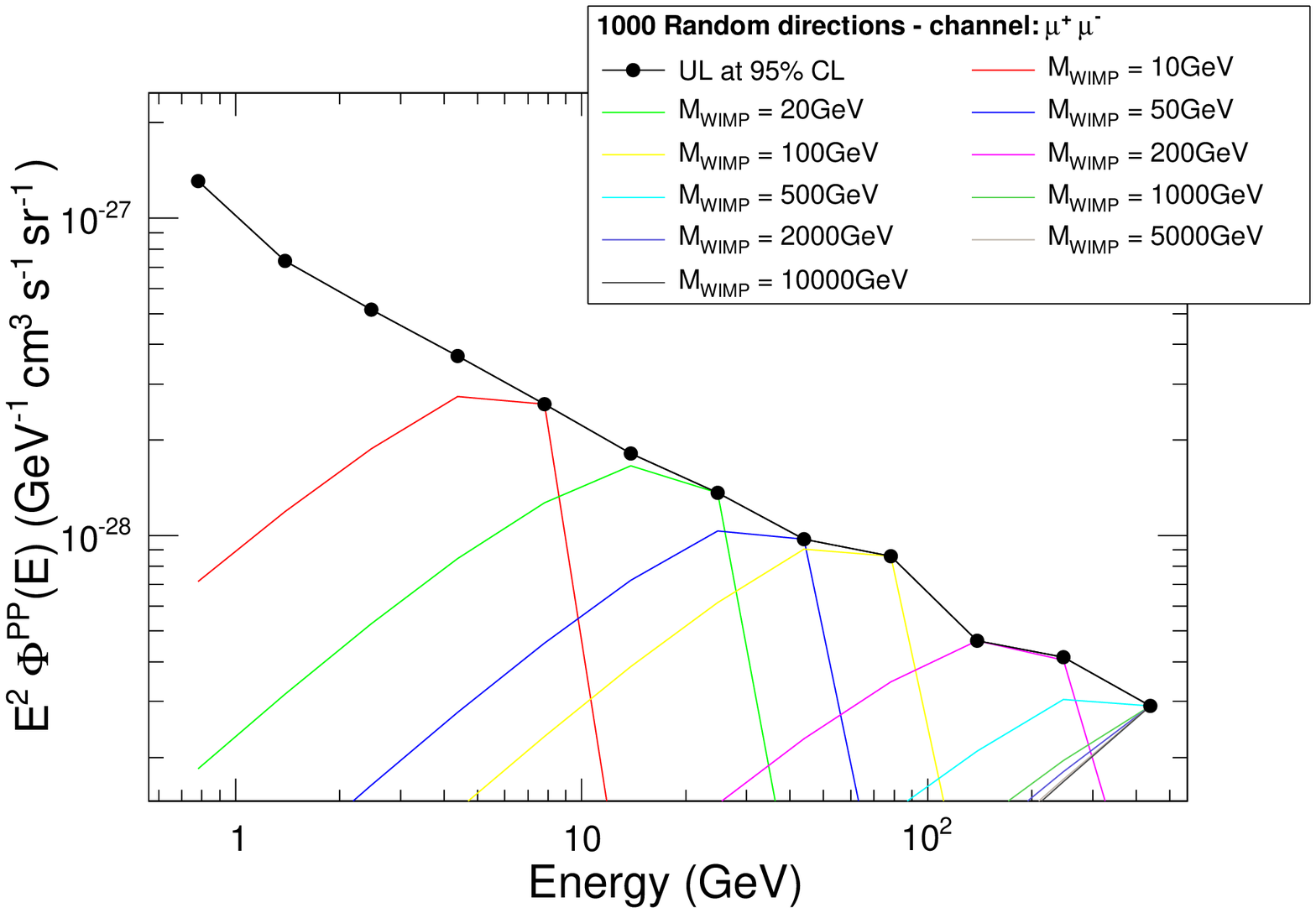}
\includegraphics[width=0.48\textwidth]{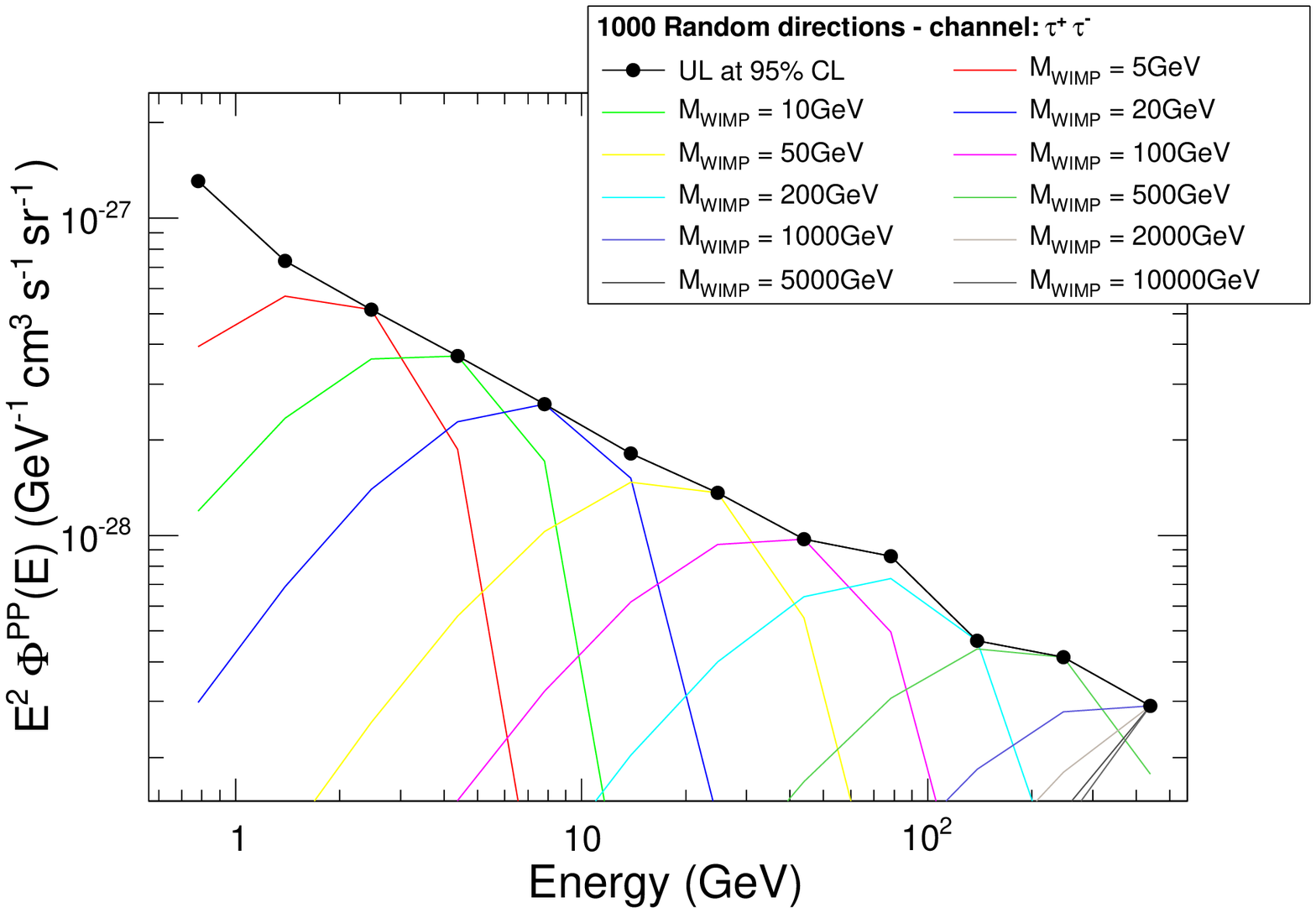}
\includegraphics[width=0.48\textwidth]{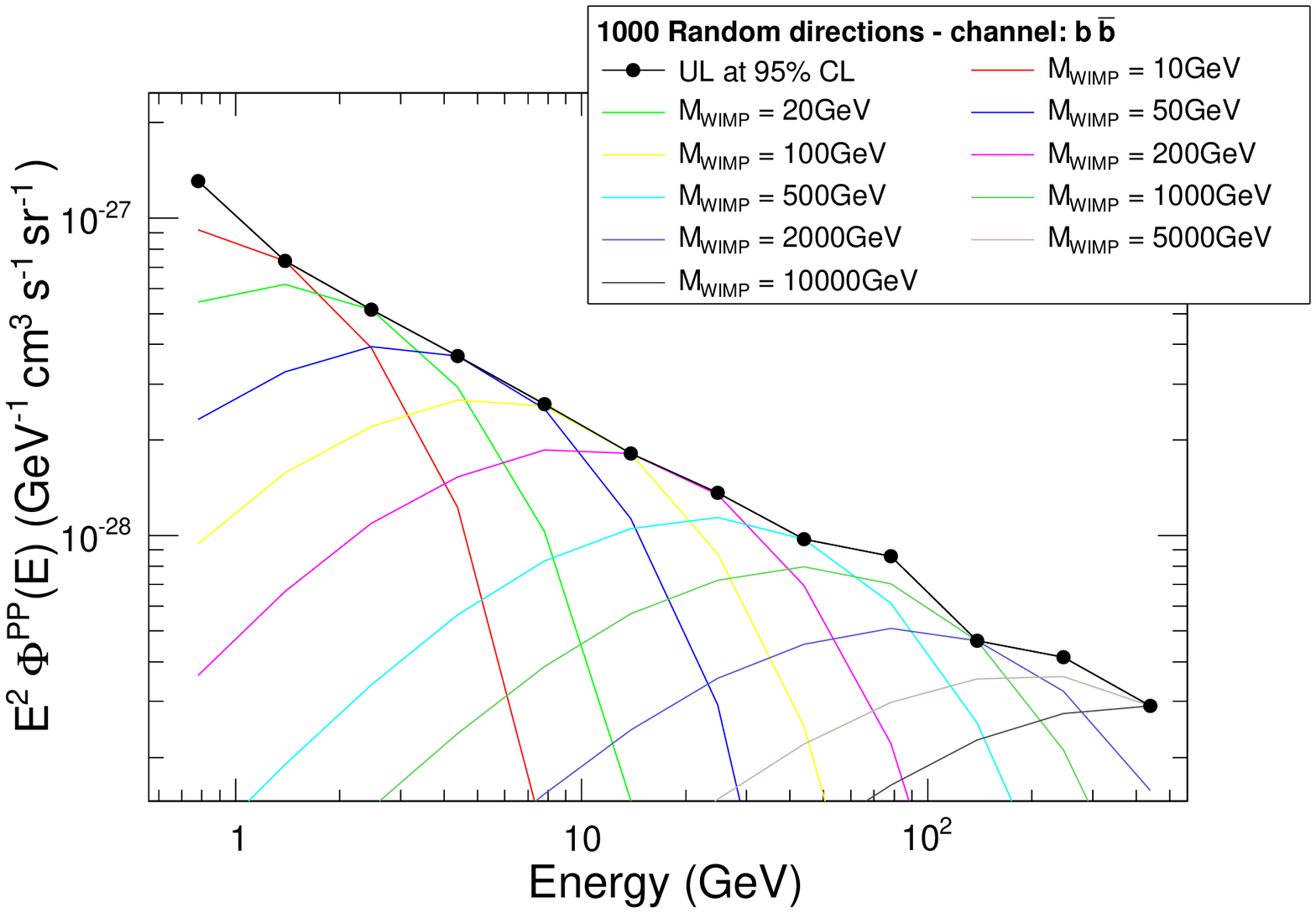}
\includegraphics[width=0.48\textwidth]{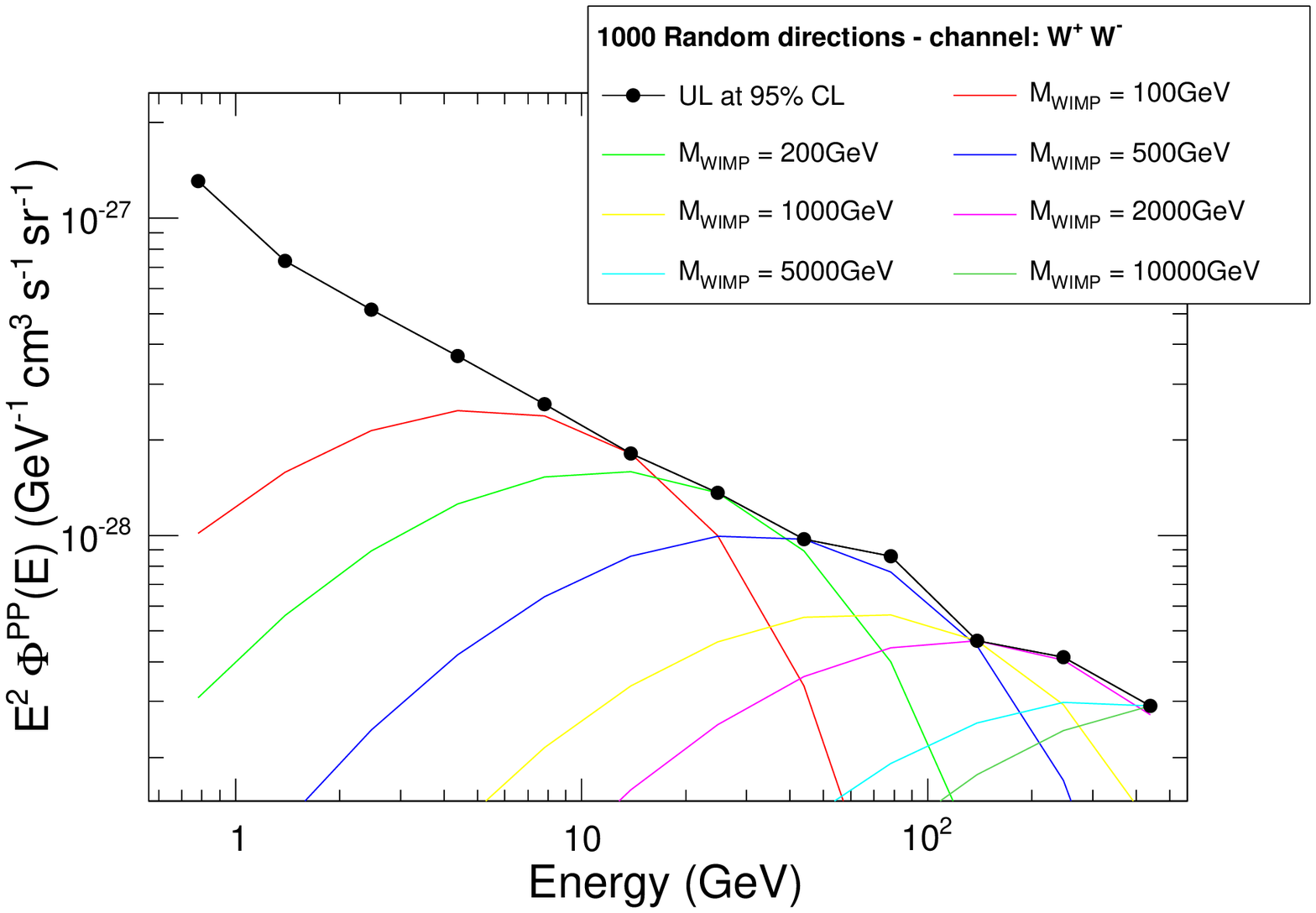}
\caption{Evaluation of the upper limits on $\langle \sigma v \rangle$ as a function of the true energy for several WIMP mass values from the Galactic halo analysis with the stacking of all the $1000$ random directions. The black lines correspond to the upper limits at $95\%$  CL on $\Phi^{PP}(E)$. The colored lines, each corresponding to a given value of the WIMP mass, indicate the maximum allowed values of $\Phi^{PP}(E)$ that do not exceed the measured upper limits. The four panels refer to the WIMP annihilations into $\mu^{+} \mu^{-}$, $\tau^{+} \tau^{-}$, $b \bar{b}$ and $W^{+} W^{-}$, respectively, as labeled.}
\label{fig:ulhalo}
\end{figure*}

\begin{figure*}
\includegraphics[width=0.48\textwidth]{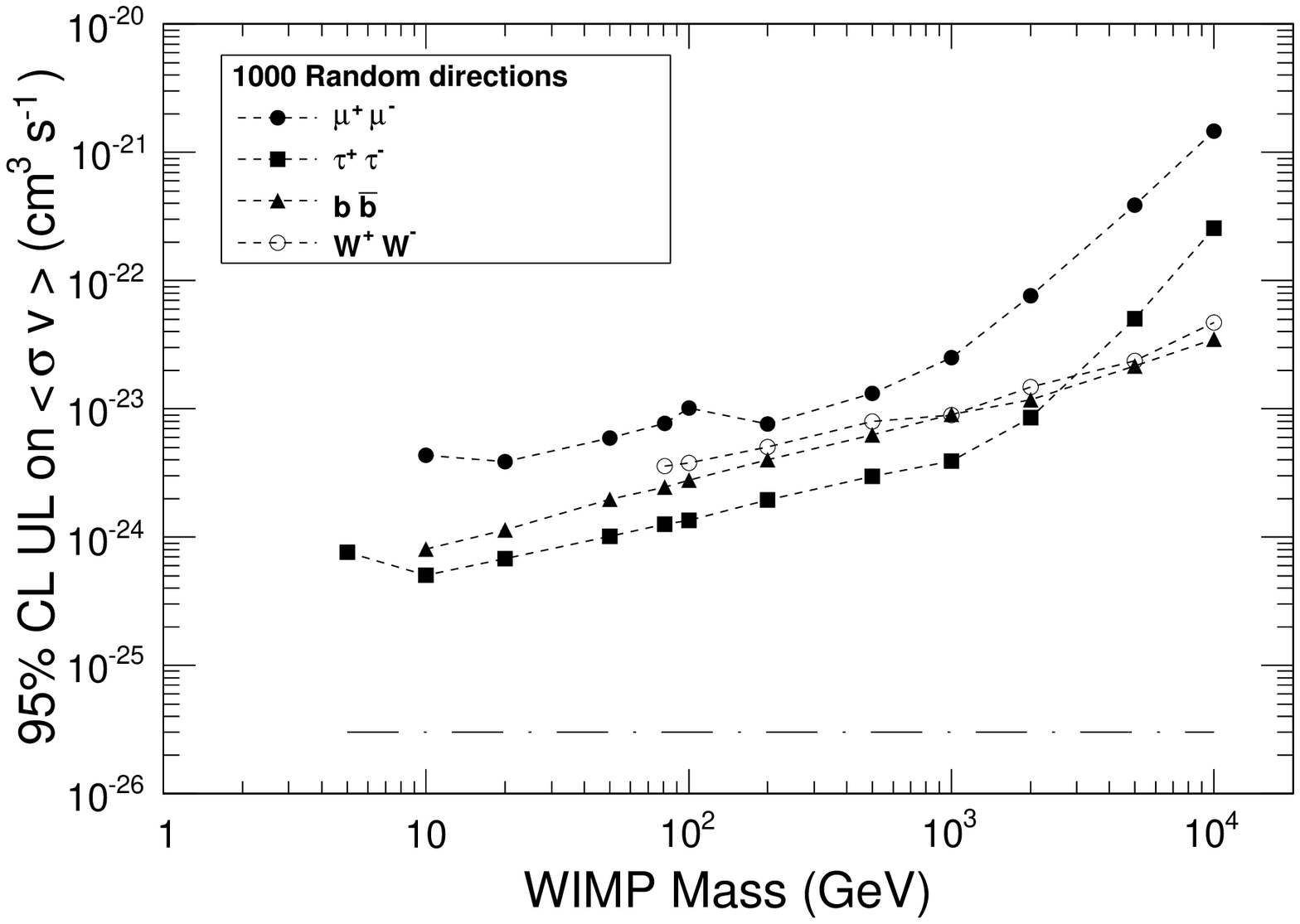}
\includegraphics[width=0.48\textwidth]{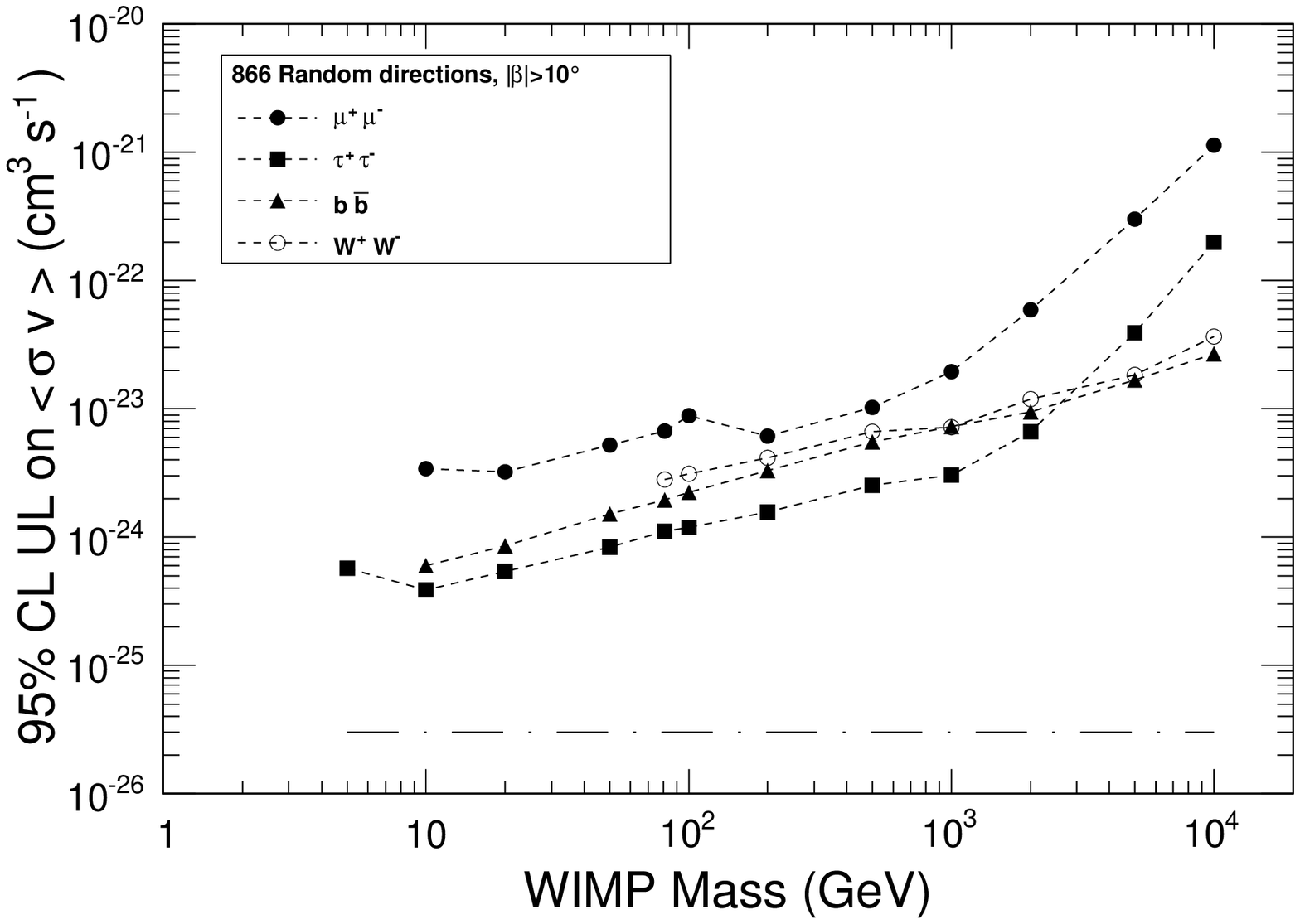}
\caption{Upper limits at $95\%$ CL on $\langle \sigma v \rangle$ as a function of the WIMP mass for the annihilation channels $\mu^{+} \mu^{-}$, $\tau^{+} \tau^{-}$, $b \bar{b}$ and $W^{+} W^{-}$. The plot shows the results obtained from the analyses of the Milky Way halo with all the $1000$ directions (left panel) and with the $866$ directions with $|\beta| > 10\degrees$ (right panel). The dashed line is the annihilation cross section of $3\times10^{-26} \units{cm^3 s^{-1}}$ in the canonical thermal relic WIMP scenario.}
\label{fig:halosigmavlimits}
\end{figure*}

The stacking analysis was performed as described in
\S\ref{sec:stacking}. In our model we assumed that all photons
originate from DM, and the upper limits on the signal counts were
evaluated using the PDF in Eq.~\ref{eq:nobkgsnglsignalpdf}. We also
performed the composite analysis of the random sources following
the procedure described in \S\ref{sec:composite}, and in this case 
the results were equivalent to those obtained from the stacking
analysis. 

Fig.~\ref{fig:4ul_comparison} shows the upper limits at $95\%$ CL on $\Phi^{PP}(E)$ for the Milky Way halo evaluated using all the $1000$ random directions and only the $866$ directions with $|\beta|>10\degrees$. The upper limits obtained from the directions with the lowest and the highest J-factors are also shown. The directions with the highest $J$-factor yield the more constraining upper limits on $\Phi^{PP}(E)$. This result is not completely obvious, since the directions with the higher $J$-factors are in the region of the Galactic Center, where a high number of photons is expected, while the directions with the lower $J$-factors are in the region of the Galactic Anti-center, where a lower number of photons is expected. The upper limits obtained from the analysis of the direction with the highest $J$-factor are also more constraining than the ones from the stacking analysis, with the exception of the high energy regime, where the constraints from the stacking analysis are tighter. This is due to the fact that in the high-energy regime the number of events in the signal regions is low, and consequently the upper limits on the fluxes decrease with increasing live time.         

Fig.~\ref{fig:ulhalo} shows the procedure used to convert the measured upper limits on $\Phi^{PP}(E)$ into upper limits on $\langle \sigma v \rangle$ in the case of the stacking analysis of all the $1000$ random sources. As in the case of the dSph analysis, we imposed the requirement that the flux values predicted from the DM annihilation scenarios must not exceed the measured upper limits in any energy bin. 

In Fig.~\ref{fig:halosigmavlimits} the upper limits at $95\%$ CL on $\langle \sigma v \rangle$ as a function of the WIMP mass are shown for WIMP annihilations into $\mu^{+} \mu^{-}$, $\tau^{+} \tau^{-}$, $b \bar{b}$ and $W^{+}W^{-}$ for the Milky Way halo. These results have been obtained from the analysis of all the $1000$ random directions (left plot) and from the analysis of the $866$ directions with $|\beta|>10\degrees$ (right plot) (see Fig.~\ref{fig:4ul_comparison}). If only the direction with the highest $J$-factor were analyzed, the upper limits on $\langle \sigma v \rangle$ in the low WIMP-mass regime would be about a factor $10$ more constraining than those obtained from the combined analyses.

\section{Discussion}
  
In this work we used the gamma-ray data collected by the Fermi LAT during its first $3$ years of operation to set constraints on the parameter $\langle \sigma v \rangle$ 
assuming DM annihilation into various channels. 

We studied a set of $10$ Milky Way dSph satellite galaxies and the Milky Way halo. The dSph galaxies were analyzed both individually and collectively, implementing dedicated stacking and composite analysis procedures. The Milky Way halo was studied by randomly sampling a set of $1000$ sky directions well-separated from all the gamma-ray sources of the 2FGL Catalog and performing a stacking analysis. The data analysis was performed using a model-independent method that allows upper limits to be set on the gamma-ray fluxes starting from the observed events using a Bayesian approach. The constraints on $\langle \sigma v \rangle$ were derived requiring that the predicted fluxes from the models must not exceed the measured ones. 

The analysis of the dSph galaxies yields upper limits on $\langle
\sigma v \rangle$ that are lower with respect to the predictions from
a canonical thermal WIMP scenario for the $\tau^{+} \tau^{-}$ and $b
\bar{b}$ final states up to masses of few tens of $\units{GeV}$. This is found in the stacking and in the composite analysis
results, but also in the results of the individual analyses of the
dSph galaxies with the highest $J$-factors. The uncertainties in the
$J$-factor calculation and on the effective area of the LAT were also
included in the present analysis, and do not affect significantly the
upper limits. Our results are consistent with recent analyses
\cite{lat_DMdwarf_paper, geringer_DMdwarf} performed using different
approaches. However, we emphasize that the upper limits on
the parameter $\langle \sigma v \rangle$ depend strongly on the values
of the $J$-factor. In particular, since no evidence of a gamma-ray flux
is observed from any dSph galaxy, the upper limits on $\Phi^{PP}(E)$ 
and consequently those on  $\langle \sigma v \rangle$ will scale with 
the $J$-factor (see Fig.~\ref{fig:ULFluxOverJ}).

For comparison, the analysis of the dSph galaxies was also performed using the P6\_V3\_DIFFUSE IRFs, 
as were used in Ref.~\cite{lat_DMdwarf_paper}, and the results were found to be in agreement with the ones already presented here,
which were obtained using the P7SOURCE\_V6 IRFs.

The analysis of the Milky Way halo, performed looking at a set of
$1000$ clean sky directions, yields upper limits on $\langle \sigma v
\rangle$ that range from $10^{-25}$ to $10^{-23} \units{cm^3~s^{-1}}$
for WIMP masses below $10 \units{TeV}$ or more for the $b \bar{b}$ and
$W^{+}W^{-}$ annihilation channels. More constraining limits on $\langle \sigma v \rangle$ 
can be obtained in the low-energy region if the analysis is limited to the direction with the highest $J$-factor. 
These limits were evaluated assuming a
NFW profile with a DM density at the solar circle of $0.3 \units{GeV cm^{-3}}$, and their
values depend on the DM mass density profile. A recent analysis suggested a revised value of the DM density at the solar circle
of $0.43 \units{GeV cm^{-3}}$~\cite{salucci}; under
this assumption the upper limits on $\langle \sigma v \rangle$ would
improve by a factor of $2$.  Nevertheless we note
that the sky directions used for the present analysis are sufficiently far
away from the Galactic Center that the $J$-factors
evaluated with different DM density profiles would likely yield similar
results (see Ref.~\cite{berg}). The current results are consistent
with the results obtained by Ref.~\cite{papucci}, where an
analysis of the all-sky Fermi LAT data was performed with a
different model-independent technique. The Milky Way halo can also be
studied following a different approach, in which a model is assumed
for the Galactic diffuse gamma-ray emission and a fit of the DM signal
together with the diffuse component is performed
(e.g., see~\cite{cuoco}). Our results are also consistent with those
obtained from that analysis. 

Both from the limits on the dSph galaxies
(Fig.~\ref{fig:sigmavlimits}) and those on the Milky Way halo
(Fig.~\ref{fig:halosigmavlimits}), it is possible to restrict the range
of allowed WIMP masses assuming the standard thermal relic scenario. 
We also note that the DMFIT package may underestimate the 
gamma-ray fluxes for WIMP masses above $1\units{TeV}$, since
it does not include radiative electroweak corrections;
hence, the limits on $\langle \sigma v \rangle$ for large masses can be viewed as
more conservative than those in the low mass region.

\section{Conclusions}

We developed a model-independent approach to set upper limits on the energy spectra of both individual and multiple gamma-ray sources using the data collected by the Fermi LAT\@. In this paper we presented the results obtained from the application of this technique to the study of a set of dSph galaxies and to the study of the Milky Way halo. These results were used to derive constraints on DM annihilation cross sections into different channels. We emphasize that the analysis techniques presented in this paper are general, and are suitable for applications where the study of a class of sources, even faint sources, with common features have to be studied.

The data analysis technique illustrated in the present paper allows us to set robust upper limits on the energy spectra of candidate gamma-ray sources. The upper limits on the gamma-ray fluxes are in fact derived starting from the data, without assuming any model for the background and for the source spectral shapes. The signal is evaluated selecting events from a cone centered on the source position, while the background is evaluated selecting events in a region close to the source under investigation. Both signal and background fluctuations are described in the framework of Poisson statistics, and the upper limits on the signal counts, and consequently on the flux, are computed following the Bayesian approach. A stacking analysis and a composite analysis procedure have also been developed to perform the collective study of multiple candidate sources with common features. 

The analysis methods presented in this paper can also be applied when several measurements of a given physical quantity, each one resulting into a confidence interval, have to be combined into a unique confidence interval taking all the results into account. The systematic uncertainties can also be incorporated in the analysis by introducing proper nuisance parameters in the probability distribution functions.

\section*{Acknowledgements}

The Fermi LAT Collaboration acknowledges generous ongoing support
from a number of agencies and institutes that have supported both the
development and the operation of the LAT as well as scientific data analysis.
These include the National Aeronautics and Space Administration and the
Department of Energy in the United States, the Commissariat \`a l'Energie Atomique
and the Centre National de la Recherche Scientifique / Institut National de Physique
Nucl\'eaire et de Physique des Particules in France, the Agenzia Spaziale Italiana
and the Istituto Nazionale di Fisica Nucleare in Italy, the Ministry of Education,
Culture, Sports, Science and Technology (MEXT), High Energy Accelerator Research
Organization (KEK) and Japan Aerospace Exploration Agency (JAXA) in Japan, and
the K.~A.~Wallenberg Foundation, the Swedish Research Council and the
Swedish National Space Board in Sweden.

Additional support for science analysis during the operations phase is gratefully
acknowledged from the Istituto Nazionale di Astrofisica in Italy and the Centre National d'\'Etudes Spatiales in France.

Some of the results in this paper have been derived using the HEALPix package~\cite{healpix}.

The authors thank James Chiang and Jennifer Siegal-Gaskins for their valuable
contributions during the preparation of the manuscript.

\appendix

\section{Derivation of Equation~\ref{eq:cdef1}}
\label{app:cdef}

When performing the stacking analysis of a set of multiple sources, the
counts from the individual signal and background regions are
added. The total counts in the signal and background region are then
given by:

\begin{eqnarray}
n  =  \sum_{i} n_{i} \\
m  =  \sum_{i} m_{i}
\end{eqnarray}
where $n_{i}$ and $m_{i}$ are respectively the counts in the signal and in the
background region of the $i$-th source.

According to the assumptions in \S\ref{sec:individualanalysis}, $n_i$
and $m_i$ are both Poissonian with expectation values
$s_{i}+c_{i}b_{i}$ and $b_{i}$ (in the following we shall use the
notation $n_{i} \sim \mathcal{P}(s_{i}+c_{i}b_{i})$ and $m_{i} \sim
\mathcal{P}(b_{i})$). Hence, from the definitions of $n$ and $m$ it
follows that:

\begin{eqnarray}
n \sim \mathcal{P}\left( \sum_{i} (s_{i}+c_{i}b_{i}) \right) \\
m \sim \mathcal{P}\left( \sum_{i} b_{i} \right). 
\end{eqnarray}

In the stacking analysis the true values of the signal and of the background counts are defined as:

\begin{eqnarray}
s  =  \sum_{i} s_{i} \\
b  =  \sum_{i} b_{i}.
\end{eqnarray}
 
This definition automatically implies that $m \sim \mathcal{P}(b)$. 
However, to ensure that $n \sim \mathcal{P}(s+cb)$, the coefficient $c$ must be
defined as:

\begin{equation}
\label{eq:cdef2}
c = \frac{\sum_{i} c_{i} b_{i}}{\sum_{i} b_{i}}.
\end{equation}
The values of $b_i$ are not known, but they can be replaced with their
best estimators $b_i^{*} = m_i+1$~\cite{dagostini}. In this way, Eq.~\ref{eq:cdef2} reduces to Eq.~\ref{eq:cdef1}.

\end{document}